\documentclass[
amsmath,amssymb,
article,
aps,twocolumn,float, 
nofootinbib]{revtex4}
\usepackage{tikz,color,cancel,acronym}
\usepackage[colorlinks,linkcolor=blue,citecolor=blue,urlcolor=blue ]{hyperref}
\usepackage{amsmath}
\usepackage{multirow}
\usepackage[papersize={10in,11in}]{geometry}
\allowdisplaybreaks[4] 

\usepackage{graphicx}
\usepackage{changes}
\usepackage[capitalize]{cleveref}
\newcommand{\beq}{\begin{equation}}
\newcommand{\beqa}{\begin{eqnarray}}
\newcommand{\eeq}{\end{equation}}
\newcommand{\eeqa}{\end{eqnarray}}

\newcommand{\TRC}{School of Physics and Astronomy,  Sun Yat-sen University (Zhuhai Campus), Zhuhai 519082, China}

\usepackage{amsmath,graphicx,mathrsfs,theorem}
\usepackage{subfigure}
\usepackage{float}

\newcommand{\tmmathbf}[1]{\ensuremath{\boldsymbol{#1}}}
\newcommand{\tmop}[1]{\ensuremath{\operatorname{#1}}}

{\theorembodyfont{\rmfamily}\newtheorem{remark}{Remark}}

\begin{document}

\title{Detecting anisotropies of the stochastic gravitational wave background with TianQin}

\author{Kun Zhou}

\thanks{Corresponding author: \href{mailto:zhouk29@mail2.sysu.edu.cn}{zhouk29@mail2.sysu.edu.cn}}

\author{Jun Cheng}


\author{Liangliang Ren}



\affiliation{\TRC}

\begin{abstract}

The investigation of the anisotropy of the stochastic gravitational wave background (SGWB) using the TianQin detector plays a crucial role in studying the early universe and astrophysics. In this work, we examine the response of the $AET$ channel of the TianQin Time Delay Interferometry (TDI) to the anisotropy of the SGWB. We calculate the corresponding angular sensitivity curves and find that TianQin is capable of detecting the anisotropy of the SGWB, with an angular sensitivity reaching $10^{-10}$ for quadrupoles. Due to the fixed $z$-axis of TianQin pointing towards J0806, its overlap reduction functions (ORFs) exhibit specific symmetries, enabling the resolution of different multipole moments $\ell m$. The detection sensitivity is optimal for the $(2, 0)$ mode, with a sensitivity reaching $10^{-10}$. Using the Fisher matrix approach, we estimate the parameters and find that in the power-law spectrum model, higher logarithmic amplitudes lead to more effective reconstruction of the spectral index for all multipole moments. Under the optimal scenario with a signal amplitude of $\Omega_{\mathrm{GW}} (f = f_{\mathrm{c}}) h^2 = 10^{-9}$, the spectral indices can be reconstructed with uncertainties of $10^{-3}$, $10$, and $10^{-3}$ for $\ell = 0$, $1$, and $2$ multipole moments, respectively. For the cases of $(\ell, m) = (0, 0)$, $(1, 1)$, $(2, 0)$, and $(2, 2)$, the spectral indices can be reconstructed with uncertainties of $10^{-3}$, $10$, $10^{-3}$, and $10$, respectively.

\end{abstract}

\maketitle

\acrodef{GW}{gravitational wave}
\acrodef{GCB}{Galactic ultra compact binaries}
\acrodef{SGWB}{Stochastic gravitational wave background}
\acrodef{SNR}{signal-to-noise ratio}
\acrodef{DWD}{double white dwarf}
\acrodef{MBHB}{massive black hole binary}
\acrodefplural{MBHB}[MBHBs]{massive black hole binaries}
\acrodef{SBBH}{stellar-mass black hole binary}
\acrodefplural{SBBH}[SBBHs]{stellar-mass black hole binaries}
\acrodef{EMRI}{extreme mass ratio inspiral}
\acrodef{PTA}{pulsar timing array}
\acrodef{CE}{cosmic explorer}
\acrodef{ET}{Einstein telescope}
\acrodef{LISA}{laser interferometer space antenna}
\acrodef{O1}{first observing run}
\acrodef{O3}{third observing run}
\acrodef{LVK}{the LIGO Scientific Collaboration, the Virgo Collaboration and the KAGRA Collaboration}
\acrodef{PCA}{principal component analysis}
\acrodef{CBC}{compact binary coalescences}
\acrodef{TDI}{time delay interferometry}
\acrodef{PSD}{power spectral density}
\acrodef{SNR}{signal-to-noise ratio}
\acrodef{PDF}{probability distribution function}
\acrodef{MCMC}{Markov Chain Monte Carlo}
\acrodef{NS}{nested sampling}


\

\section{INTRODUCTION}

The stochastic gravitational wave background (SGWB) is formed by the superposition of gravitational wave signals generated by a large number of weak, independent, and indistinguishable sources \cite{Cornish:2015pda}. The SGWB commonly studied by researchers is assumed to be stationary, Gaussian, non-polarized, and isotropic. However, in actual observations, the SGWB exhibits anisotropy.

From a classification based on the origin of the SGWB, it can be categorized into cosmological origin and astrophysical origin \cite{Regimbau:2011rp, Maggiore:2018sht, Caprini:2018mtu}, and both may exhibit anisotropy. Due to the stochastic nature of the SGWB, whether it originates from cosmological sources or astrophysical sources (referred to as astrophysical stochastic gravitational-wave background, ASGWB), isotropic scenarios can be characterized by isotropic energy densities, while anisotropic scenarios can be described using spatial angular power spectra. Theoretical analyses of the anisotropic origin of the SGWB can be conducted using the Boltzmann equation approach \cite{Contaldi:2016koz, Bartolo:2019oiq, Bartolo:2019yeu, Cusin:2018avf, Pitrou:2019rjz}. This approach enables the distinction of the effects of anisotropy during GW generation and the anisotropy arising from GW propagation in an inhomogeneous universe. The anisotropy of the SGWB contains a wealth of information relevant to cosmology and astrophysics, and its detection holds significant physical implications.

Regarding the anisotropy of the cosmological origin of the SGWB, it can also originate from early universe phenomena such as inflation, primordial black hole phase transitions, and the formation of topological defects \cite{Regimbau:2011rp, Maggiore:2018sht, Caprini:2018mtu}. For example, expected anisotropy exists in scenarios like preheating in scale-invariant models \cite{Bethke:2013aba, Bethke:2013vca}, where the SGWB in this model typically peaks at higher frequencies beyond the detection range of experiments like TianQin and LISA \cite{Figueroa:2017vfa}. Regarding phase transitions, if only considering SGWB produced by cosmological adiabatic perturbations, the corresponding energy density is expected to be small \cite{Geller:2018mwu, Kumar:2021ffi}. For GWs generated by topological defects, studies in \cite{Jenkins:2018nty, Kuroyanagi:2016ugi, Olmez:2011cg} have calculated the anisotropy caused by Nambu-Goto cosmic string loop networks. The results show that the angular power spectrum $C_{\ell}$, which characterizes the multipole decomposition of the SGWB spectrum, depends on the model of the loop network. The anisotropy is generated by local Poisson fluctuations in the number of loops, and the resulting angular power spectrum is white in frequency, meaning it is constant for a given $\ell$, without the need to consider a specific loop distribution \cite{Jenkins:2018nty}.

For the anisotropy of the astrophysical origin of the ASGWB, it is generated by the incoherent superposition of signals emitted by a large number of resolvable and unresolved astrophysical sources. Various astrophysical sources contribute to the ASGWB in different frequency bands, including supermassive black hole binaries \cite{Kelley:2017lek}, rotating neutron stars \cite{Surace:2015ppq, Talukder:2014eba, Lasky:2013jfa}, stellar core collapse \cite{Crocker:2017agi, Crocker:2015taa}, triple systems of stars \cite{Regimbau:2011rp}, and the merger of stellar-mass black holes or binary neutron stars \cite{2016PhRvL.116m1102A, Regimbau:2016ike, Mandic:2016lcn, Bavera:2021wmw, Dvorkin:2016okx, Nakazato:2016nkj, Dvorkin:2016wac, Evangelista:2014oba}, among others. Recent studies have provided analytical derivations of the anisotropic energy density of the SGWB \cite{Cusin:2017fwz, Contaldi:2016koz, Cusin:2017mjm, Cusin:2018avf, Pitrou:2019rjz, Bertacca:2019fnt} and made predictions for the angular power spectrum of the energy density in the Hertz and millihertz frequency bands \cite{Cusin:2018rsq, Jenkins:2018uac, Jenkins:2018kxc, Cusin:2019jpv, Cusin:2019jhg, Bertacca:2019fnt}. Particularly, it has been found that the anisotropy of the astrophysical SGWB in the Hertz band strongly depends on the galaxy and subgalactic astrophysical models \cite{Cusin:2019jhg}. The anisotropy exhibits a range of variabilities, depending on the underlying astrophysical models of stellar formation, mass distribution, collapse, and the cosmological perturbative effects considered.

Up to now, there have been studies on the detection of anisotropy in the SGWB. For ground-based detectors, the LIGO/Virgo collaboration provided upper limits on the detectable energy density in \cite{LIGOScientific:2016jlg}, while \cite{LIGOScientific:2016nwa,KAGRA:2021mth} presented constraints on the directional energy flux and energy density. On the other hand, the research by NANOGrav \cite{NANOGrav:2020bcs} suggests the possibility of detecting an SGWB signal in the nanohertz range.
For space-based detectors, early studies such as \cite{Peterseim:1997ic, Cutler:1997ta, Moore:1999zw} provided the angular resolution for space gravitational wave detectors like LISA. Research in \cite{Ungarelli:2001xu, Seto:2004np, Kudoh:2004he, Taruya:2005yf, Taruya:2006kqa} investigated the ability of LISA to detect the anisotropy in the SGWB. In 2022, the LISA team's work \cite{LISACosmologyWorkingGroup:2022kbp} analyzed and computed the statistical overlap reduction functions (ORFs) of LISA for different multipoles in the context of isotropic SGWB. They also demonstrated several symmetries of these ORFs.
These studies provide theoretical foundations and detection methods for probing the anisotropy in the SGWB. However, actual detection requires further observational data and technological advancements.

The detection of the SGWB is one of the primary scientific objectives of the TianQin space gravitational wave detector \cite{2017arXiv170200786A}. Regarding isotropic SGWB, a study by Liang et al. in 2021 \cite{Liang:2021bde} investigated the capability of TianQin to detect the SGWB.
In terms of data processing for TianQin, a work by Cheng et al. in 2022 \cite{Cheng:2022vct} demonstrated the presence of imaginary components in the noise cross-spectra, which play an important role in breaking the degeneracy of position noise in the common laser-link configuration.
These studies contribute to understanding the detection capabilities of TianQin for the SGWB. However, further research, observations, and technological advancements are needed to fully explore and exploit the potential of TianQin and other gravitational wave detectors.

This study primarily investigates the capability of TianQin to detect the anisotropy of the SGWB. In Section \ref{s:expectationvalues}, a concise overview and representation of the definition of the SGWB are provided. Section \ref{s:DASGWB_TQ} employs TianQin's time-delay interferometry $AET$ channel to calculate the overlap reduction functions (ORFs) of TianQin's auto-correlation channels ($AA$, $EE$, and $TT$) and cross-correlation channels ($AE$ and $AT$), along with analytical expressions at low frequencies. Moreover, the unique symmetries of TianQin and its potential to discern different $\ell m$ modes are demonstrated. The sensitivity of TianQin to multipole moments of the SGWB with varying $\ell$ and $\ell m$ values is constructed from the perspective of signal-to-noise ratio.
In Section \ref{s:Fisher_Est}, parameter estimations for the SGWB, satisfying power-law spectra in both $\ell$-order and $\ell m$-order, are calculated using the Fisher estimation method. Finally, Section \ref{s:Sum_Disc} provides a summary and outlook for this work.

\section{Stochastic Gravitational Wave Background and its Anisotropy}\label{s:expectationvalues}

The SGWB is formed by the superposition of weak gravitational wave signals emitted by a large number of unresolved sources. For a gravitational wave, it is possible to expand it using a chosen set of polarization tensors. In the following, we will provide a brief description of the linear polarization expansion and circular polarization expansion. We start with the most common representation using the $+,\times$ polarization modes for a gravitational wave expressed as follows:
\begin{equation}
  h_{ab}  (\tmmathbf{x}, t) = \int_{- \infty}^{+ \infty} \mathrm{d}f \int \mathrm{d}
  \Omega_{\hat{k}} \mathrm{e}^{- 2 \pi if (t - \hat{k} \cdot \tmmathbf{x})}
  \sum_A \tilde{h}_A  (f, \hat{k}) e_{ab}^A (\hat{k}) \hspace{0.27em},
  \label{eq:GW-decoab}
\end{equation}

The polarization tensors $\mathrm{e}^A_{ab} (\hat{k})$ ($A = +, \times$) are defined as
\begin{equation}
\begin{array}{l}
e_{ab}^+ (\hat{k}) = \frac{\hat{l}_a \hat{l}_b - \hat{m}_a
\hat{m}_b}{\sqrt{2}},\
e_{ab}^\times (\hat{k}) = \frac{\hat{l}_a \hat{m}_b + \hat{m}_a
\hat{l}_b}{\sqrt{2}},
\end{array} \label{eq:eabjiaca}
\end{equation}
where $\hat{n}$, $\hat{l}$, and $\hat{m}$ are the unit orthogonal basis vectors in the spherical coordinate system, forming a right-handed coordinate system. Here, $\hat{n} = -\hat{k}$, where $\hat{k}$ is the unit wave vector of the gravitational wave. The transformation between $\hat{n}$, $\hat{l}$, $\hat{m}$, and the Cartesian coordinate system is given by
\begin{equation}
\begin{array}{l}
\hat{n} = \sin \theta \cos \phi \hat{x} + \sin \theta \sin \phi \hat{y} +
\cos \theta \hat{z} \equiv \hat{r},\\
\hat{l} = \cos \theta \cos \phi \hat{x} + \cos \theta \sin \phi \hat{y} -
\sin \theta \hat{z} \equiv \hat{\theta},\\
\hat{m} = - \sin \phi \hat{x} + \cos \phi \hat{y} \equiv \hat{\phi}.
\end{array} \label{e:nlm_def}
\end{equation}
In this definition, the polarization tensors are normalized as
\begin{equation}
\mathrm{e}^A_{ab} (\hat{k}) \mathrm{e}^{A', ab} (\hat{k}) = \delta^{A A'},
\end{equation}
and the gravitational wave strain tensor $h{ab} (f, \hat{k})$ can be expressed as
\begin{equation}
h_{ab} (f, \hat{k}) = h_+ (f, \hat{k}) e^+_{ab} (\hat{k}) + h_{\times}
(f, \hat{k}) e^{\times}_{ab} (\hat{k}),
\end{equation}
where $h_+ (f, \hat{k})$ and $h_{\times} (f, \hat{k})$ are the amplitudes of the $+$ and $\times$ polarization modes, respectively.

To construct the right-handed basis vectors as described above, it is sufficient for $\hat{l}$ and $\hat{m}$ to be orthogonal to $\hat{n}$. In other words, $\hat{l}$ and $\hat{m}$ can be rotated around $\hat{n}$ by an arbitrary angle $\psi$, given by
\begin{equation}
\begin{array}{l}
\hat{p} \equiv \cos \psi \hat{l} + \sin \psi \hat{m},\\
\hat{q} \equiv - \sin \psi \hat{l} + \cos \psi \hat{m}.
 \end{array}
\end{equation}
Using the previous equations, we can derive the new polarization tensors as follows:
\begin{equation}
\begin{array}{l}
\epsilon_{ab}^+ (\hat{n}, \psi) \equiv \hat{p}_a \hat{p}_b - \hat{q}_a
\hat{q}_b,\\
\epsilon_{ab}^\times (\hat{n}, \psi) \equiv \hat{p}_a \hat{q}_b +
\hat{q}_a \hat{p}_b .
\end{array}
\end{equation}
It is straightforward to show their relationship with \eqref{eq:eabjiaca}:
\begin{equation}
\begin{array}{c}
\epsilon_{ab}^+ (\hat{n}, \psi) = \cos 2 \psi e_{ab}^+ (\hat{n}) + \sin 2
\psi e_{ab}^\times (\hat{n}),\\
\epsilon_{ab}^\times (\hat{n}, \psi) = - \sin 2 \psi e_{ab}^+ (\hat{n})
+ \cos 2 \psi e_{ab}^\times (\hat{n}).
\end{array}
\end{equation}
It is important to note that since $h_{ab} (\mathbf{x}, t)$ is a real function, the following relation holds:
\begin{equation}
\tilde{h}_A (-f, \hat{k}) = \tilde{h}_A^* (f, \hat{k}),
\end{equation}

From a statistical perspective, understanding how gravitational wave signals contribute to the formation of a SGWB is quite straightforward. The main approach involves characterizing the statistical properties of the gravitational wave through the probability distribution of metric perturbations or its different moments.
We can describe the statistical properties of the SGWB using the following quantities:
\begin{equation}
\begin{array}{l}
\langle h_{ab} (t, \vec{x}) \rangle, \quad \langle h_{ab} (t, \vec{x})
h_{cd} (t', \vec{x}') \rangle, \quad \\ \langle h_{ab} (t, \vec{x}) h_{cd} (t',
\vec{x}') h_{ef} (t'', \vec{x}'') \rangle, \quad \cdots
\end{array}
\end{equation}
Without loss of generality, we typically assume that the expectation value of the SGWB is zero:
\begin{equation}
\langle h_{ab} (t, \vec{x}) \rangle = 0 \quad \Leftrightarrow \quad \langle
h_A (f, \hat{n}) \rangle = 0,
\end{equation}
where $A \equiv {+, \times}$.

It is also commonly assumed that the gravitational wave background is in a stationary state. Stationarity implies that all statistical quantities constructed from the metric perturbations at different time points, such as $t$ and $t'$, depend only on the time difference $t - t'$ and not on the specific values of $t$ or $t'$. This assumption is reasonable considering that the typical duration of observations is on the order of 10 years, which is approximately nine orders of magnitude smaller than the age of the universe. Therefore, it is unlikely that the statistical properties of the SGWB change significantly on the timescale of the observation period.

For a Gaussian background, it is sufficient to consider the second-order moments because all higher-order moments either vanish or can be expressed in terms of the second-order moments. In this case, the statistical properties of the background are fully characterized by the mean and the covariance matrix. However, for a non-Gaussian background, it becomes necessary to consider higher-order moments. In addition to the mean and covariance, third-order moments (skewness) and even higher-order moments (kurtosis, etc.) need to be computed to fully describe the statistical properties of the non-Gaussian background. These higher-order moments capture the departure from Gaussianity and provide additional information about the underlying distribution of the gravitational wave signals.

In addition to the assumption of stationarity, the specific form of each moment typically depends on the origin of the background. For example, the cosmic background, which is formed by the superposition of a large number of independent gravitational wave signals from the early universe, is expected to be Gaussian due to the central limit theorem. It is also expected to exhibit isotropic distribution on the sky.
On the other hand, the foreground generated by the unresolved binary neutron stars in the frequency range of the TianQin detector ($10^{-4}$ Hz to $10^{-0}$ Hz) is also considered to be Gaussian and stationary due to astrophysical constraints. However, this foreground will have an anisotropic distribution following the spatial distribution of the Milky Way galaxy. Additionally, the antenna pattern of the TianQin constellation varies with its orbital motion, leading to modulations in the detector's output signal due to the SGWB.
Therefore, different origins of the SGWB will result in different statistical distributions and modulation effects in the detector's output data. All of these factors need to be considered when formulating data analysis strategies for studying the gravitational wave signals.

In the context of a stationary, Gaussian, and non-polarized random gravitational wave background, taking into account its anisotropy, the second-order moments can be described as follows:
\begin{equation}
\langle h_A (f, \hat{k}) h_{A'}^* (f', \hat{k}') \rangle = \frac{1}{4} \mathcal{P} (f, \hat{k}) \delta (f - f') \delta_{AA'} \delta^2 (\hat{k}, \hat{k}'),
\end{equation}
where $\mathcal{P} (f, \hat{k})$ is the angular power spectrum density, which describes the distribution of power of gravitational waves with frequency $f$ across the sky. The factor $\delta (f - f')$ arises from the assumption of stationarity. The factor $\delta_{AA'}$ accounts for the statistical independence of the polarization modes of gravitational waves and assumes no preferred polarization component. The factor $\delta^2 (\hat{k}, \hat{k}')$ is due to the assumption of spatial homogeneity and isotropy.
If the angular power spectrum density $\mathcal{P} (f, \hat{n})$ is expanded in terms of spherical harmonics, for a non-polarized and anisotropic random gravitational wave background with intensity $I$, it can be defined as:
\begin{eqnarray}
 \langle \tilde{h}_A (f, \hat{k}) \tilde{h}{A'} (f', \hat{k}') \rangle &=& \delta (f - f') \frac{\delta^{(2)} (\hat{k} - \hat{k}')}{4 \pi} \delta_{AA'} \nonumber\\
  &  &\sum_{\ell m} \tilde{I}_{\ell m} (f) \tilde{Y}_{\ell m} (\hat{k}),\label{eq:aver-hh-aa}   
\end{eqnarray}
where $\tilde{Y}_{\ell m} (\hat{k}) \equiv \sqrt{4 \pi} Y_{\ell m} (\hat{k})$ and $Y_{\ell m} (\hat{k})$ are the spherical harmonics normalized such that $\tilde{Y}_{00} (\hat{k}) = 1$. Consequently, measurements of the anisotropy can be translated into measurements of $\tilde{I}_{\ell m} (f)$.

In fact, for an isotropic background, the $(0,0)$ term of equation \eqref{eq:aver-hh-aa} corresponds to the second-order moment, which is entirely determined by the power spectral density $S_h(f)$:
\begin{equation}
S_h(f) = \int \mathrm{d}^2\Omega_{\hat{n}} \mathcal{P}(f,\hat{n})
\end{equation}
The gravitational wave strain power spectral density $S_h(f)$ is directly related to the differential energy density spectrum of gravitational waves:
\begin{equation}
S_h(f) = \frac{3H_0^2}{2\pi^2}\frac{\Omega_{\rm GW}(f)}{f^3}
\label{e:Sh-Omega_g}
\end{equation}
This correlation can be derived in detail from the reference provided \cite{Allen:1997ad}, where $\Omega_{\rm GW}(f) = \frac{1}{\rho_c} \frac{\mathrm{d}\rho_{\rm GW}}{\mathrm{d}\ln f}$ represents the energy density of gravitational waves contained within the frequency interval $f$ to $f + \mathrm{d}f$. Here, $\rho_c \equiv \frac{3c^2 H_0^2}{8\pi G}$ is the critical energy density of the current universe. Therefore, the total energy density in gravitational waves normalized by the critical energy density is given by:
\begin{equation}
\Omega_{\rm GW} = \int_{f=0}^{f_{\rm max}} \mathrm{d}(\ln f) \Omega_{\rm GW}(f)
\end{equation}
Here, $f_{\rm max}$ represents some maximum cutoff frequency (e.g., related to the Planck scale) beyond which our understanding of gravity breaks down. For example, $\Omega_{\rm GW}$ can be compared with the total energy density of other components such as baryons ($\Omega_b$) and dark energy ($\Omega_\Lambda$). The presence of the Hubble constant, $H_0$, in $\rho_c$ sometimes leads to the notation $H_0 = h_0 \times 100 , \text{km s}^{-1} \text{Mpc}^{-1}$, where $h_0$ is dimensionless. In this case, the factor of $h_0^2$ is absorbed into $\Omega_{\rm GW}(f)$, making $h_0^2\Omega_{\rm GW}(f)$ independent of the value of the Hubble constant. The specific functional form of $\Omega_{\rm GW}(f)$ depends on the origin of the background, which will be clarified below. As for $h_0$, the latest precision measurements from the Planck satellite \cite{Planck:2015fie} suggest $h_0 = 0.68$, and this value will be adopted in this context.

If we consider the anisotropy caused by specific sources, we need to relate $\tilde{I}_{\ell m}(f)$ to the angular power spectrum of the source. In general, this relationship can be expressed as \cite{LISACosmologyWorkingGroup:2022kbp}:
\begin{equation}
\tilde{I}_{\ell m}(f) = \frac{1}{\sqrt{4\pi}}\frac{3H_0^2}{4\pi^2}\frac{\Omega_{\rm GW}(f)}{f^3}\delta_{\rm GW,\ell m}
\label{I-to-delta}
\end{equation}
Here, $\delta_{\rm GW,\ell m}$ represents the spherical harmonic component of the density contrast:
\begin{equation}
\delta_{\rm GW}(f,\hat{k}) = \sum_{\ell}\sum_{m=-\ell}^{\ell}\delta_{\rm GW,\ell m}(f)Y_{\ell m}(\hat{k})
\label{e:dec}
\end{equation}
In general, we have:
\begin{equation}
\langle\delta_{\rm GW,\ell m}\delta_{\rm GW,\ell'm'}^*\rangle = C_{\ell m}^{\rm GW}(f)\delta_{\ell\ell'}\delta_{mm'}
\label{Cellm-def}
\end{equation}
Furthermore, assuming that for the same $\ell$ but different $m$, the values of $C_{\ell m}^{\rm GW}(f)$ are the same, we have:
\begin{equation}
\langle\delta_{\rm GW,\ell m}\delta_{\rm GW,\ell'm'}^*\rangle = C_{\ell}^{\rm GW}(f)\delta_{\ell\ell'}\delta_{mm'}
\label{Cell-def}
\end{equation}
$C_{\ell m}^{\rm GW}(\eta_0,q)$ and $C_{\ell}^{\rm GW}(\eta_0,q)$ are known as the angular power spectra of the source. Detailed discussions on them can be found in the reference provided \cite{LISACosmologyWorkingGroup:2022kbp}.

\section{Detection of Anisotropies in the SGWB by TianQin}\label{s:DASGWB_TQ}

When considering the detection of actual random gravitational wave backgrounds, it is crucial to take noise into account. One viable approach is to employ Time Delay Interferometry (TDI){\cite{Tinto:2001ii,Tinto:2002de,Hogan:2001jn,Tinto:2004wu,Christensen:1992wi,Adams:2010vc,Romano:2016dpx}}, as discussed in Section \ref{sec:TDI_Noise}, to mitigate the effects of laser frequency noise in the detectors. By constructing the response function for the $AET$ channel of the TianQin mission, the overlap reduction functions (ORFs) can be computed. Subsequently, the sensitivity can be derived from a signal-to-noise ratio perspective, enabling us to determine which anisotropies can be detected by TianQin and what the corresponding detection limits are.

To construct the response function for the $AET$ channel, it is necessary to first construct the response function for the $XYZ$ channels and then transform them into the $AET$ channel. For space-based detectors like TianQin, it is common to simplify their configuration as an equilateral triangle. With these assumptions in place, the computation of the response functions can be carried out.

\subsection{TianQin's ORFs for Anisotropies in the SGWB}\label{sec:tq-anni-renspons}

For the TianQin constellation, it has an orbital radius of 100,000 kilometers. The three satellites composing the constellation continuously orbit around the Earth, while their orbital plane remains fixed, pointing towards J0806 \cite{TianQin:2015yph}. This means that the response function of TianQin evolves with time in the coordinate system of the galactic center and the ecliptic.
In practical observations, the data obtained can be regarded as the modulation of the SGWB by the time-dependent response function of the detector in the frequency domain.

\begin{figure}[H]
\begin{center}
{\resizebox{200pt}{140pt}{\includegraphics{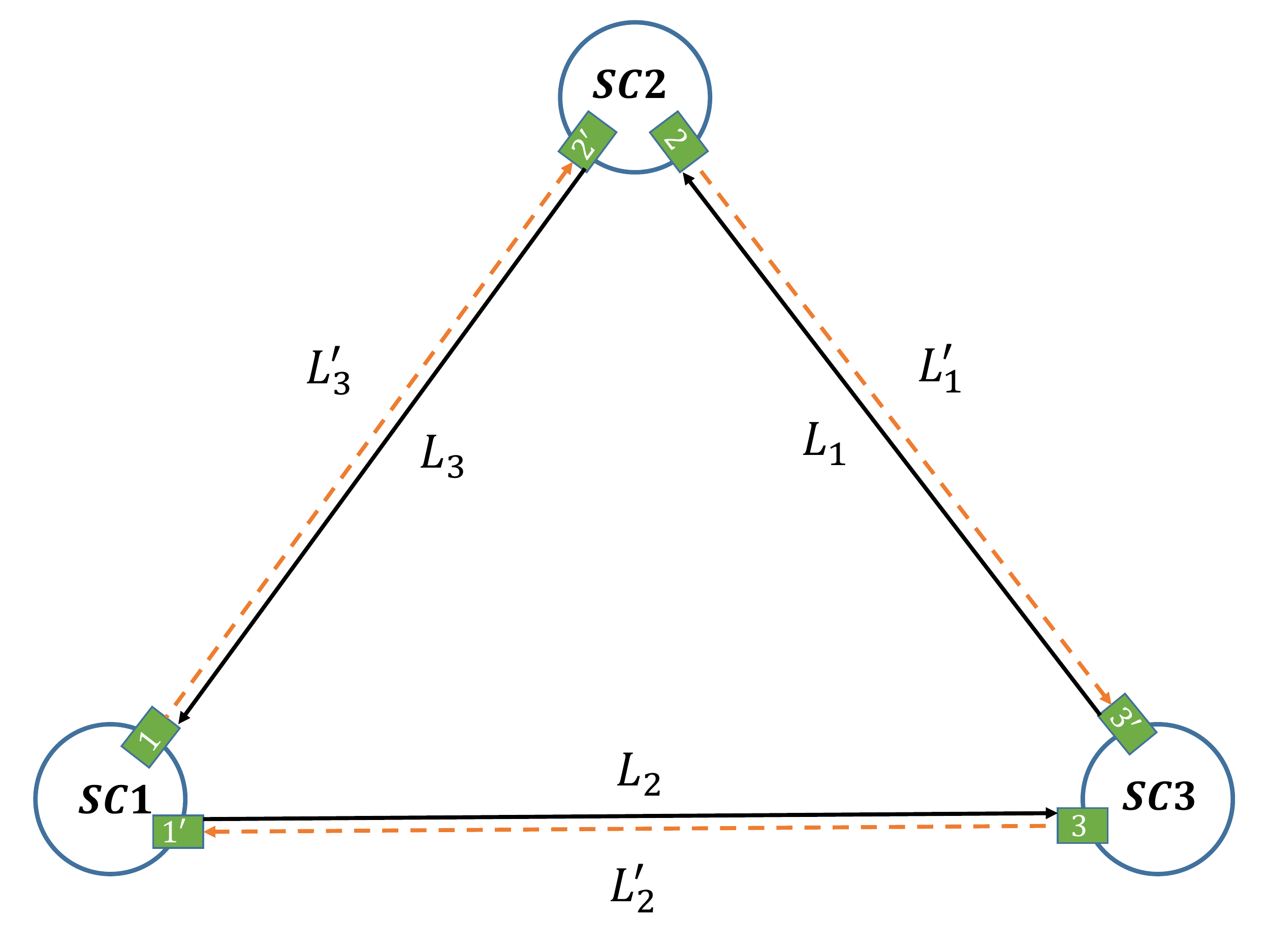}}}
\end{center}
\caption{\label{fig:TDIstruc}TDI Loop in a Space-based Detector}
\end{figure}

For the equilateral triangle configuration shown in Figure \ref{fig:TDIstruc}, considering a particular one-way link, we denote the length of each arm in the absence of gravitational waves as $L$. $\hat{l}_{ij}$ and $\hat{l}_{ik}$ are unit vectors pointing from the interferometer's vertex to the two endpoint masses, located at position vectors $\vec{x}_i$ and $\vec{x}_j = \vec{x}j + L \hat{l}_{ij}$, respectively. In the presence of a planar gravitational wave propagating along the direction $\hat{k} = -\hat{n}$,

The relative variation in the light propagation path of a photon emitted at $\vec{x}i$ and received at $\vec{x}_j$ at time $t$ is given by reference \cite{1975GReGr...6..439E,Romano:2016dpx}:
\begin{equation}
  \Delta T_{ij} (t) = \frac{1}{2}  \hat{l}_{ij}  \hat{l}_{ij}  \int_{s = 0}^L
  \mathrm{d} s h_{ab} (t (s), \vec{x} (s)), \label{e:deltaT}
\end{equation}
where $i, j, k = {1, 2, 3}$ correspond to the three satellites of the detector, and $i < j < k$. The corresponding Doppler shift is given by:

\begin{eqnarray}
h_{\text{doppler}} (t) &\equiv& \frac{\Delta \nu (t)}{\nu} =  - \frac{\mathrm{d}}{\mathrm{d}t} \Delta T_{ij} (t) \nonumber\\
  & = & \int_{- \infty}^{\infty} \mathrm{d}f \int \mathrm{d}^2 \Omega_{\hat{k}} R^{ab} (f, \hat{k}) h_{ab} (f, \hat{k}) e^{i 2 \pi ft},\nonumber\\ \label{e:hddefine}
\end{eqnarray} 
where $h_{\text{doppler}} (t)$ represents the Doppler shift, $\Delta \nu (t)$ is the frequency shift, $\nu$ is the reference frequency, $R^{ab} (f, \hat{k})$ is the Response Function (RF), $h_{ab} (f, \hat{k})$ is the gravitational wave strain amplitude, and $\hat{k}$ is the direction of the incoming gravitational wave.

Similarly, we can construct a bidirectional link, which includes the process of the laser returning from $\vec{x}_j$ to $\vec{x}_i$ in addition to the one-way link $\vec{x}_i \to \vec{x}_j$. To calculate the timing residual $\Delta T (t)$ for a bidirectional link in a space-based detector $\vec{x}_i \to \vec{x}_j \to \vec{x}_i$, we need to generalize the calculations from the previous section to include the photon returning from $\vec{x}_j$ to $\vec{x}_i$. This can be achieved by simply summing the expressions for the one-way timing residuals:
\begin{equation}
  \Delta T_{i (j)} (t) = \Delta T_{ij}  (t - L) + \Delta T_{ji} (t),
\end{equation}
where $\Delta T_{ij} (t - L)$ represents the timing residual for the link $\vec{x}_i \to \vec{x}_j$ delayed by $L$ seconds, and $\Delta T_{ji} (t)$ represents the timing residual for the link $\vec{x}_j \to \vec{x}_i$.
\begin{equation}
  \Delta T_{i (j)} (t) = \Delta T_{ij}  (t - L) + \Delta T_{ji} (t),
\end{equation}

Furthermore, considering the Michelson interferometer for the equilateral triangle configuration shown in Figure \ref{fig:TDIstruc}, the laser propagation paths for a specific arm can be described as$\vec{x}_i \to \vec{x}_j \to \vec{x}_i$ and $\vec{x}_i \to \vec{x}_k \to \vec{x}_i$. The corresponding timing residual is the difference between the round-trip photon propagation times in the adjacent arms. Let $\vec{L}_{ij}$ and $\vec{L}_{ik}$ represent the vectors for the two arms of the detector. We have:
\begin{equation}
  \Delta T_{i (jk)} (t) \equiv T_{i (j)} (t) - T_{i (k)} (t) = \Delta T_{i
  (j)} (t) - \Delta T_{i (k)} (t), \label{eq:michDeltaT}
\end{equation}
where the last equality holds for an equal-arm interferometer. In the previous section, we calculated the round-trip durations $\Delta T_{ij} (t)$ for these single arms. Therefore, for the response function of the equal-arm Michelson, we can write:
\begin{equation}
  R^{ab}  (f, \hat{k}, \hat{l}_{ij}, \hat{l}_{ik}) = \frac{1}{2} 
  [\hat{l}_{ij}^a  \hat{l}_{ij}^b \mathcal{T}(f, \hat{k} \cdot \hat{l}_{ij})-
  \hat{l}_{ik}^a  \hat{l}_{ik}^b \mathcal{T}(f, \hat{k} \cdot \hat{l}_{ik})],
  \label{eq:michRAtiming-New}
\end{equation}
The corresponding transfer function is defined as follows:
\begin{eqnarray}
  \mathcal{T} (f, \hat{k} \cdot \hat{l}_{ij}) &=& e^{- \frac{if}{2 f_{\ast}}  (3 - \hat{k} \cdot \hat{l}_{ij})} \mathrm{sinc} \left( \frac{f}{2 f_{\ast}} [1 - \hat{k} \cdot \hat{l}_{ij}] \right) + \nonumber\\
  &  & e^{- \frac{if}{2 f_{\ast}}  (1 +
  \hat{k} \cdot \hat{l}_{ji})} \mathrm{sinc} \left( \frac{f}{2 f_{\ast}} [1 -
  \hat{k} \cdot \hat{l}_{ji}] \right),\nonumber\\
  &  & \label{eq:twoway-Transfunctoion-New}
\end{eqnarray}
where the subscripts have been omitted. For the $XYZ$ channel, the time-domain observable is given by:
\begin{eqnarray}
X &\equiv& [\Delta T_{i (j)} (t) - \Delta T_{i (k)} (t)] -\nonumber\\
  &  & [\Delta T_{i (j)} (t - 2 L_j) - \Delta T_{i (k)} (t - 2 L_k)].\label{eq:XTA99-New}
\end{eqnarray}
The response function for the $XYZ$ channel is given by:
\begin{equation}
  R_X^{ab}  (f, \hat{k}, \hat{l}_{ij}, \hat{l}_{ik}) = (1 - e^{- i 2 f /
  f_{\ast}}) R^{ab}  (f, \hat{k}, \hat{l}_{ij}, \hat{l}_{ik})
  \label{eq:Rab-XYZ-define},
\end{equation}
This allows us to combine Equation \eqref{eq:michRAtiming-New} and Equation \eqref{eq:Rab-XYZ-define}, and express the response functions for the Michelson and $XYZ$ channels together as:
\begin{equation}
R_i^{ab}  (f, \hat{k}, \hat{l}_{ij}, \hat{l}_{ik}) = U (f) R^{ab}  (f, \hat{k}, \hat{l}_{ij}, \hat{l}_{ik}) \label{eq:Ri-A}, 
\end{equation}
The factor $U(f)$ is defined as follows:
\begin{equation}
  U (f) = \left\{ \begin{array}{l}
    1 \hspace{0.27em} \hspace{0.27em}, \hspace{0.27em} \hspace{0.27em}
    \hspace{0.27em} \hspace{0.27em} \quad \quad \quad \quad \textrm{for
    {\hspace{0.27em}} Michelson,}\\
    1 - e^{- i 2 f / f_{\ast}} \hspace{0.27em}, \hspace{0.27em}
    \hspace{0.27em} \hspace{0.27em} \hspace{0.27em} \textrm{for
    {\hspace{0.27em}} XYZ.}
  \end{array} \right.
\end{equation}
For the Michelson or $XYZ$ channels, the corresponding Doppler shift can be written as:
\begin{eqnarray}
    \Delta d_{i (jk)} (t) &=& \int_{- \infty}^{\infty} \frac{if}{f_{\ast}} \mathrm{d}f \int \mathrm{d}^2 \Omega_{\hat{k}} \nonumber\\
  & &R_i^{ab}  (f, \hat{k}, \hat{l}_{ij}, \hat{l}_{ik})
  h_{ab}  (f, \hat{k}) e^{i 2 \pi ft} \label{TDI-integral-ab},
\end{eqnarray}
where $i$ represents the Michelson or $XYZ$ channel.
We can simplify it as:
\begin{equation}
\Delta d_i (t) \equiv \Delta d_{i (i + 1, i + 2)} \hspace{0.27em},
\hspace{0.27em} \hspace{0.27em} \hspace{0.27em} i = 1, 2, 3, \label{short-i}
\end{equation}
By using a tensor basis decomposition, we have:
\begin{equation}
  R^A_i  (f, \hat{k}, \hat{l}_{ij}, \hat{l}_{ik}) = \frac{1}{2} e_{ab}^A 
  [\hat{l}_{ij}^a  \hat{l}_{ij}^b \mathcal{T}(f, \hat{k} \cdot \hat{l}_{ij}) -\hat{l}_{ik}^a  \hat{l}_{ik}^b \mathcal{T}(f, \hat{k} \cdot \hat{l}_{ik})],
  \label{eq:mich-frenq-A}
\end{equation}
The corresponding expression is:
\begin{equation}
  \Delta d_i (t) = \int_{- \infty}^{\infty} \frac{if}{f_{\ast}} \mathrm{d}f \int \mathrm{d}^2
  \Omega_{\hat{k}} R_i^A  (f, \hat{k}, \hat{l}_{ij}, \hat{l}_{ik}) h_A  (f,
  \hat{k}) e^{i 2 \pi ft} \label{TDI-integral-A} .
\end{equation}
For a SGWB that satisfies assumptions such as stationary, Gaussian, unpolarized, and isotropic, the two-point correlation function is given by:
\begin{eqnarray}
  \langle \Delta d_i (t) \Delta d_j (t) \rangle & = & 4 \sum_{\ell m}
  \int_0^{\infty} \mathrm{d}f \left| \frac{f}{f_{\ast}} U (f) \right|^2 
  \tilde{R}_{ij}^{\ell m} (f)  \tilde{I}_{\ell m} (f) \nonumber\\
  & \equiv & \sum_{\ell m} \int_0^{\infty} \mathrm{d}fR_{ij}^{\ell m} (f) 
  \tilde{I}_{\ell m} (f) \hspace{0.27em},  \label{eq:DFDF}
\end{eqnarray}
The expression for $|U(f)|^2$ is:
\begin{equation}
  |U (f) |^2 = \left\{ \begin{array}{l}
    1 \hspace{0.27em} \hspace{0.27em}, \hspace{0.27em} \hspace{0.27em}
    \hspace{0.27em} \hspace{0.27em} \quad \quad \quad \quad \textrm{for
    {\hspace{0.27em}} Michelson},\\
    4 \sin^2 \left( \frac{f}{f_{\ast}} \right) \hspace{0.27em}
    \hspace{0.27em}, \hspace{0.27em} \hspace{0.27em} \hspace{0.27em}
    \hspace{0.27em} \hspace{0.27em} \textrm{for {\hspace{0.27em}} XYZ} .
  \end{array} \right.
\end{equation}
Equation \eqref{eq:DFDF} represents the correlation between TDI measurements in \eqref{TDI-integral-ab} or \eqref{TDI-integral-A}. This introduces the overlap reduction functions (ORFs) of an equilateral triangle space gravitational wave detector to an anisotropic SGWB.
\begin{equation}
  \tilde{R}_{ij}^{\ell m} (f) \equiv \frac{1}{4 \pi}  \int \mathrm{d}^2  \hat{k} 
  \tilde{Y}_{\ell m} (\hat{k}) R_{ij}  (f, \hat{k}) \label{response},
\end{equation}
where
\begin{equation}
  R_{ij}  (f, \hat{k}) \equiv \frac{1}{2}  \sum_A R_i^A  (f, \hat{k},
  \hat{l}_{ip}, \hat{l}_{iq}) R_j^{A \ast}  (f, \hat{k}, \hat{l}_{jm},
  \hat{l}_{jn}) \label{e:GammaIJ-New} .
\end{equation}
The function $R^A$ is given in equation \eqref{eq:Ri-A}. In the case of isotropy, the response function in equation \eqref{response} is consistent with equation (A.21) in the reference \cite{Flauger:2020qyi}.

The motion of the detector constellation can be decomposed into the motion of the constellation center around the Sun and the rotation of the constellation around its own center. The quantities $\hat{k}$, $\hat{l}_{ij}$, and $\hat{l}_{ik}$ in the response function $R^A (f, \hat{k}, \hat{l}_{ij}, \hat{l}_{ik})$ vary with the changing positions of the detector constellation. This means that the response function $R^A (f, \hat{k}, \hat{l}_{ij}, \hat{l}_{ik})$ depends on time. As a result, the subsequent computation of the ORFs also becomes time-dependent. The ORFs given by equation \eqref{response} will vary continuously with the changing positions of the constellation, which introduces additional challenges for the calculations.

\subsubsection{Symmetry of the ORFs}\label{sec:ORFsymmetry}

Although the $\tilde{R}_{ij}^{\ell m}(f)$ actually depend on time, it is important to note that for space-based gravitational wave detectors like TianQin or LISA, which have an ideal configuration of an equilateral triangle, they exhibit a high degree of symmetry. This symmetry is manifested in the $\tilde{R}_{ij}^{\ell m}(f)$ in a specific form. Exploiting these symmetries not only facilitates calculations but also provides unique observational capabilities. In the following, we present these symmetry relations directly from the work by LISA{\cite{LISACosmologyWorkingGroup:2022kbp}}, where they are discussed in detail.

For any arbitrary rigid rotation $\mathscr{R}$, it holds that\begin{equation}
  \tilde{R}_{\mathscr{R} i \mathscr{R} j}^{\ell m} (f) = \sum_{m' = -
  \ell}^{\ell} [D_{mm'}^{(\ell)} (R)]^{\ast}  \tilde{R}_{ij}^{\ell m'} (f)
  \hspace{0.27em}, \label{rot-R}
\end{equation}
where $D_{mm'}^{(\ell)}$ represents the matrix elements of the Wigner D-matrix.

Using the symmetry of the ORFs, we can define another quantity that does not depend on time, which is dependent on $\ell$. We define the $\ell$-dependent ORFs as:
\begin{equation}
  \tilde{R}_{ij}^{\ell} (f) \equiv \left( \sum_{m = - \ell}^{\ell} |
  \tilde{R}_{ij}^{\ell m} (f) |^2 \right)^{1 / 2} \hspace{0.27em}, \label{Rl}
\end{equation}
For a rotation around the $z$-axis with an angle $\alpha$, we have:
\begin{equation}
  \tilde{R}_{\mathscr{R}_z (\alpha) i, \mathscr{R}_z (\alpha) j}^{\ell m} (f)
  = \mathrm{e}^{im \alpha} \tilde{R}_{ij}^{\ell m} (f) . \label{rotz-R}
\end{equation}
Regarding the TianQin mission, its constellation plane has a fixed normal vector pointing towards J0806, which can be considered as an infinitely distant point compared to the Sun-Earth distance. In other words, the normal vector of the TianQin constellation plane is constant, and the TianQin constellation can be regarded as a rotation around the fixed $z$-axis. This clever design allows the quantity $[\tilde{R}_{ij}^{\ell m}(f)]^2$ for TianQin to be independent of time.

The difference between TianQin and LISA leads to the ability of TianQin to distinguish different $\ell$ and $m$ modes compared to LISA. However, the ORFs, as given by Equation \eqref{response}, possesses several important symmetries, including:

For any $(\ell, m)$, we have:
\begin{equation}
  \tilde{R}_{ji}^{\ell m} (f) = (- 1)^{\ell}  \tilde{R}_{ij}^{\ell m} (f) .
  \label{transpose-R}
\end{equation}

If $\ell = 2 n + 1, n \in N$, then
\begin{equation}
  \tilde{R}_{ii}^{\ell m} (f) = 0 \hspace{0.27em} \hspace{0.27em} .
  \label{R-prop-m}
\end{equation}

If $\ell + m = 2 n + 1, n \in N$, then
\begin{equation}
  \tilde{R}_{ij}^{\ell m} (f) = 0. \label{R-prop-lm}
\end{equation}

Using the property \eqref{rotz-R}, it can be concluded that for an equilateral triangle configuration, it exhibits symmetry under rotations around the $z$-axis by angles $\frac{2}{3}
n \pi, n \in N$. If the three satellites are placed on the $xy$-plane, the ORFs satisfies
\begin{equation}
  \tilde{R}_{i' j'}^{\ell m} = \mathrm{e}^{im \frac{2}{3}  (i' - i) \pi}
  \tilde{R}_{ij}^{\ell m} (f) . \label{rotz-Rii1}
\end{equation}
where $i'$ and $j'$ are the rotated indices, and $i$ and $j$ are the original indices. Exploiting these symmetries can significantly simplify the calculations. For a more detailed discussion on these symmetries, please refer to the reference \cite{LISACosmologyWorkingGroup:2022kbp}.

\subsubsection{The ORFs in the TianQin $AET$ channel combination}

In general, we consider the linear combination of the $\Delta d_i$ measurements obtained so far and express it in a more compact form as follows:
\begin{equation}
  \Delta d_O \equiv c_{Oi} \Delta d_i \hspace{0.27em}, \hspace{0.27em}
  \hspace{0.27em} O \in \{ A, E, T \} \hspace{0.27em}, \hspace{0.27em}
  \hspace{0.27em} i \in \{ 1, 2, 3 \} \hspace{0.27em} . \label{DFO}
\end{equation}
where $c_{Oi}$ are the elements of the matrix $c$ given by:
\begin{equation}
  c = \left(\begin{array}{ccc}
    - \frac{1}{\sqrt{2}} & 0 & \frac{1}{\sqrt{2}}\\
    \frac{1}{\sqrt{6}} & - \frac{2}{\sqrt{6}} & \frac{1}{\sqrt{6}}\\
    \frac{1}{\sqrt{3}} & \frac{1}{\sqrt{3}} & \frac{1}{\sqrt{3}}
  \end{array}\right)
\end{equation}

Under the assumption that the detector constellation forms an equilateral triangle with identical instruments at the vertices, these combinations diagonalize the noise variance (normalized as described in reference \cite{Flauger:2020qyi}), and the rotation matrices associated with these transformations are orthogonal. To assess the impact of these linear combinations on the detection, particularly for detectors like LISA, where the constellation plane's normal vector is time-varying, one can define ORFs dependent on $\ell$ for each channel combination. These ORFs are time-dependent quantities and can be defined as follows:
\begin{equation}
  \tilde{R}_{OO'}^{\ell} (f) \equiv \left( \sum_{m = - \ell}^{\ell} | c_{Oi}
  c_{O' j}  \tilde{R}_{ij}^{\ell m} (f) |^2 \right)^{1 / 2} \hspace{0.27em},
  \label{Rl-OOp}
\end{equation}
For the $AET$ channel combination in the case of TianQin, one can define the ORFs dependent on $\ell$ and $m$ as follows:
\begin{equation}
  \tilde{R}_{OO'}^{\ell m} (f) = | c_{Oi} c_{O' j}  \tilde{R}_{ij}^{\ell m}
  (f) | \hspace{0.27em} . \label{Rlm-OOp}
\end{equation}
Based on the analysis in Section \ref{sec:ORFsymmetry}, for TianQin, it can be assumed that equation \eqref{Rlm-OOp} does not evolve with time. Additionally, it holds that:
\begin{equation}
  \tilde{R}_{OO'}^{\ell} (f) = \left( \sum_{m = - \ell}^{\ell}
  \tilde{R}_{OO'}^{\ell m} (f) \right)^{1 / 2} \hspace{0.27em} .
  \label{Rlm-Rl}
\end{equation}

Using the symmetries discussed in Section \ref{sec:ORFsymmetry}, it can be shown that the specific forms of the ORFs are related to the parity of $\ell$.

For odd values of $\ell$, the following relations hold:

\begin{eqnarray}
  \tilde{R}_{AA}^{\ell m} (f) &=& \tilde{R}_{EE}^{\ell m} (f) =
  \tilde{R}_{TT}^{\ell m} (f) = 0 \hspace{0.27em},  \label{R-AET-lodd} \nonumber\\
  \tilde{R}_{AE}^{\ell m} (f) &=& \left\{ \frac{1}{3}  \left[ 1 + 2 \cos
  \left( \frac{2 m \pi}{3} \right) \right]^2 | \tilde{R}_{12}^{\ell m} (f) |^2
  \right\}^{1 / 2} \hspace{0.27em},  \label{R-AET-lodd1} \nonumber\\
  \tilde{R}_{AT}^{\ell m} (f) &=& \tilde{R}_{ET}^{\ell m} (f) = \left\{ 2
  \sin^2 \left( \frac{m \pi}{3} \right) | \tilde{R}_{12}^{\ell m} (f) |^2
  \right\}^{1 / 2} \hspace{0.27em} . \nonumber\\& & \label{R-AET-lodd2}
\end{eqnarray}

For even values of $\ell$, the following relations hold:
\begin{widetext}
\begin{eqnarray}
   \tilde{R}_{AA}^{\ell m} (f) &=& \tilde{R}_{EE}^{\ell m} (f) = \left\{
  \frac{1}{4}  \left| \left( 1 + \mathrm{e}^{- \frac{4}{3} im \pi} \right) 
  \tilde{R}_{11}^{\ell m} (f) - 2 \tilde{R}_{12}^{\ell m} (f) \right|^2
  \right\}^{1 / 2} \hspace{0.27em},  \label{R-AET-leven}\nonumber\\
\tilde{R}_{TT}^{\ell m} (f) &=& \left\{ \frac{1}{9}  \left[ 1 + 2 \cos
  \left( \frac{2 m \pi}{3} \right) \right]^2 | \tilde{R}_{11}^{\ell m} (f) + 2
  \tilde{R}_{12}^{\ell m} (f) |^2 \right\}^{1 / 2} \hspace{0.27em}, 
  \label{R-AET-leven1}\nonumber\\
  \tilde{R}_{AE}^{\ell m} (f) &=& \left\{ \frac{1}{3} \sin^2 \left(
  \frac{m \pi}{3} \right)  \left| \left( 1 + \mathrm{e}^{\frac{2 im \pi}{3}}
  \right)  \tilde{R}_{11}^{\ell m} (f) - 2 \tilde{R}_{12}^{\ell m} (f)
  \right|^2 \right\}^{1 / 2} \hspace{0.27em},  \label{R-AET-leven2}\nonumber\\
  \tilde{R}_{AT}^{\ell m} (f) &=& \tilde{R}_{ET}^{\ell m} (f) = \left\{
  \frac{2}{3} \sin^2 \left( \frac{m \pi}{3} \right)  \left| \left( 1 +
  \mathrm{e}^{\frac{2 im \pi}{3}} \right)  \tilde{R}_{11}^{\ell m} (f) +
  \tilde{R}_{12}^{\ell m} (f) \right|^2 \right\}^{1 / 2} . \nonumber\\
  & &\label{R-AET-leven3}
\end{eqnarray}
\end{widetext}

\begin{figure*}
  \begin{center}
    {\resizebox{220pt}{165pt}{\includegraphics{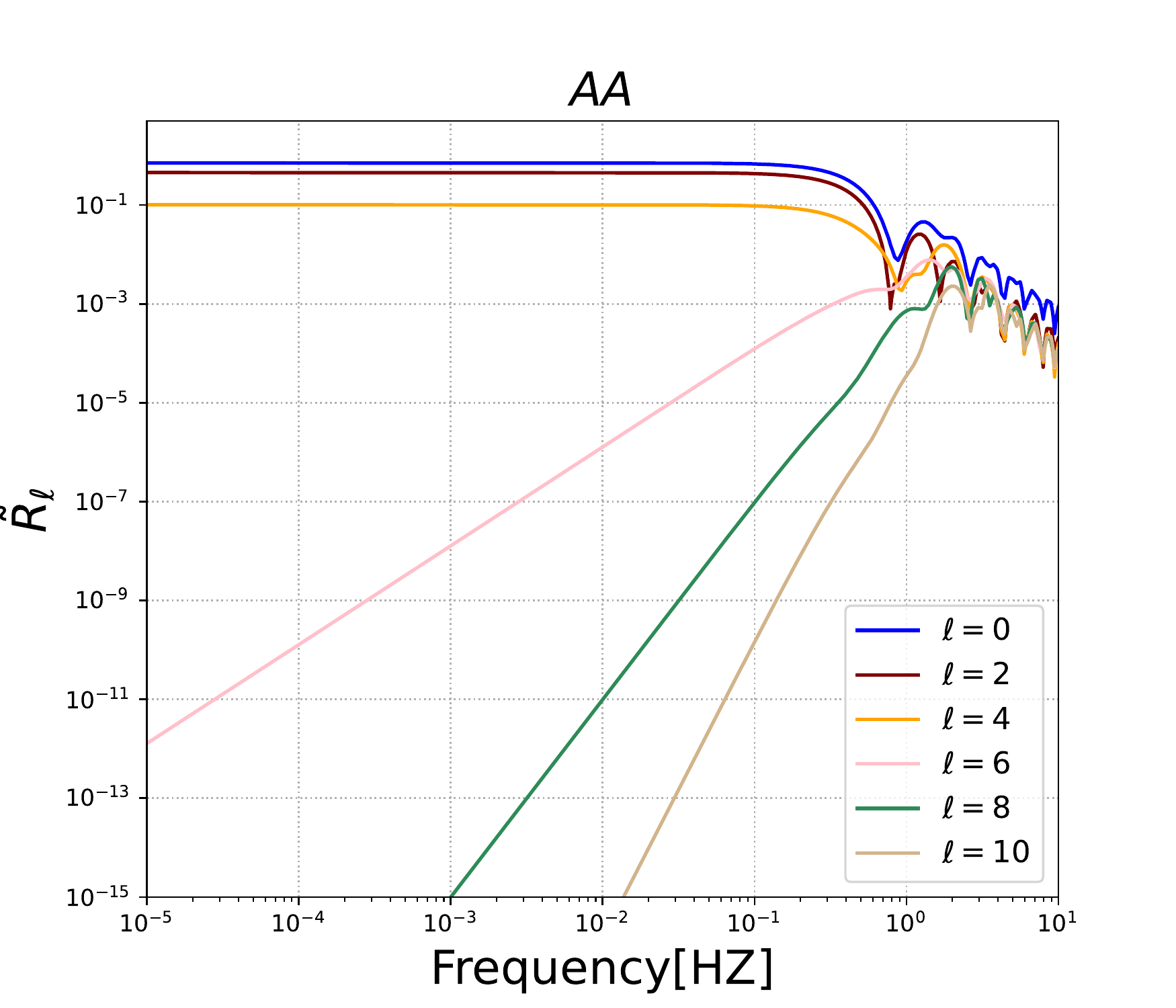}}}{\resizebox{220pt}{165pt}{\includegraphics{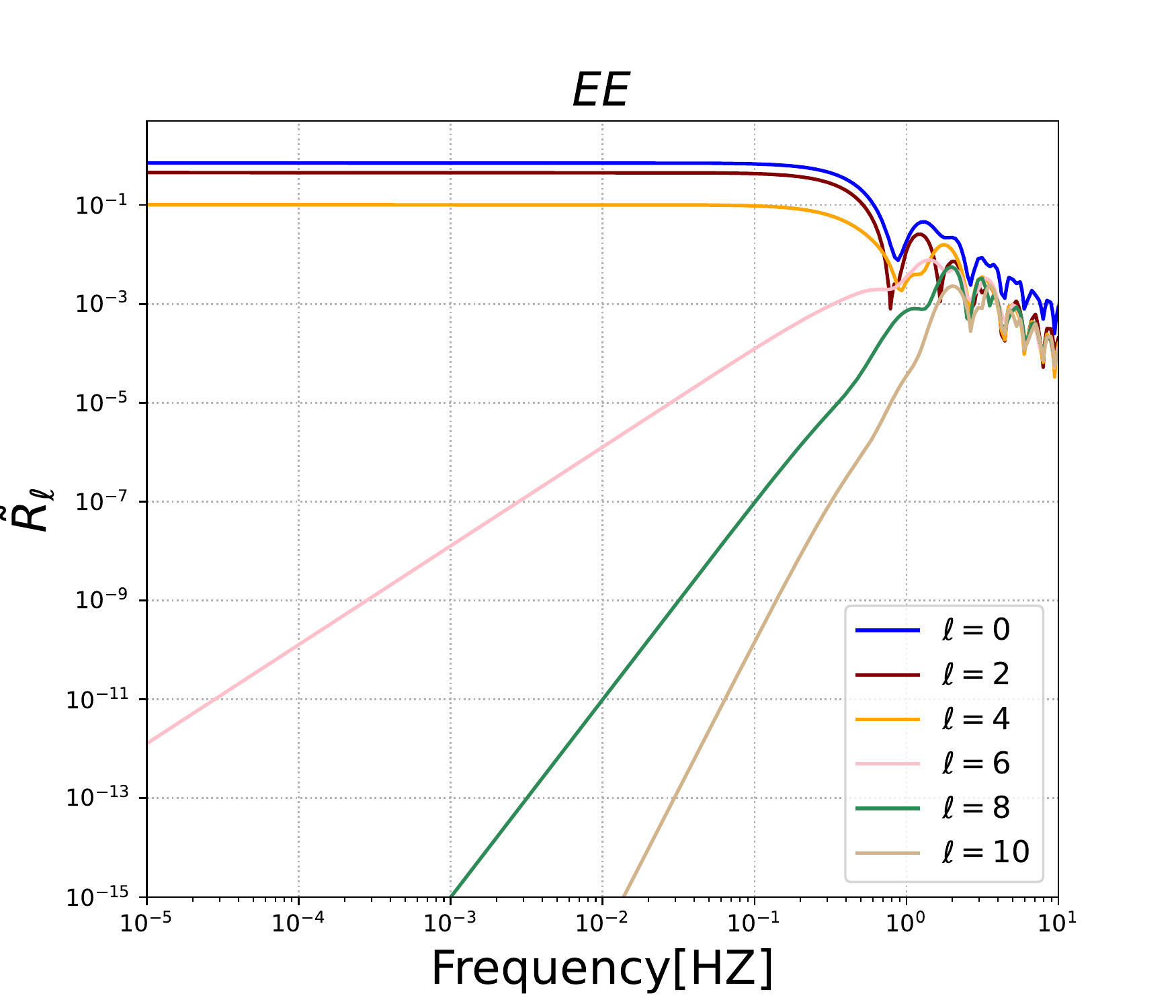}}}
    
    {\resizebox{220pt}{165pt}{\includegraphics{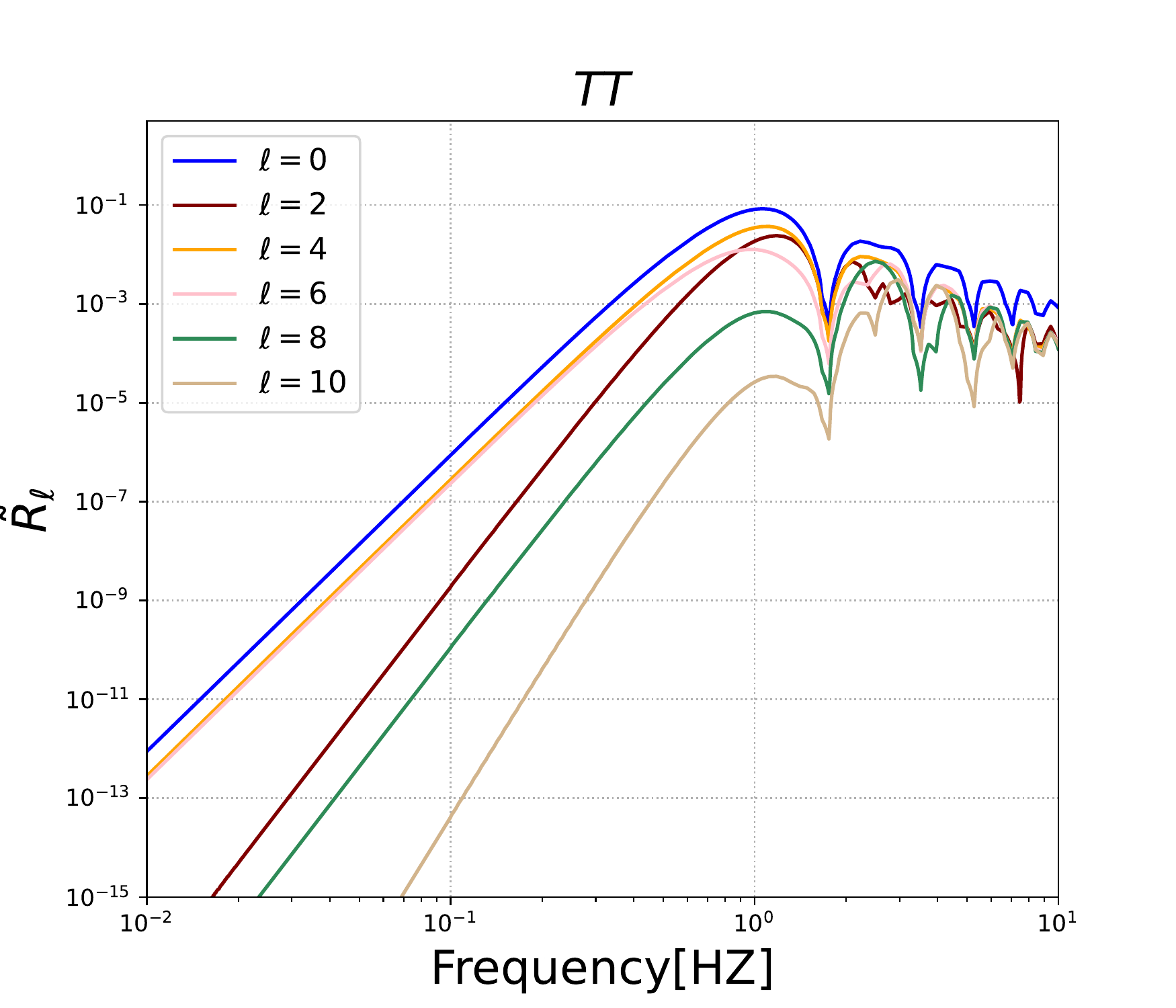}}}{\resizebox{220pt}{165pt}{\includegraphics{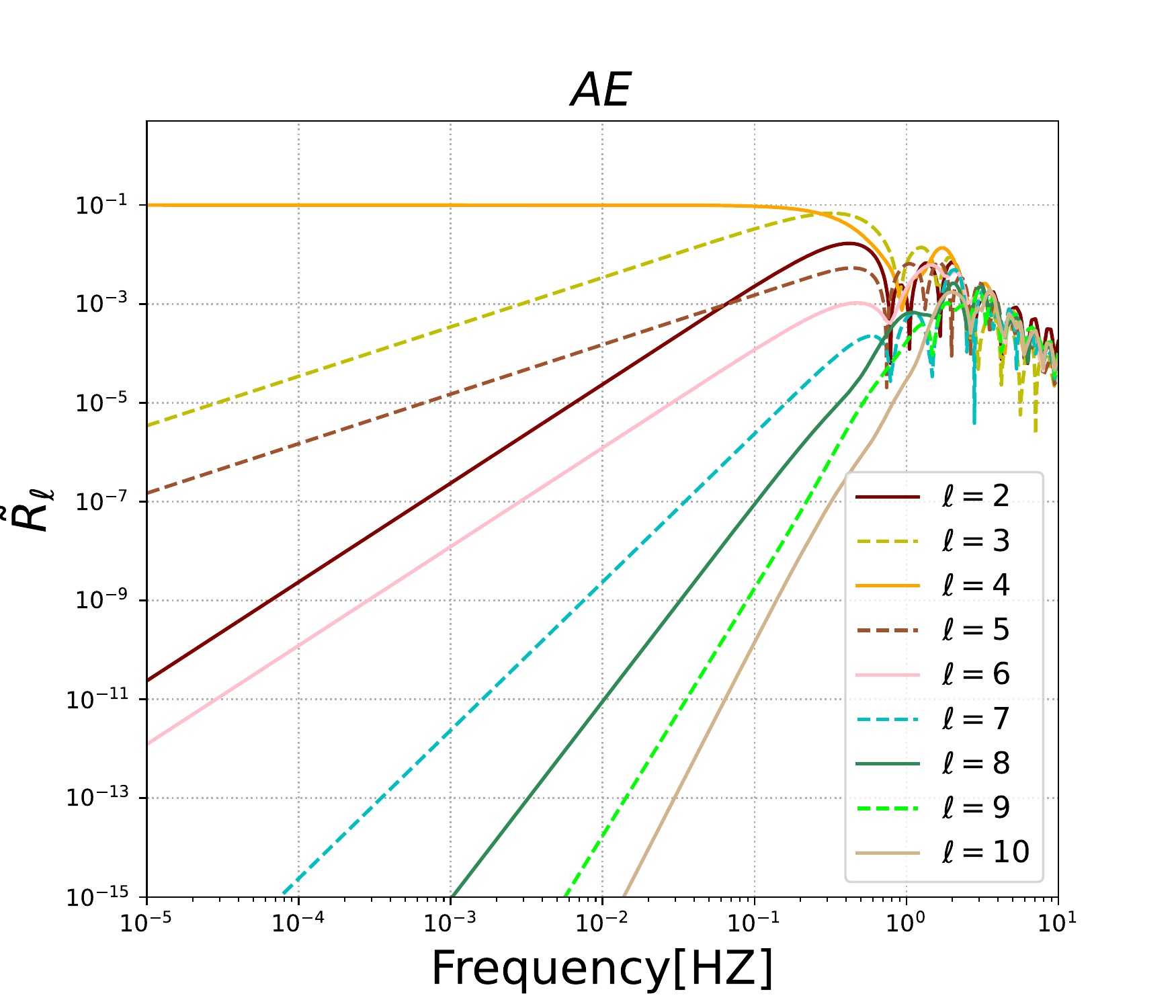}}}
    
    {\resizebox{220pt}{165pt}{\includegraphics{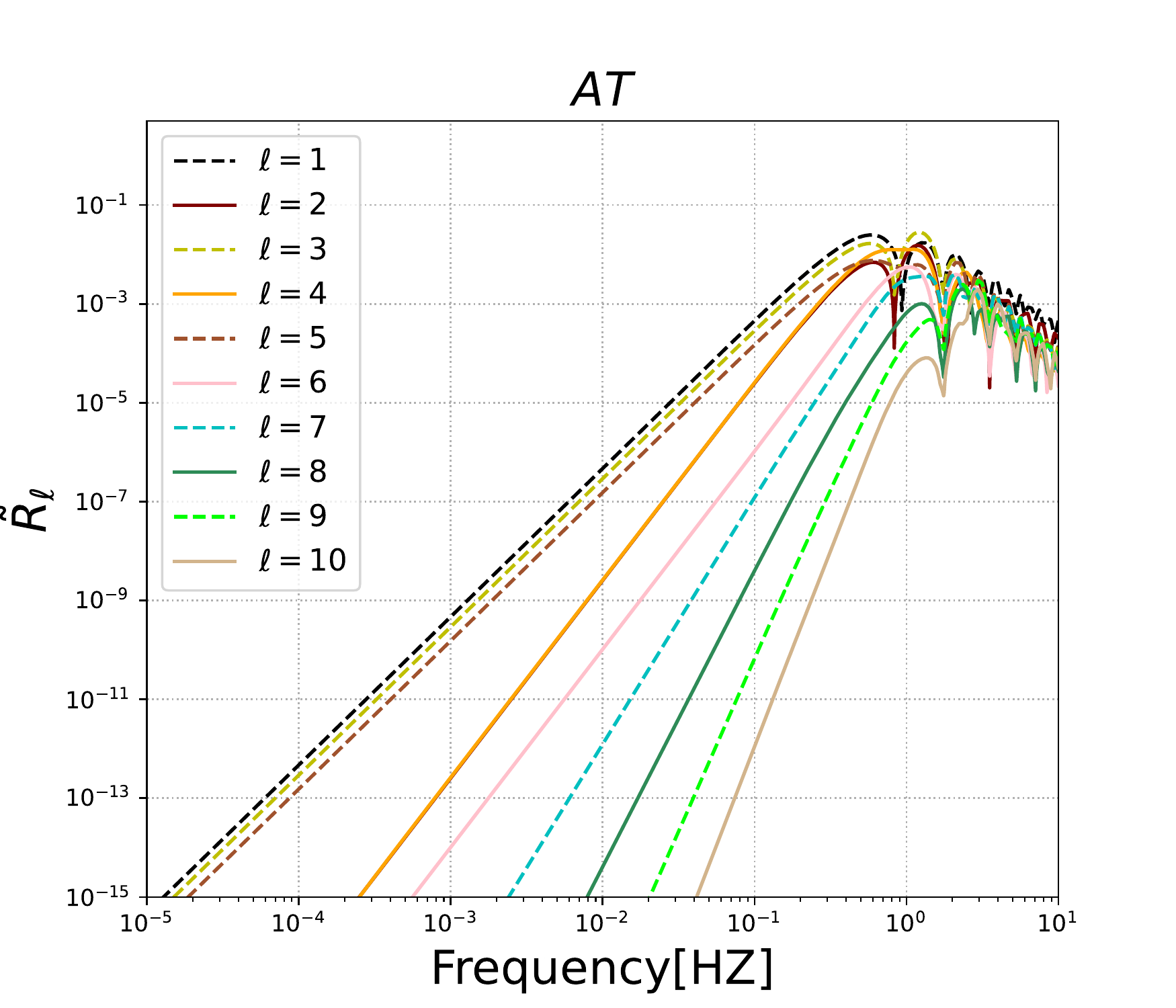}}}{\resizebox{220pt}{165pt}{\includegraphics{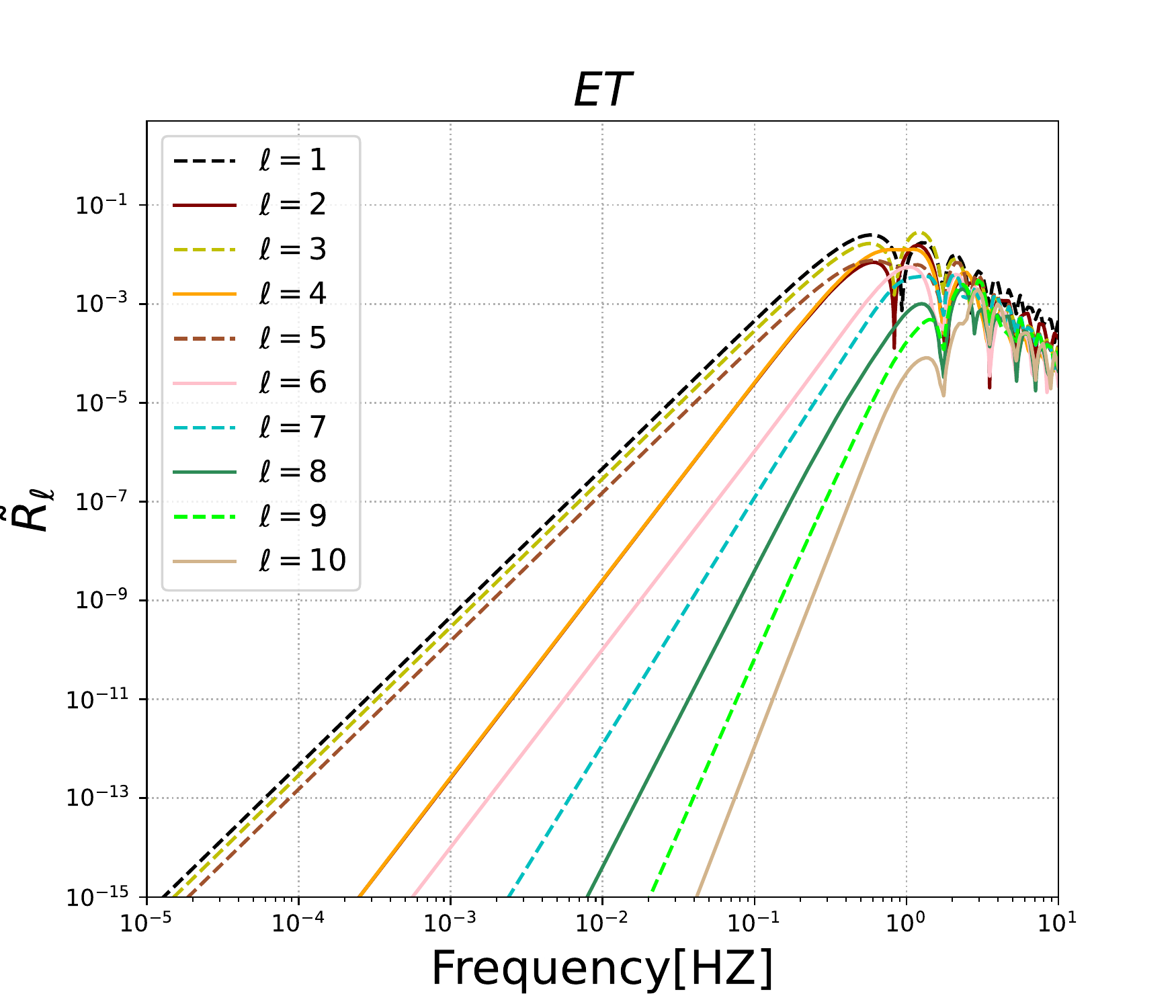}}}
  \end{center}
  \caption{\label{fig:TianQin-Respond-l} ORFs for different values of $\ell$ in the TianQin detector, as given by equations \eqref{Rlm-Rl}, and equations \eqref{R-AET-lodd} to \eqref{R-AET-leven3}.}
\end{figure*}

\begin{figure*}
  \begin{center}
    {\resizebox{220pt}{165pt}{\includegraphics{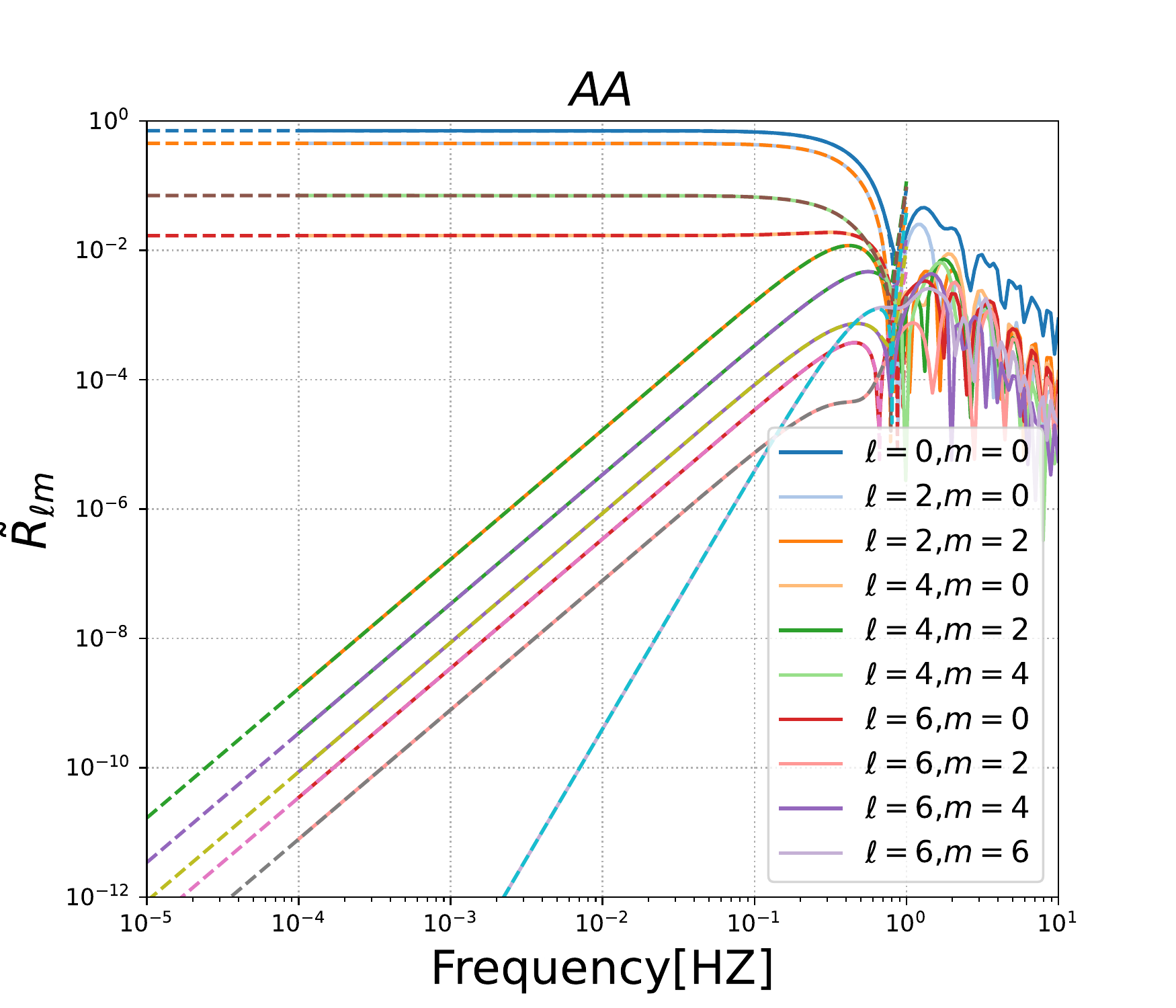}}}{\resizebox{220pt}{165pt}{\includegraphics{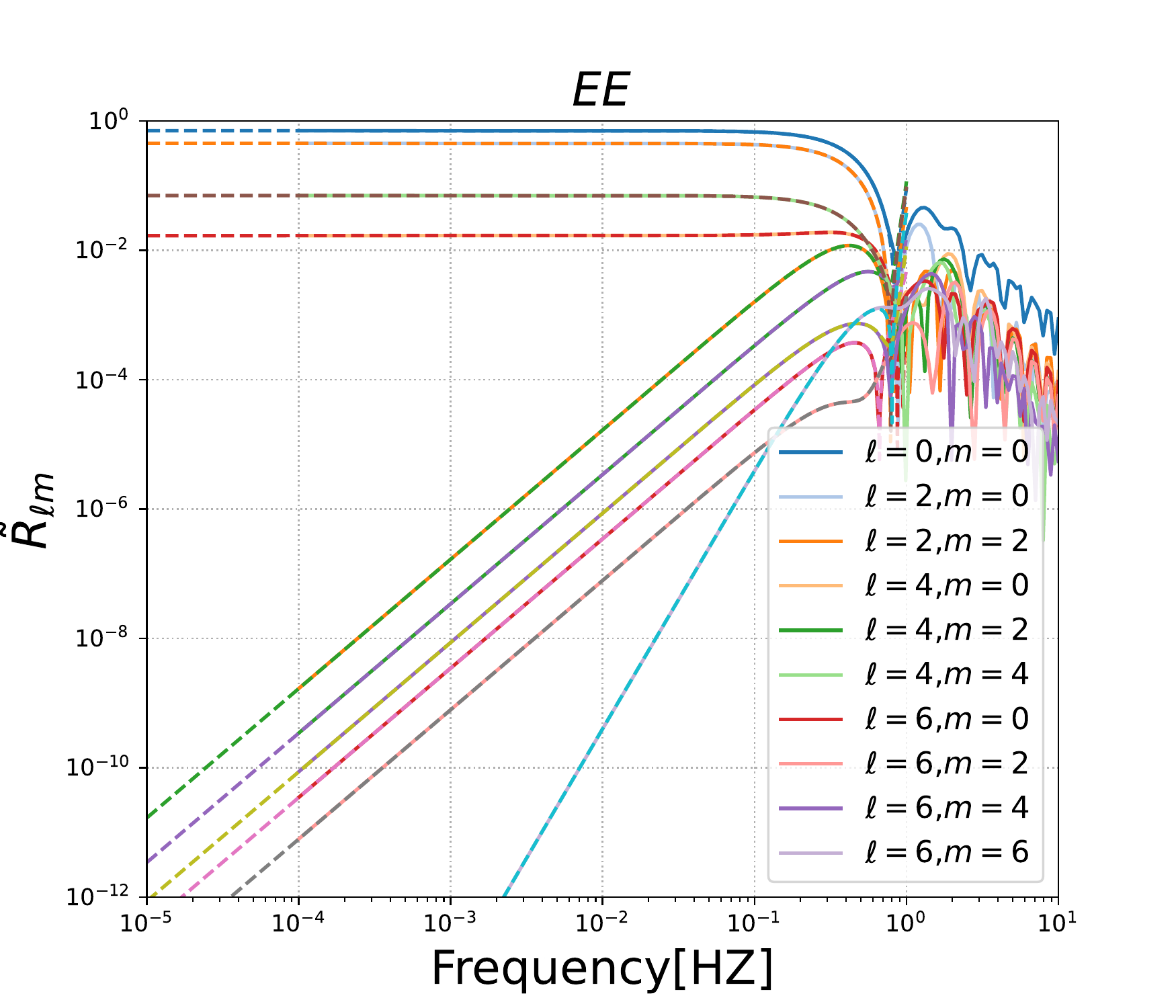}}}
    
    {\resizebox{220pt}{165pt}{\includegraphics{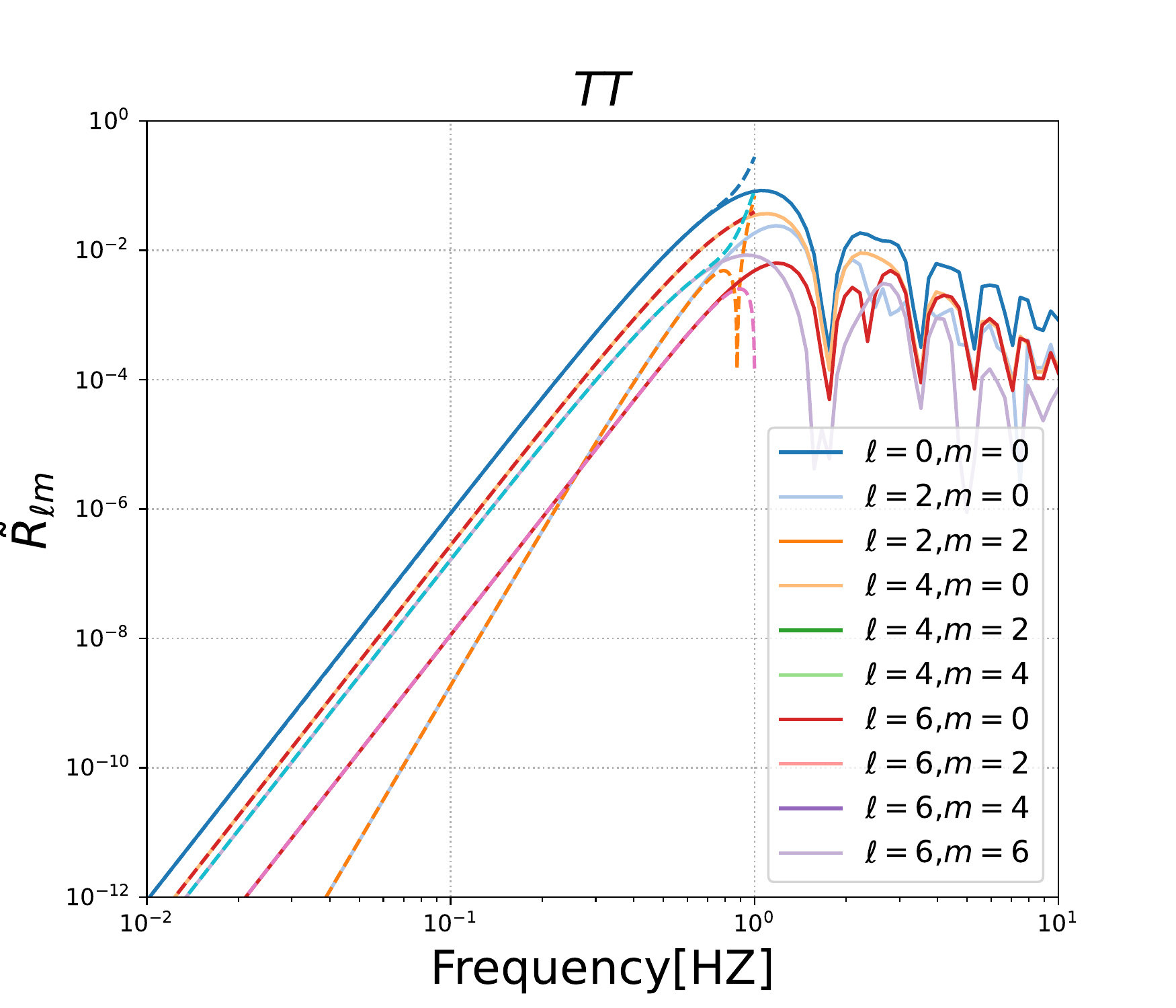}}}{\resizebox{220pt}{165pt}{\includegraphics{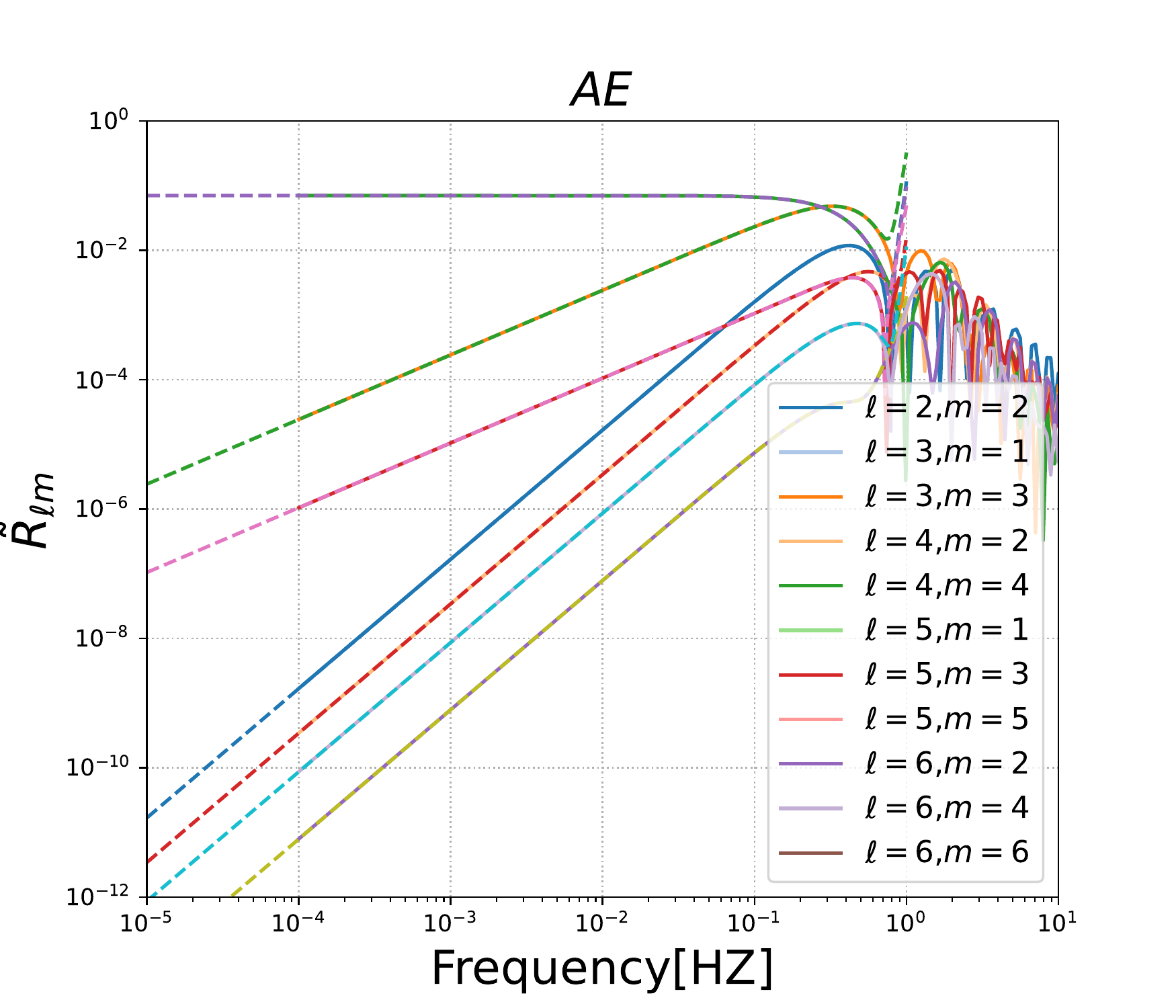}}}
    
    {\resizebox{220pt}{165pt}{\includegraphics{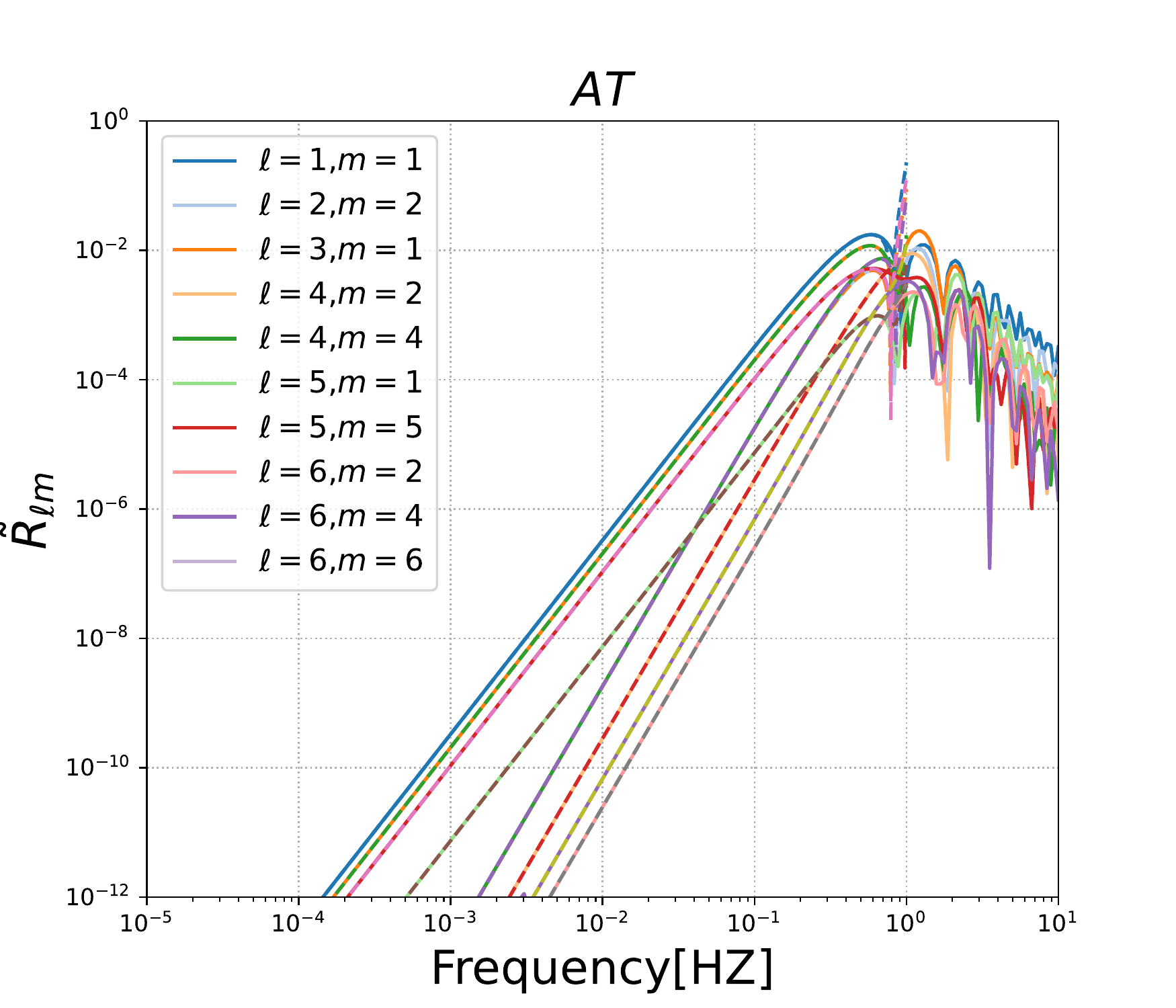}}}{\resizebox{220pt}{165pt}{\includegraphics{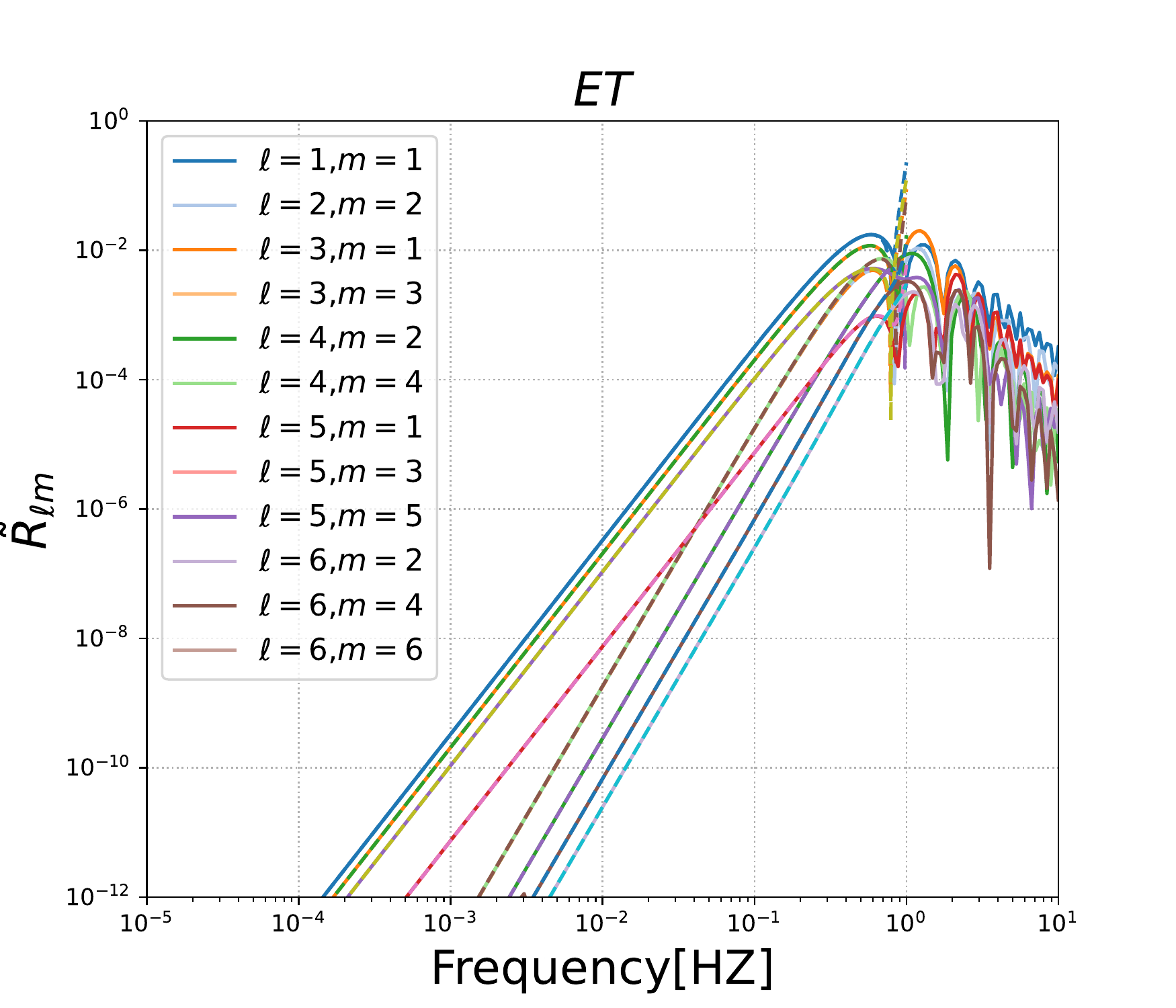}}}
  \end{center}
  \caption{\label{fig:TianQin-Respond-lser}ORFs for different values of $\ell$ and $m$ in the TianQin detector. The dashed line represents the result obtained by expanding $\tilde{R}_{11}^{\ell
m}$ and $\tilde{R}_{12}^{\ell m}$ using the small parameter $a = f / f_{\ast}$, while the solid line represents the numerical result. See equations \eqref{R-AET-lodd} to \eqref{R-AET-leven3}.}
\end{figure*}

From equations \eqref{Rl-OOp} to\eqref{R-AET-leven3}, it can be seen that the calculation of the ORFs for the $AET$ channel combination of the detector can be simplified to the calculation of $\tilde{R}_{11}^{\ell m}$ and $\tilde{R}_{12}^{\ell m}$. The calculation of $\tilde{R}{11}^{\ell m}$ and $\tilde{R}{12}^{\ell m}$ involves the spherical harmonic decomposition of $R_{ij} (f, \hat{k})$. Due to the singularity of the transfer function given in equation \eqref{eq:twoway-Transfunctoion-New}, the computation of the ORFs is challenging. In contrast, for ground-based detectors, the transfer function can be approximated as 1 at low frequencies, simplifying the expression \cite{Romano:2016dpx}. As shown in Figure \ref{fig:TianQin-Respond-lser}, the ORFs $\tilde{R}_{OO'}^{\ell m} (f)$ were calculated using numerical methods and series expansions. The dashed lines represent the results of expanding $\tilde{R}_{11}^{\ell m}$ and $\tilde{R}_{12}^{\ell m}$ as a series in $a = f/f{\ast}$, while the solid lines represent the numerical results. It can be observed that for sufficiently high orders of expansion, the series approximation provides a good fit in the low-frequency regime. In practical calculations, this approximation can be used instead of numerical computations to improve accuracy and computational speed. Furthermore, from Figure \ref{fig:TianQin-Respond-lser} and Figure \ref{fig:TianQin-Respond-l}, it can be seen that the auto-correlations of $AA$ and $EE$ have higher magnitudes compared to the cross-correlations of other channel combinations. The cross-correlation of $AE$ is intermediate, while the cross-correlation of $TT$ is the lowest.

\subsection{Angular Sensitivity}

Similar to the isotropic case, we can also define relevant sensitivities for anisotropy. The first step is to construct the signal-to-noise ratio (SNR) for the $AET$ channels. We begin by considering the Fourier transform of the signal in Equation \eqref{DFO}, using the short-time Fourier transform:
\begin{equation}
  \tilde{\Delta} d_O (f, t) \equiv \int_{t - \tau / 2}^{t + \tau / 2} \mathrm{d}t'
  \Delta d_O (t') e^{- 2 i \pi ft'} \hspace{0.27em} . \label{Fourier-signal}
\end{equation}
This signal, if present, is then added to the instrumental noise in the measurement:
\begin{equation}
  \tilde{m}_O  (f, t) \equiv \tilde{\Delta} d_O (f, t) + \tilde{n}_O  (f, t)
  \hspace{0.27em} . \label{eq:data_noise_model}
\end{equation}
Assuming the noise is Gaussian, it is diagonal in the $A$, $E$, and $T$ bases, given by:
\begin{equation}
  \langle n_O (f) n_{O'} (f) \rangle \equiv \frac{1}{2} \delta (f - f')
  \delta_{OO'} N_O (f) \hspace{0.27em}, \label{NO-def}
\end{equation}
where the explicit expression for $N_O(f)$ is provided in Appendix \ref{sec:TDI_Noise}. We define the estimator as:
\begin{widetext}
\begin{equation}
  \mathcal{C} \equiv \sum_{O, O'} \int_0^T \mathrm{d}t \int_{- \infty}^{+ \infty} \mathrm{d}f
  [\tilde{m}_O  (f, t)  \tilde{m}_{O'}^{\ast}  (f, t) - \langle \tilde{n}_O 
  (f, t)  \tilde{n}_{O'}^{\ast}  (f, t) \rangle]  \tilde{Q}_{OO'} (t, f)
  \hspace{0.27em}, \label{C-def}
\end{equation}
\end{widetext}
where $\tilde{Q}_{OO'} (t,f)$ is the chosen filtering function to maximize the signal-to-noise ratio (SNR) of this measurement \cite{Smith:2019wny}. The measurement time is denoted by $T$, and for simplicity, we integrate over equal times, neglecting correlations between measurements taken at different times. In the estimation process, the expectation value of the instrument noise $\tilde{n}_O$ correlated with the measurement $\tilde{\Delta} d_O$ is subtracted, yielding an unbiased estimate of the SGWB. Assuming noise dominance, the signal-to-noise ratio (SNR) can be obtained by computing $\langle \mathcal{C} \rangle$ and $\langle |\mathcal{C}|^2 \rangle$ \cite{LISACosmologyWorkingGroup:2022kbp}. The SNR is given by:
\begin{eqnarray}
  \mathrm{SNR} &=& \frac{\langle C \rangle}{\sqrt{\langle |\mathcal{C}|^2
  \rangle}} \nonumber\\
  & = & \frac{\sum_{OO'} \int_0^{\infty} \mathrm{d}f \int_0^T \mathrm{d}t \gamma_{OO'} (f,
  t) \mathcal{Q}_{OO'} (t, f)}{\sqrt{\sum_{OO'} \int_0^T dt \int_0^{+ \infty}
  df | \mathcal{Q}_{OO'} (t, f) |^2}} \hspace{0.27em}, \nonumber\\
  &  &\label{SNR-app}    
\end{eqnarray}
where the utilized $\gamma_{OO'} (f, t)$ is defined as:
\begin{eqnarray}
  \gamma_{OO'} (f, t) &\equiv& \frac{3 H_0^2}{4 \pi^2  \sqrt{4 \pi}} 
  \frac{\Omega_{\mathrm{GW}} (f)}{f^3}  \hspace{0.27em}\nonumber\\
  & & \frac{\sum_{\ell, m}
  \delta_{\tmop{GW}, \ell m} (f) \mathcal{R}_{OO'}^{\ell m} (f)}{\sqrt{N_O (f)
  N_{O'} (f)}} .    
\end{eqnarray}
Thus, the SNR is maximized by selecting $\mathcal{Q}_{OO'} (t, f) = c \times \gamma_{OO'}^{\ast} (f, t)$, where $c$ is a constant, resulting in the optimal SNR:
\begin{widetext}
\begin{equation}
  \mathrm{SNR} = \frac{3 H_0^2}{4 \pi^2  \sqrt{4 \pi}}  \sqrt{\sum_{O, O'}
  \int_0^{\infty} \mathrm{d}f \int_0^T \mathrm{d}t \frac{\Omega_{\tmop{GW}}^2 (f)}{f^6 N_{OO}
  (f) N_{O' O'} (f)} \left| \sum_{\ell, m} \delta_{\tmop{GW}, \ell m} (f)
  R_{OO'}^{\ell m} (f) \right|^2} \hspace{0.27em} . \label{SNR-ell-m}
\end{equation}
\end{widetext}
where $\delta_{\mathrm{GW}, \ell m}$ is defined in \eqref{e:dec}.

\subsubsection{Sensitivity to $\ell$-th Multipole}\label{subs:sensitivity-l}

Equation \eqref{SNR-ell-m} can be used to estimate the SNR of the SGWB, which represents the total SNR considering contributions from all possible multipoles. Although it is not explicitly expressed, as discussed in Section \ref{sec:tq-anni-renspons}, for detectors like LISA, the quantity $\mathcal{R}_{OO'}^{\ell m} (f)$ also depends on time due to its dependence on the satellite positions.
Let's assume that only one multipole dominates the SGWB for each $\ell$, and for a given $\ell$, different multipoles with different $m$ values follow the same Gaussian distribution. This corresponds to statistically isotropic SGWB with respect to $\ell$, and the correlation functions have been provided by Equation \eqref{Cell-def}. Considering these assumptions and for the purpose of comparison with LISA, we can express the SNR in Equation \eqref{SNR-ell-m} as a sum over various multipoles:
\begin{equation}
  \langle \mathrm{SNR} \rangle \equiv \sqrt{\sum_{\ell} \langle \mathrm{SNR}
  \rangle_{\ell}^2} \hspace{0.27em},
\end{equation}
where for each multipole $\ell$, we have:
\begin{widetext}
\begin{equation}
  \langle \mathrm{SNR} \rangle_{\ell} = \frac{3 H_0^2}{4 \pi^2  \sqrt{4 \pi}} 
  \sqrt{\sum_{O, O'} \int_0^{\infty} \mathrm{d}f \int_0^T \mathrm{d}t
  \frac{\Omega_{\mathrm{GW}}^2 (f)}{f^6 N_O (f) N_{O'} (f)}
  C_{\ell}^{\mathrm{GW}} [R_{OO'}^{\ell} (f)]^2} \label{SNR-l}
\end{equation}
\end{widetext}
where the response function $R_{OO'}^{\ell} (f)$ is defined in Equation \eqref{Rl-OOp}, but Equation \eqref{Rl-OOp} is scaled. As shown in Equations \eqref{NA,ETQ} and \eqref{NTTQ}, the noise functions can be rescaled accordingly based on the scaling of $R_{OO'}^{\ell} (f)$. Additionally, as mentioned earlier, the response function $R_{OO'}^{\ell} (f)$ is time-independent for statistically isotropic signals, allowing us to simplify the time integral in Equation \eqref{SNR-l} to a factor of $T$, resulting in the SNR increasing linearly with the square root of the observation time.
Finally, dividing the Hubble constant by its dimensionless quantity factor $h$ and considering the combination $\Omega h^2$, we obtain:
\begin{widetext}
\begin{equation}
  \langle \mathrm{SNR} \rangle_{\ell} = \frac{3 (H_0 / h)^2}{4 \pi^2  \sqrt{4
  \pi}}  \sqrt{T \sum_{O, O'} \int_0^{\infty} \mathrm{d}f \frac{\Omega_{\mathrm{GW}}^2
  (f) h^4}{f^6  \tilde{N}_O (f)  \tilde{N}_{O'} (f)} C_{\ell}^{\mathrm{GW}}
  [\tilde{R}_{OO'}^{\ell} (f)]^2} . \label{SNR-l-2}
\end{equation}
\end{widetext}
Based on this expression, we can define the ``channel-channel" sensitivity as:
\begin{equation}
  \Omega_{\tmop{GW}, OO', n}^{\ell} (f) h^2 \equiv \frac{4 \pi^2  \sqrt{4
  \pi}}{3 (H_0 / h)^2}  \frac{f^3  \sqrt{\tilde{N}_O (f)  \tilde{N}_{O'}
  (f)}}{\tilde{R}_{OO'}^{\ell} (f)} \hspace{0.27em}, \label{sensitivity-l-OOp}
\end{equation}
and the optimal weighted sum over the three channels as:

\begin{figure*}
  \begin{center}
    {\resizebox{440pt}{360pt}{\includegraphics{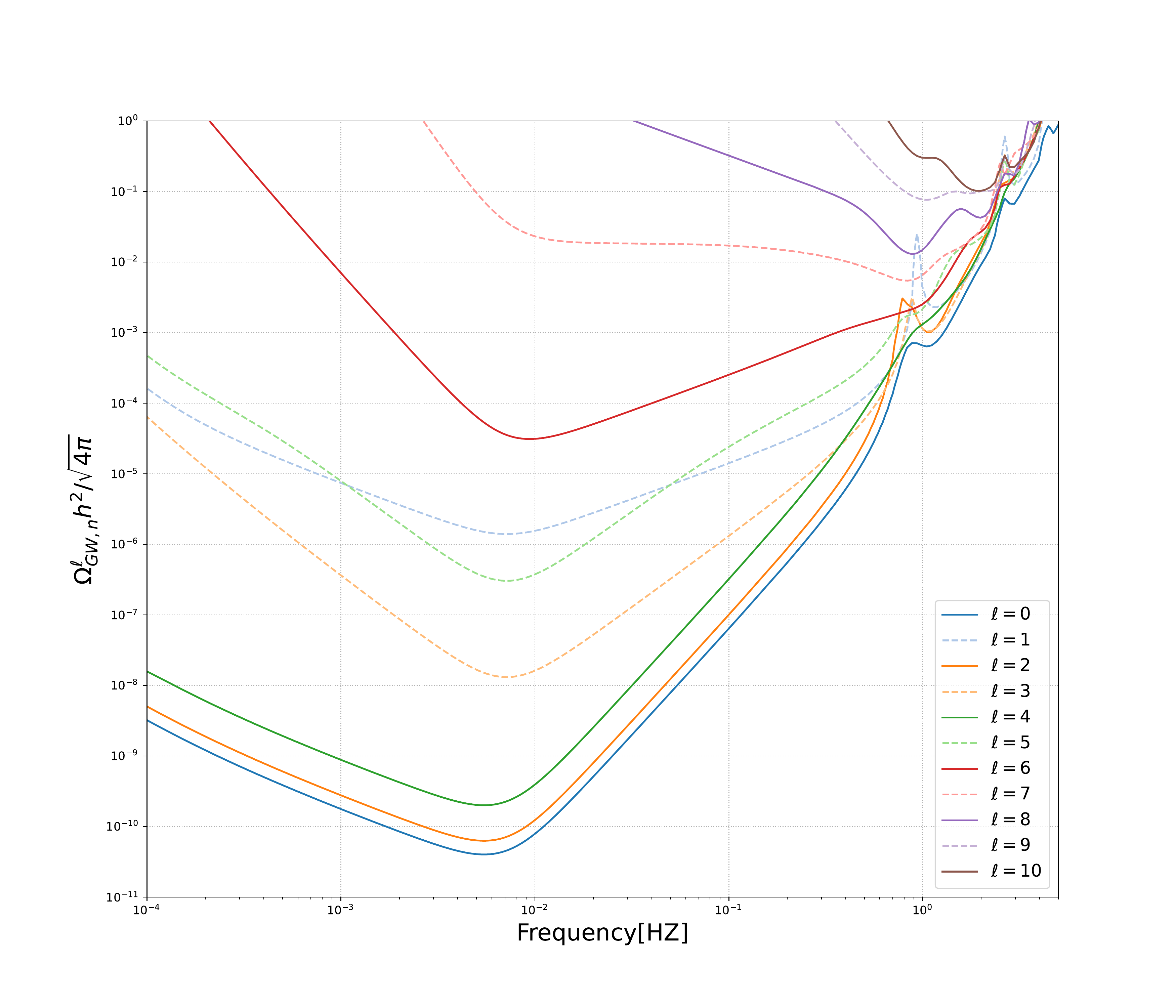}}} 
  \end{center}
  \caption{\label{fig:TianQin-sensitivity-l}Sensitivity curve of TianQin for a given $\ell$-th multipole, with $\ell_{\max} = 10$. The expressions used are given in Eq. \eqref{sensitivity-l-OOp} and \eqref{sensitivity-l}.}
\end{figure*}
\begin{equation}
  \Omega_{\tmop{GW}, n}^{\ell} (f) h^2 \equiv \left\{ \sum_{O, O'} \left[
  \frac{1}{\Omega_{\tmop{GW}, OO', n}^{\ell} (f) h^2} \right]^2 \right\}^{- 1
  / 2} \hspace{0.27em} . \label{sensitivity-l}
\end{equation}
We can immediately obtain:
\begin{equation}
  \langle \mathrm{SNR} \rangle_{\ell}^2 = T \int_0^{\infty} \mathrm{d} f \left[
  \frac{\sqrt{C_{\ell}^{\mathrm{GW}}} \Omega_{\mathrm{GW}} (f)
  h^2}{\Omega_{\tmop{GW}, n}^{\ell} (f) h^2} \right]^2 \hspace{0.27em} .
  \label{eq:ang_snr}
\end{equation}
As shown in Figure \ref{fig:TianQin-sensitivity-l}, the angular sensitivity for multipole orders $\ell = 0, 1, \ldots, 10$ has been computed. It is important to note that the curves shown have been rescaled using $Y_{00} = 1 / \sqrt{4 \pi}$. When $\ell = 0$, the result aligns with the sensitivity curve obtained for the isotropic SGWB scenario using the $AET$ channel.

\subsubsection{Sensitivity for $\ell m$-mode Multipoles}\label{subs:sensitivity-lm}

In Section \ref{s:expectationvalues}, it was discussed that a comprehensive analysis of the individual contributions from various multipoles requires a component separation process. This involves extracting the multipole amplitudes $\tilde{I}_{\ell m}(f)$ by deconvolving the time-correlated data streams measured by the satellite. In the case of the TianQin constellation, since the $z$-axis is fixed to the direction of J0806 relative to the distance from Earth to the Sun, J0806 can be regarded as an infinitely distant point. This means that the $z$-axis of TianQin remains fixed, allowing for the differentiation of different $\ell m$ multipoles. As a result, TianQin has the capability to distinguish and separate the contributions from different $\ell m$ multipoles.

The expression \eqref{SNR-ell-m} can be rewritten as:
\begin{equation}
  \langle \mathrm{SNR} \rangle \equiv \sqrt{\sum_{\ell} \sum^{m = \ell}_{m = -
  \ell} \langle \mathrm{SNR} \rangle_{\ell m}^2} \hspace{0.27em},
\end{equation}
where, for each multipole,
\begin{widetext}
\begin{equation}
  \langle \mathrm{SNR} \rangle_{\ell m} = \frac{3 H_0^2}{4 \pi^2  \sqrt{4
  \pi}}  \sqrt{\sum_{O, O'} \int_0^{\infty} \mathrm{d}f \int_0^T \mathrm{d}t
  \frac{\Omega_{\mathrm{GW}}^2 (f)}{f^6 N_O (f) N_{O'} (f)} C_{\ell
  m}^{\mathrm{GW}} [R_{OO'}^{\ell m} (f)]^2} \hspace{0.27em}, \label{SNR-lm}
\end{equation}
\end{widetext}
where the response function $R_{OO'}^{\ell}(f)$ is defined by equation \eqref{Rlm-OOp}, with the note that equation \eqref{Rlm-OOp} has been scaled. As shown in equations \eqref{NA,ETQ} and \eqref{NTTQ}, the noise functions can be rescaled accordingly based on the scaling of $R_{OO'}^{\ell}(f)$. Additionally, as mentioned before, the response function $R_{OO'}^{\ell}(f)$ is time-independent for statistically isotropic signals, so the integration over time in equation \eqref{SNR-lm} leads to the general characteristic that the SNR grows with the square root of the observation time. Finally, by dividing the uncertainty in the Hubble constant by its redefined value $h$ and considering the combination $\Omega h^2$, we obtain:
\begin{widetext}
\begin{equation}
  \langle \mathrm{SNR} \rangle_{\ell m} = \frac{3 (H_0 / h)^2}{4 \pi^2 
  \sqrt{4 \pi}}  \sqrt{T \sum_{O, O'} \int_0^{\infty} \mathrm{d}f
  \frac{\Omega_{\mathrm{GW}}^2 (f) h^4}{f^6  \tilde{N}_O (f)  \tilde{N}_{O'}
  (f)} C_{\ell m}^{\mathrm{GW}} [\tilde{R}_{OO'}^{\ell m} (f)]^2}
  \hspace{0.27em} . \label{SNR-lm-2}
\end{equation}
\end{widetext}
According to this expression, we can define the ``channel-channel" sensitivity as:
\begin{equation}
  \Omega_{\tmop{GW}, OO', n}^{\ell m} (f) h^2 \equiv \frac{4 \pi^2  \sqrt{4
  \pi}}{3 (H_0 / h)^2}  \frac{f^3  \sqrt{\tilde{N}_O (f)  \tilde{N}_{O'}
  (f)}}{\tilde{R}_{OO'}^{\ell m} (f)} \hspace{0.27em},
  \label{sensitivity-lm-OOp}
\end{equation}
and the optimal weighted sum over the three channels is given by

\begin{figure*}
  \begin{center}
    {\resizebox{440pt}{360pt}{\includegraphics{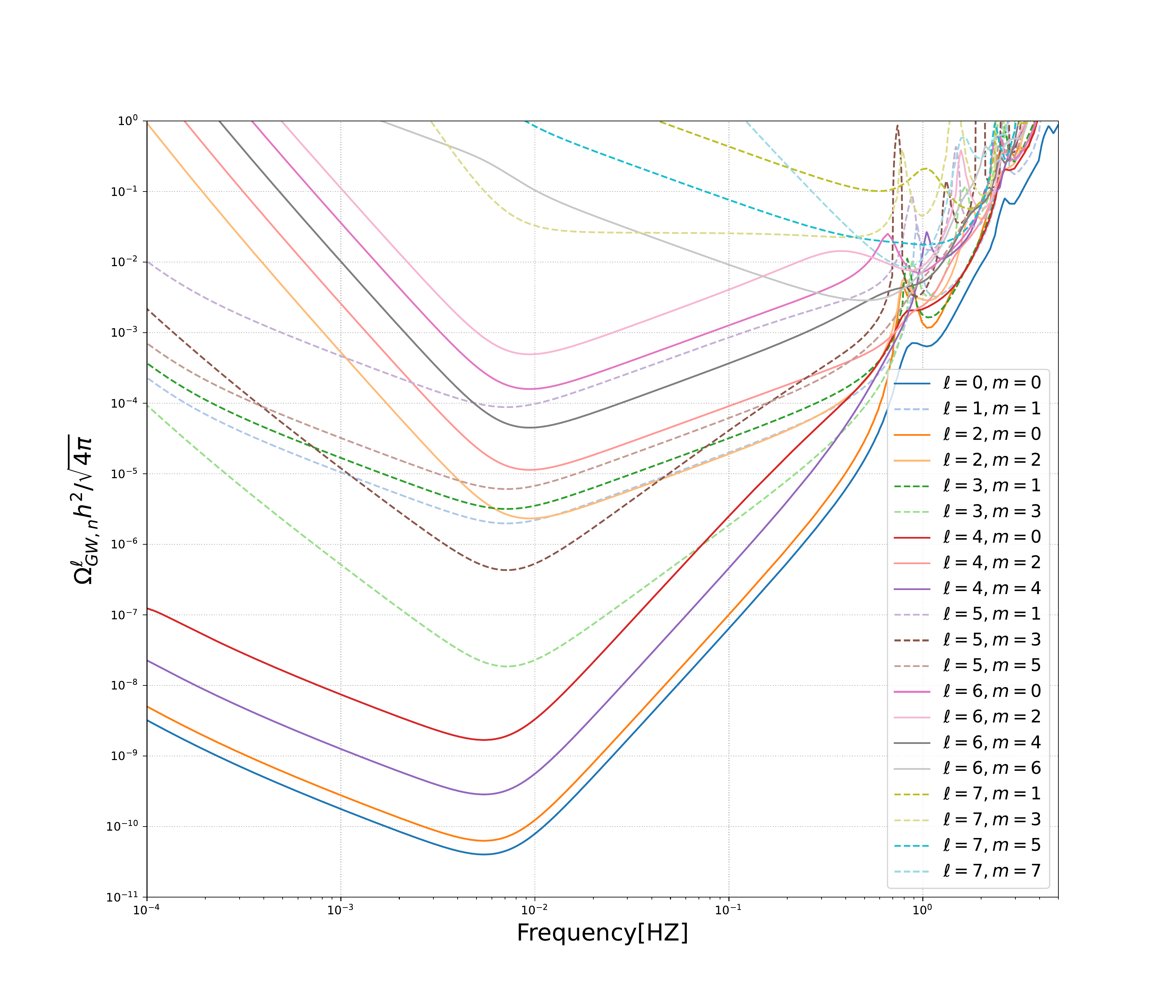}}} 
  \end{center}
  \caption{\label{fig:TianQin-sensitivity-lm}Sensitivity curves for different $\ell$ and $m$ modes with TianQin. $\ell_{\max} = 7$. The expressions used are given in equations \eqref{sensitivity-lm-OOp} and \eqref{sensitivity-lm}.}
\end{figure*}
\begin{equation}
  \Omega_{\tmop{GW}, n}^{\ell m} (f) h^2 \equiv \left\{ \sum_{O, O'} \left[
  \frac{1}{\Omega_{\tmop{GW}, OO', n}^{\ell m} (f) h^2} \right]^2 \right\}^{-
  1 / 2} \hspace{0.27em} . \label{sensitivity-lm}
\end{equation}
We can also immediately obtain the expression for the squared signal-to-noise ratio (SNR) for each $\ell m$ mode:
\begin{equation}
  \langle \mathrm{SNR} \rangle_{\ell m}^2 = T \int_0^{\infty} \mathrm{d} f \left[
  \frac{\sqrt{C_{\ell m}^{\mathrm{GW}}} \Omega_{\tmop{GW}} (f)
  h^2}{\Omega_{\tmop{GW}, n}^{\ell m} (f) h^2} \right]^2 \hspace{0.27em} .
  \label{eq:ang_snr-lm}
\end{equation}
where $C_{\ell m}^{\mathrm{GW}}$ represents the gravitational wave angular power spectrum for the specific $\ell m$ mode.
As shown in Figure \ref{fig:TianQin-sensitivity-lm}, the sensitivity curves for different $\ell$ and $m$ modes are displayed. It can be observed that TianQin is capable of distinguishing different $\ell m$ modes.

\section{Fisher Estimation}\label{s:Fisher_Est}

In this section, we employ the Fisher matrix method to statistically predict the detectability of the lowest multipole components of the angular power spectrum of the SGWB by the TianQin constellation. We consider a total observation time of $t_{\mathrm{obs}} = 5$ years with an operational efficiency of 50\% (corresponding to the TianQin 3+3 working mode). The frequency resolution is set to $\Delta f = 10^{-6}$, which corresponds to dividing the TDI data stream into time blocks of 11.5 days, and the average value of the block spectra is used as the final spectrum.

\subsection{Likelihood Function}

\subsubsection{Likelihood Function Dependent on $\ell$}

Assuming statistical independence among different multipoles and assuming that all $m$ modes for each $\ell$ follow the same distribution, we can consider each multipole individually to extract the information contained in each multipole. Under this assumption, we can obtain measurements for each multipole.
Based on the results obtained in equation \eqref{eq:ang_snr}, we define the $\ell$-th multipole SGWB power spectrum as follows:
\begin{equation}
  \Omega_{\tmop{GW}}^{\ell} (f) h^2 = \sqrt{C^{\tmop{GW}}_{\ell}}
  \Omega_{\tmop{GW}} (f) h^2,
\end{equation}
where $C_{\ell}^{\mathrm{GW}}$ is the angular power spectrum of the GW density contrast defined in equation \eqref{Cell-def}. For each multipole and channel combination $OO'$, we can assume that the data is described by a Gaussian likelihood function $\mathcal{L}_{\ell}$, given by:
\begin{equation}
  \ln \mathcal{L}_{\ell} = - \frac{N_c}{2}  \sum_{OO'} \sum_k
  \frac{(\mathscr{D}_{OO', \ell}^{(k)} - \mathscr{D}_{OO', \ell}^{(k),
  \mathrm{th}})^2}{\sigma_{OO', \ell}^{(k) 2}}, \label{eq:fisher:logl}
\end{equation}
where $N_{\mathrm{c}}$ is the number of data segments in the analysis. The summation is performed over frequencies (or frequency bins) $f_k$, $\mathscr{D}_{OO', \ell}$ represents the average signal in the data segments for the channel combination $OO'$, and $\mathscr{D}_{OO', \ell}^{\mathrm{th}}$ is the theoretical value of the data.
\begin{equation}
  \mathscr{D}^{(k), \mathrm{th}}_{OO', \ell} = \tilde{R}_{OO', \ell} (f_k)
  \Omega_{\tmop{GW}}^{\ell} (f_k) h^2 + \tilde{N}_{OO'}^{\Omega} (f_k) .
\end{equation}
Here, $\tilde{R}_{OO', \ell}$ represents the overlap reduction functions (ORFs) of the detector, and $\tilde{N}_{OO'}^{\Omega}$ is the noise defined in the previous section, expressed in terms of the $\Omega$ units.
Since we assume Gaussianity, the variance can be expressed as:
\begin{equation}
  \sigma_{OO', \ell}^{(k) 2} = \left( \mathscr{D}^{(k), \text{th}}_{OO', \ell}
  \right)^2 .
\end{equation}
The variance represents the uncertainty in the measured data, taking into account both the SGWB signal and the instrumental noise contributions.

\subsubsection{Likelihood Function Dependent on $\ell m$}

Considering the fixed pointing of TianQin towards J0806, TianQin can distinguish different $\ell m$ multipole components. Therefore, we can define the $\ell m$-order SGWB power spectrum $\Omega_{\tmop{GW}}^{\ell m} (f) h^2$ using Equation \eqref{eq:ang_snr-lm} as follows:
\begin{equation}
  \Omega_{\tmop{GW}}^{\ell m} (f) h^2 = \sqrt{C^{\tmop{GW}}_{\ell m}}
  \Omega_{\tmop{GW}} (f) h^2,
\end{equation}
where $C_{\ell m}^{\mathrm{GW}}$ is defined in Equation \eqref{Cellm-def}.
For each multipole component $\ell m$ and channel combination $OO'$, we can assume that the data follows a Gaussian likelihood function $\mathcal{L}_{\ell m}$, given by:
\begin{equation}
  \ln \mathcal{L}_{\ell m} = - \frac{N_c}{2}  \sum_{OO'} \sum_k
  \frac{(\mathscr{D}_{OO', \ell m}^{(k)} - \mathscr{D}_{OO', \ell m}^{(k),
  \mathrm{th}})^2}{\sigma_{OO', \ell m,}^{(k) 2}}, \label{eq:fisher:loglm}
\end{equation}
where $\tilde{R}_{OO', \ell m}$ is the overlap reduction functions (ORFs) of the detector for the $\ell m$ multipole, and $\tilde{N}_{OO'}^{\Omega}$ is the noise expressed in the $\Omega$ units, as defined in the previous sections. The variance can be expressed as:
\begin{equation}
  \sigma_{OO', \ell m}^{(k) 2} = \left( \mathscr{D}^{(k), \text{th}}_{OO',
  \ell m} \right)^2 .
\end{equation}

\subsection{Fisher Matrix for Power Law Spectrum}

\subsubsection{Fisher Matrix for $\ell$ Dependence}

To generalize the analysis for the detection of $\ell$-dependent multipole components, we consider an SGWB power law spectrum that peaks only at a reference multipole order $L$ with the following form:
\begin{equation}
  \Omega_{\tmop{GW}}^{\ell} (f) h^2 = \delta_{\ell, L} 10^{\log_{10}
  A_{\mathrm{c}}} \left( \frac{f}{f_{\mathrm{c}}} \right)^{\alpha} .
  \label{eq:fisher:signal}
\end{equation}
where $\log_{10}
A_{\mathrm{c}}$ represents the logarithmic amplitude and $\alpha$ is the spectral index. The pivot frequency $f_{\mathrm{c}} = 4.4
\times 10^{- 3}
\tmop{Hz}$ is chosen as the approximate frequency at which the multipole sensitivity peaks, as shown in Figures \ref{fig:TianQin-sensitivity-l} and \ref{fig:TianQin-sensitivity-lm}.

In practical observations, we consider a single data vector and compare it with the effective noise combination defined in Equation \eqref{sensitivity-l} shown in Figure \ref{fig:TianQin-sensitivity-l}, without summing over the channel indices $OO'$. Assuming a fixed noise model, the Fisher information matrix for the likelihood defined in Equation \eqref{eq:fisher:logl}, considering the amplitude index model, can be written as:

\begin{figure*}
  \begin{center}
    {\resizebox{220pt}{165pt}{\includegraphics{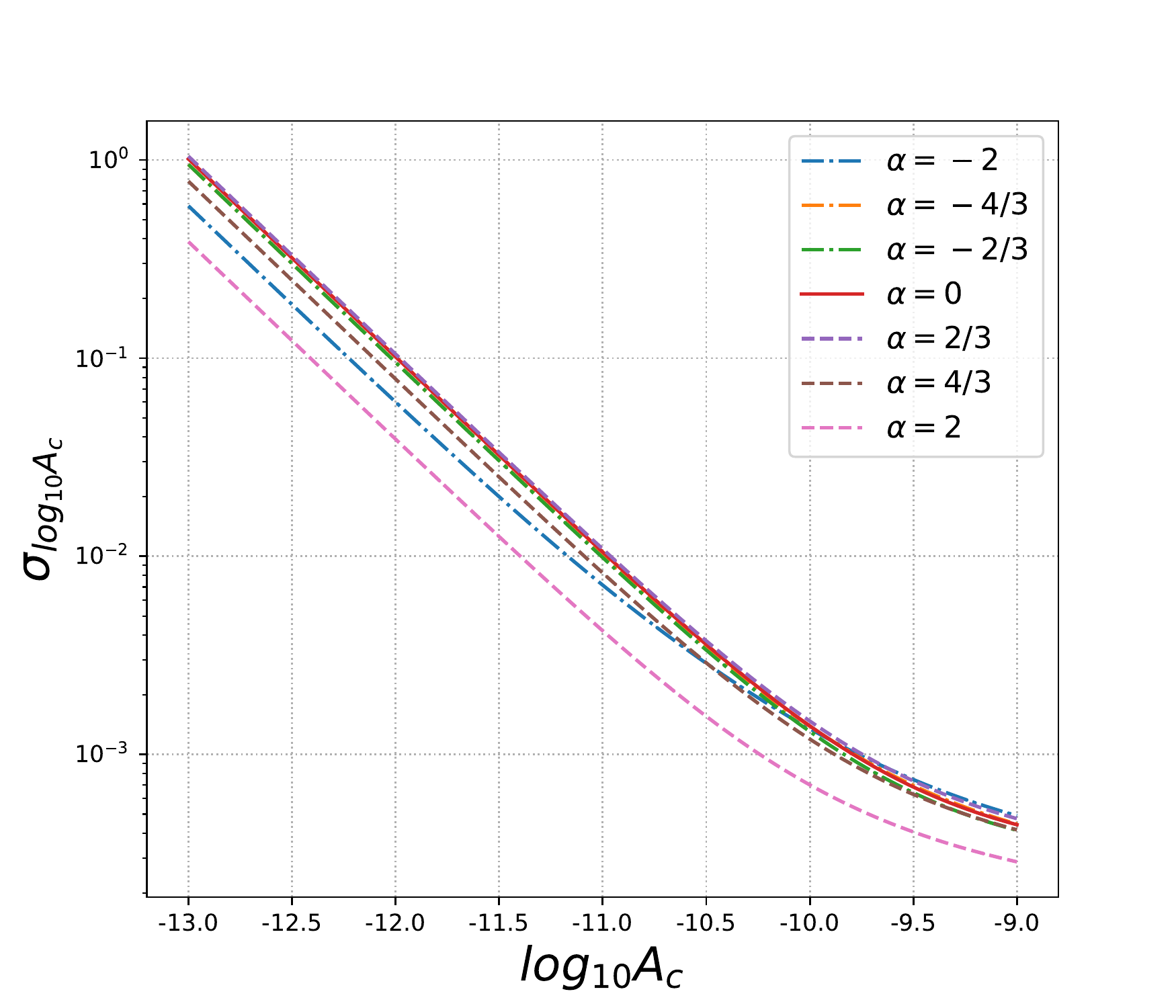}}}{\resizebox{220pt}{165pt}{\includegraphics{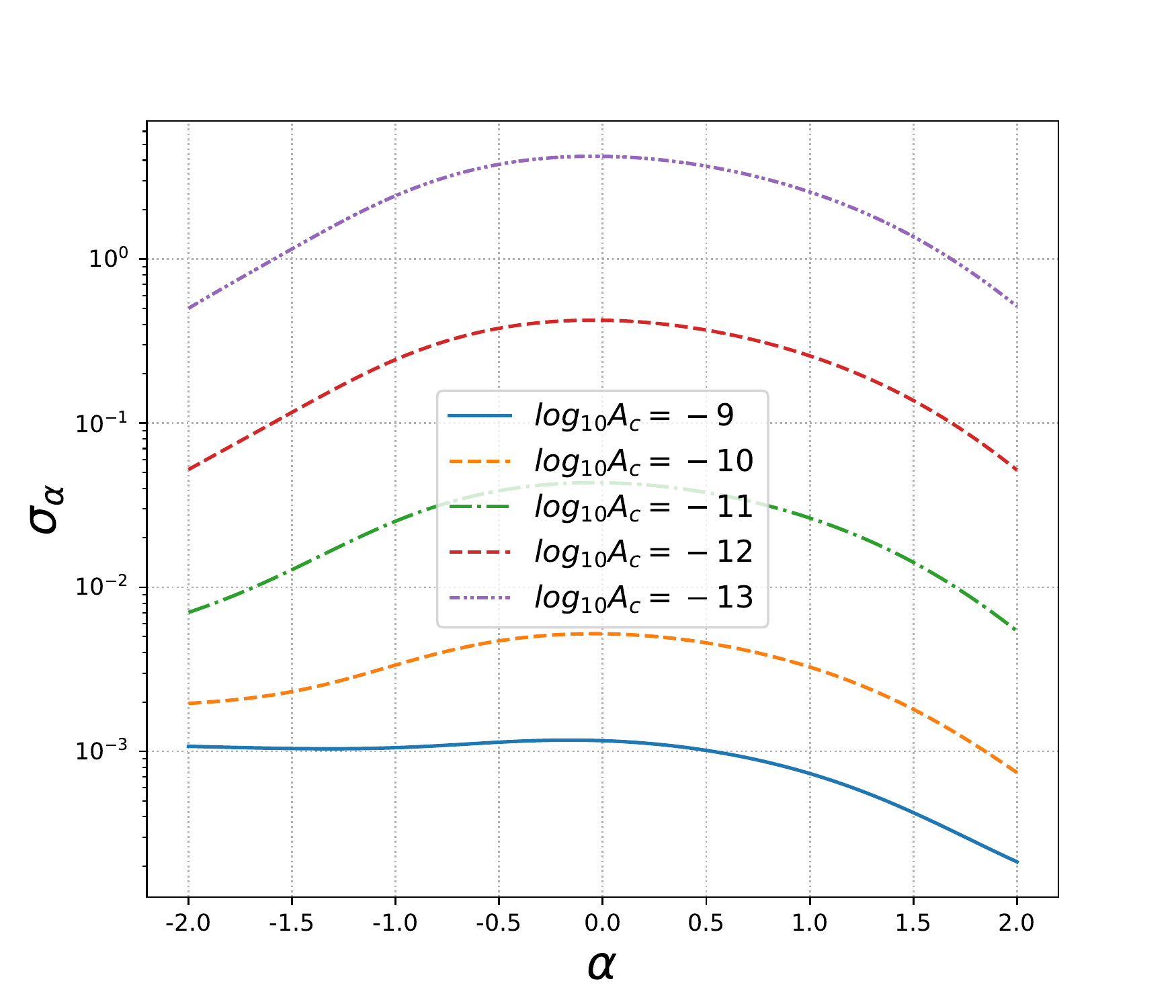}}}
    
    {\resizebox{220pt}{165pt}{\includegraphics{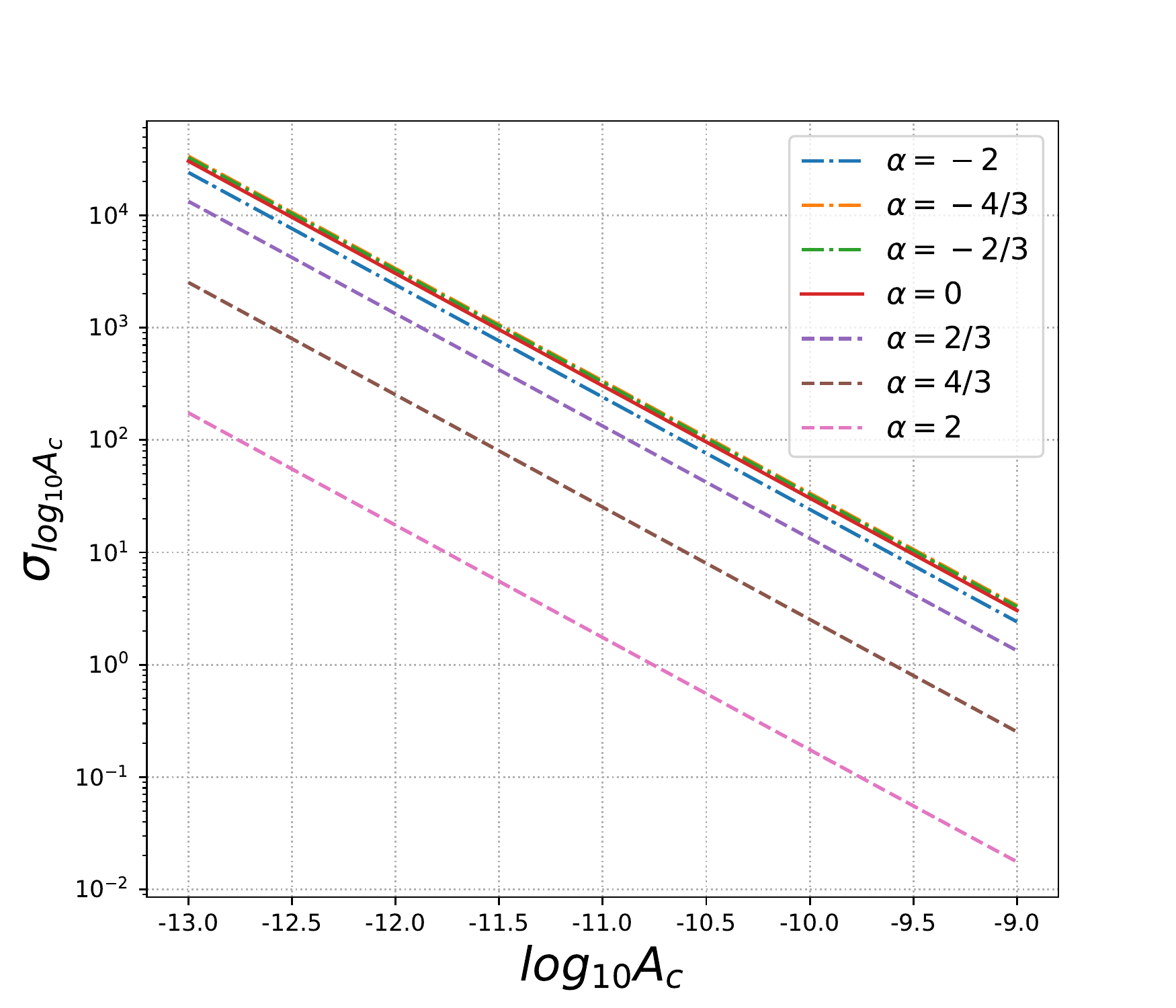}}}{\resizebox{220pt}{165pt}{\includegraphics{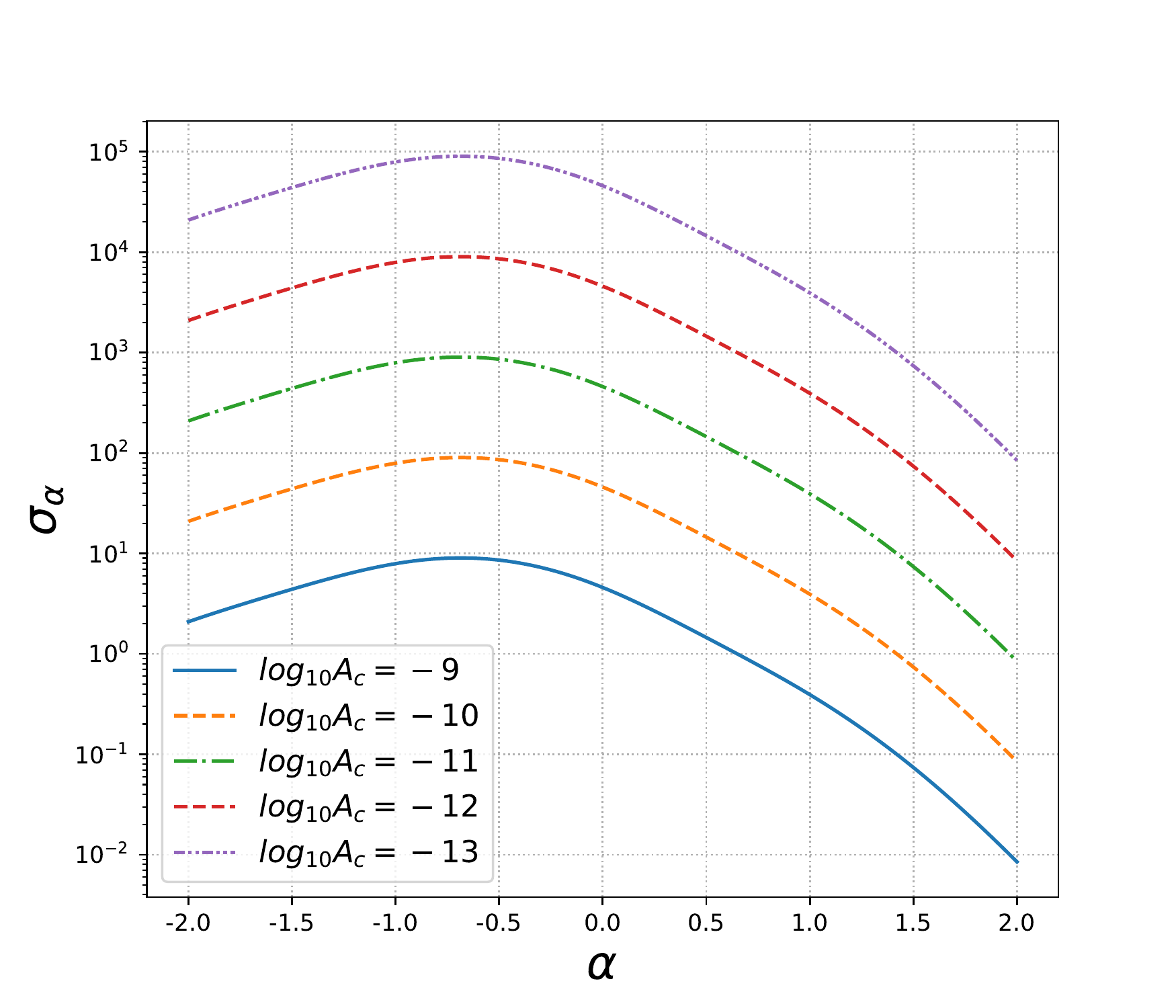}}}
    
    {\resizebox{220pt}{165pt}{\includegraphics{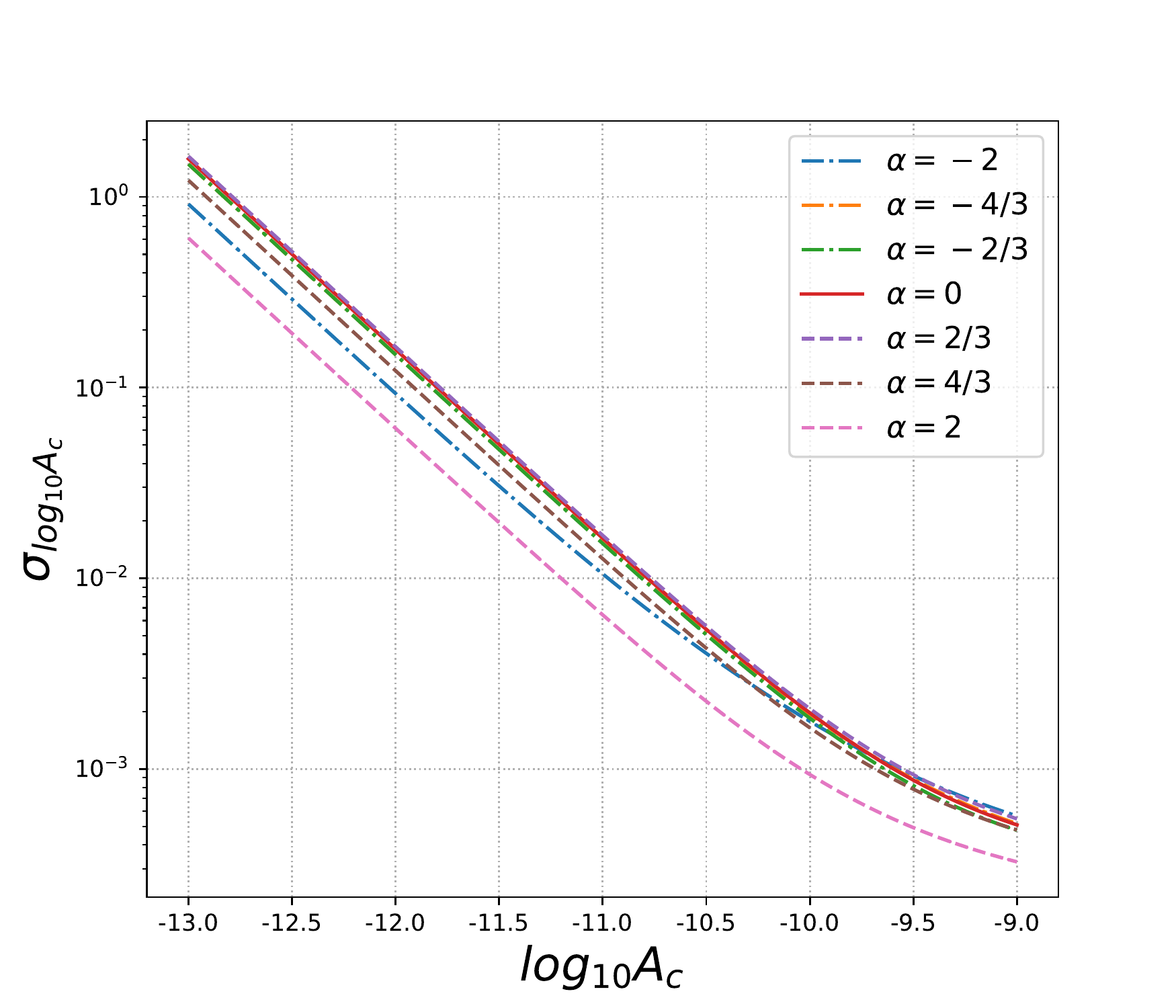}}}{\resizebox{220pt}{165pt}{\includegraphics{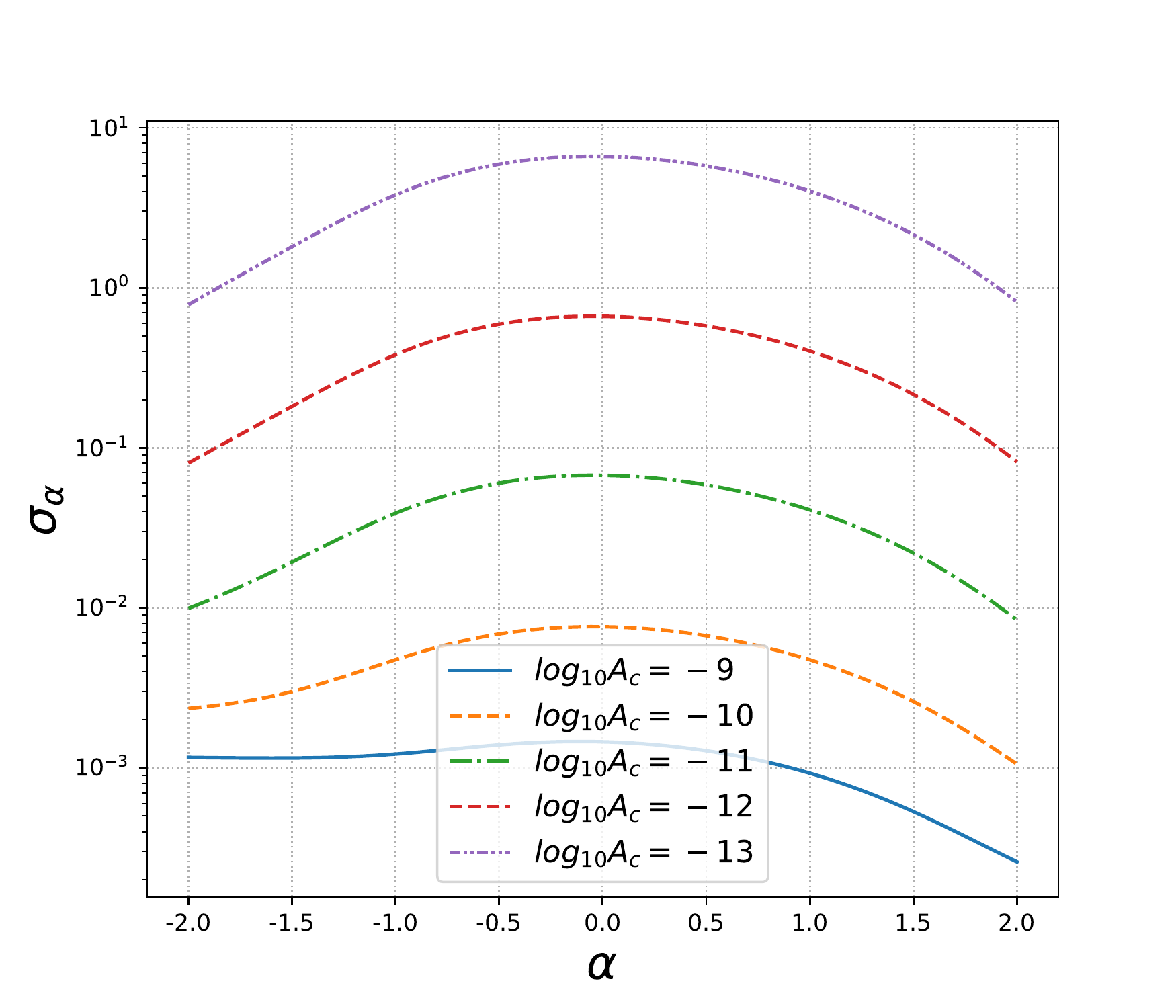}}}
  \end{center}
  \caption{\label{fig:forecasted_sigma_l}A series of reference values for the amplitude $\log_{10} A_{\mathrm{c}}$ and spectral index $\alpha$ of the SGWB for different multipole orders $\ell = 0, 1,$ and $2$ (top, middle, and bottom panels). The left panels show the $1\sigma$ predicted limits for $\log_{10} A_{\mathrm{c}}$, while the right panels show the $1\sigma$ predicted limits for $\alpha$. The limits are obtained by marginalizing over the other parameter, as described in Equations \eqref{eq:fisher:signal} and \eqref{eq:Cthetaphi}.}
\end{figure*}
\begin{eqnarray}
  \mathcal{C}_{\theta \rho}^{l - 1} &\equiv& \mathcal{F}^l_{\theta \rho} = -
  \partial_{\theta} \partial_{\rho} \ln \mathcal{L}|_{\textrm{best fit}} \nonumber\\&=&
  N_{\mathrm{c}}  \sum_k \tilde{R}_{OO', \ell} (f_k)  (\partial_{\theta}
  \Omega_{\tmop{GW}}^{\ell} (f_k) h^2) \nonumber\\& & (\partial_{\rho}
  \Omega_{\tmop{GW}}^{\ell} (f_k) h^2) \frac{1}{\sigma_{\ell}^{(k) 2}},
  \label{eq:Cthetaphi}
\end{eqnarray}
where $\theta$ and $\rho$ represent the combination of the signal model parameters $\log_{10} A_{\mathrm{c}}$ and $\alpha$. The corresponding partial derivatives are given by:
\begin{eqnarray}
  \partial_{\log_{10} A_{\mathrm{c}}} \Omega_{\tmop{GW}}^{\ell} h^2 &=& \log
  (10) \Omega_{\tmop{GW}}^{\ell} h^2, \nonumber\\ \partial_{\alpha}
  \Omega_{\tmop{GW}}^{\ell} h^2 &=& \log \left( \frac{f}{f_c} \right)
  \Omega_{\tmop{GW}}^{\ell} h^2 . \label{eq:partial-wgwl}
\end{eqnarray}
The Fisher estimates for the single-multipole power-law SGWB, defined in Equation \eqref{eq:fisher:signal}, are shown in Figure \ref{fig:forecasted_sigma_l} for the monopole ($\ell = 0$), dipole ($\ell = 1$), and quadrupole ($\ell = 2$) cases. The parameters considered are the SGWB amplitude $\log_{10} A_{\mathrm{c}}$ and the spectral index $\alpha$. In all cases, the standard deviations of each parameter correspond to the diagonal elements of the covariance matrix obtained from the Fisher information matrix, $\mathcal{C}_{\theta \rho}$, which is the inverse of the Fisher matrix.

As shown in Figure \ref{fig:forecasted_sigma_l} for $\ell = 0$ and $\ell = 2$, the solution for the logarithmic amplitude is independent of the sign of the spectral index because the pivot frequency is chosen to approximate the sensitivity peak of both multipoles. On the other hand, for $\ell = 1$, a positive spectral index can enhance the solution for the amplitude. This is partly because the corresponding sensitivity peak occurs at slightly higher frequencies relative to $f_c$. Additionally, as shown in Figure \ref{fig:TianQin-sensitivity-l}, the slope of the sensitivity curve for ``$\ell = 0, 2$ " is smaller at high frequencies. Therefore, for lower values of $|\alpha|$, the power-law for $\ell = 1$ is closer to the high-frequency noise spectrum compared to the power-law for $\ell = 0$ and $\ell = 2$.

For all multipoles, higher logarithmic amplitudes lead to more effective solutions for the spectral index. In the best-case scenario with a signal amplitude of $\Omega_{\mathrm{GW}}(f = f_{\mathrm{c}}) h^2 = 10^{-9}$, the spectral index can be resolved with uncertainties of approximately $10^{-3}$, $10$, and $10^{-3}$ for $\ell = 0, 1, 2$ multipoles, respectively. In a more pessimistic scenario with $\Omega_{\mathrm{GW}}(f = f_{\mathrm{c}}) h^2 = 10^{-13}$, the uncertainties in resolving the spectral index for $\ell = 0, 2$ can be greater than $0.1$ or even higher for most positive or negative values. However, for an SGWB spectrum with a spectral index between $-1$ and $1$, the corresponding uncertainty approaches the order of $10$. Therefore, in such low-amplitude scenarios, the detector will be more sensitive to models with SGWB spectra exhibiting strong variations. Additionally, due to the same reasons mentioned in the previous paragraph, the dipole $\ell = 1$ is more sensitive to positive spectral indices, while for $\ell = 0, 2$, the precision is almost symmetrically equal.

\subsubsection{Fisher Matrix for $\ell m$ Dependence}

If we consider the detection of $\ell m$-order multipoles, we can assume an SGWB power-law spectrum that peaks at $(L, M)$ as follows:
\begin{equation}
  \Omega_{\tmop{GW}}^{\ell m} (f) h^2 = \delta_{\ell, L} \delta_{\ell, M}
  10^{\log_{10} A_{\mathrm{c}}} \left( \frac{f}{f_{\mathrm{c}}}
  \right)^{\alpha} . \label{eq:fisher:signal-lm}
\end{equation}
With this assumption, we can obtain the Fisher information matrix corresponding to equation \eqref{eq:fisher:loglm}.
\begin{eqnarray}
  \mathcal{C}_{\theta \rho}^{\ell m - 1} &\equiv& \mathcal{F}^{\ell m}_{\theta
  \rho} = - \partial_{\theta} \partial_{\rho} \ln \mathcal{L}|_{\textrm{best
  fit}} \nonumber\\&=& N_{\mathrm{c}}  \sum_k \tilde{R}_{OO', \ell} (f_k) 
  (\partial_{\theta} \Omega_{\tmop{GW}}^{\ell m} (f_k) h^2) \nonumber\\& & (\partial_{\rho}
  \Omega_{\tmop{GW}}^{\ell m} (f_k) h^2) \frac{1}{\sigma_{\ell m}^{(k) 2}},
  \label{eq:Cthetaphi-lm}
\end{eqnarray}
\begin{figure*}
  \begin{center}
    {\resizebox{220pt}{165pt}{\includegraphics{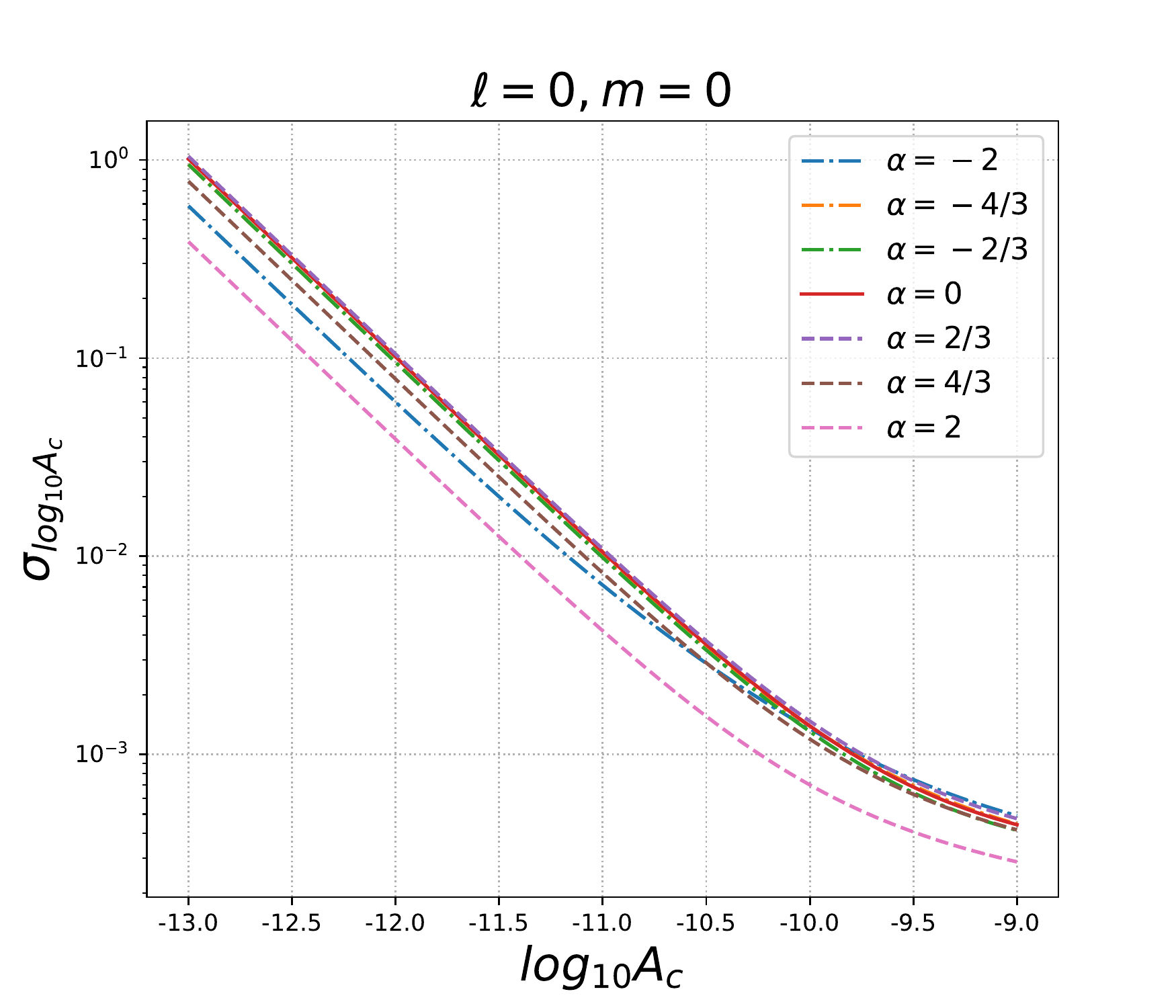}}}{\resizebox{220pt}{165pt}{\includegraphics{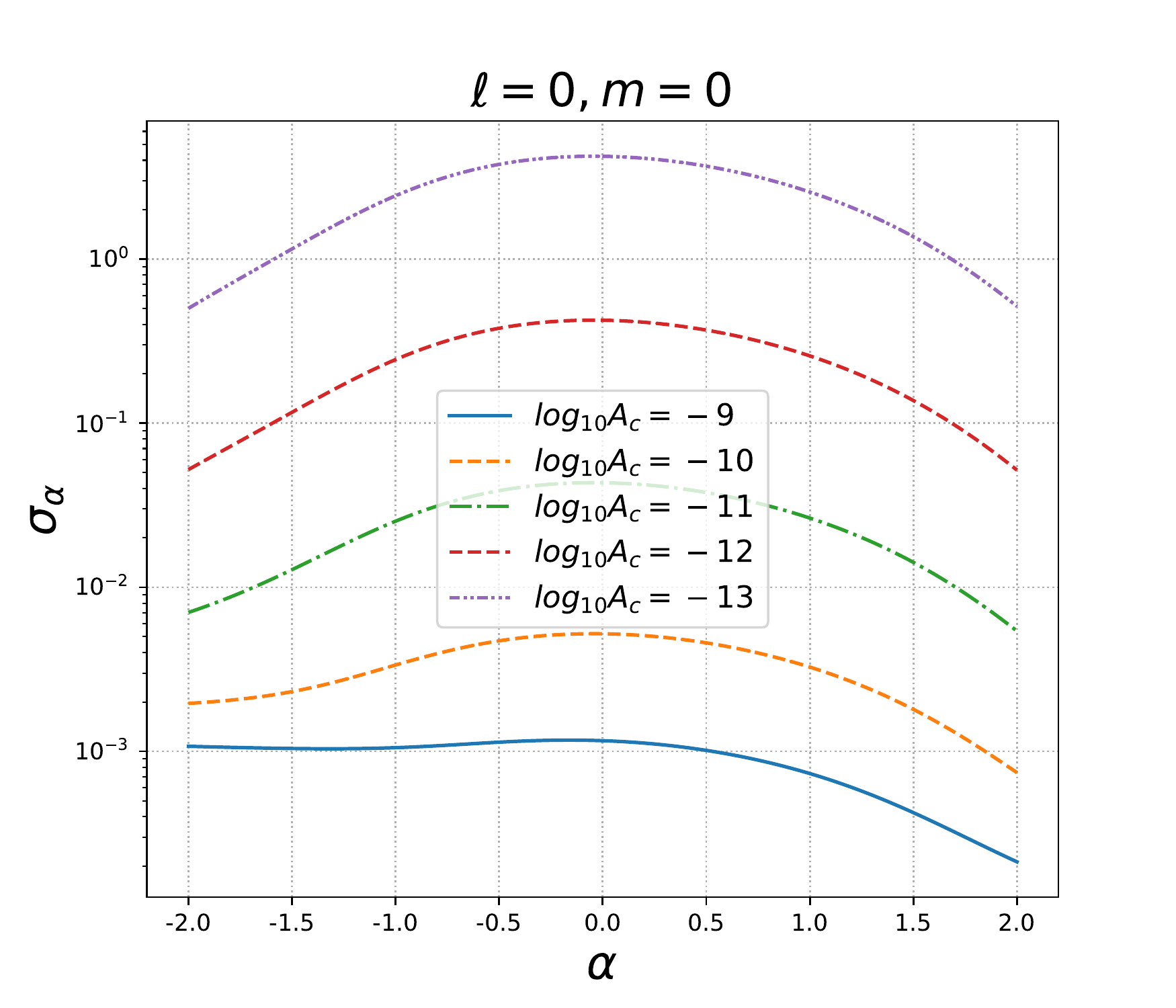}}}
    
    {\resizebox{220pt}{165pt}{\includegraphics{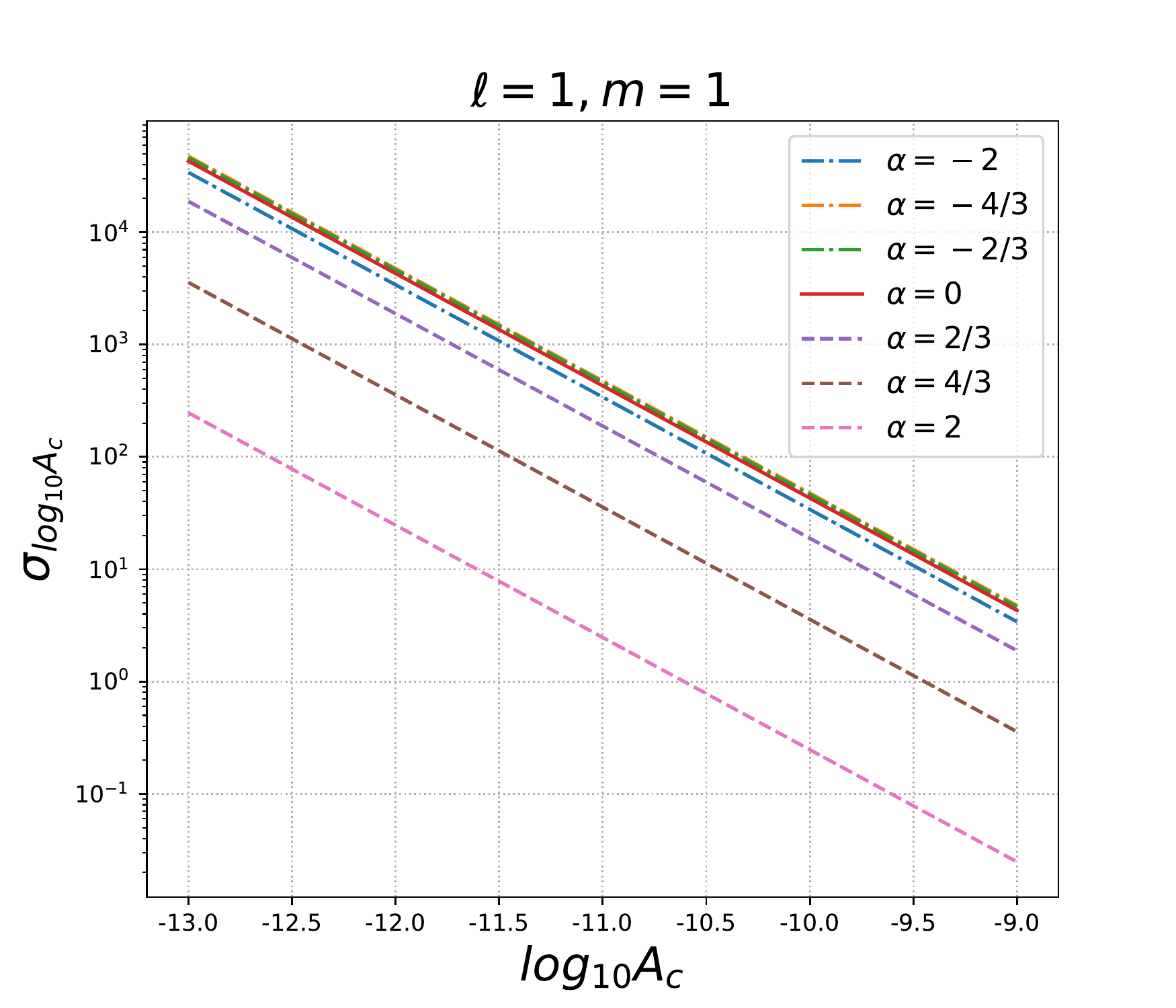}}}{\resizebox{220pt}{165pt}{\includegraphics{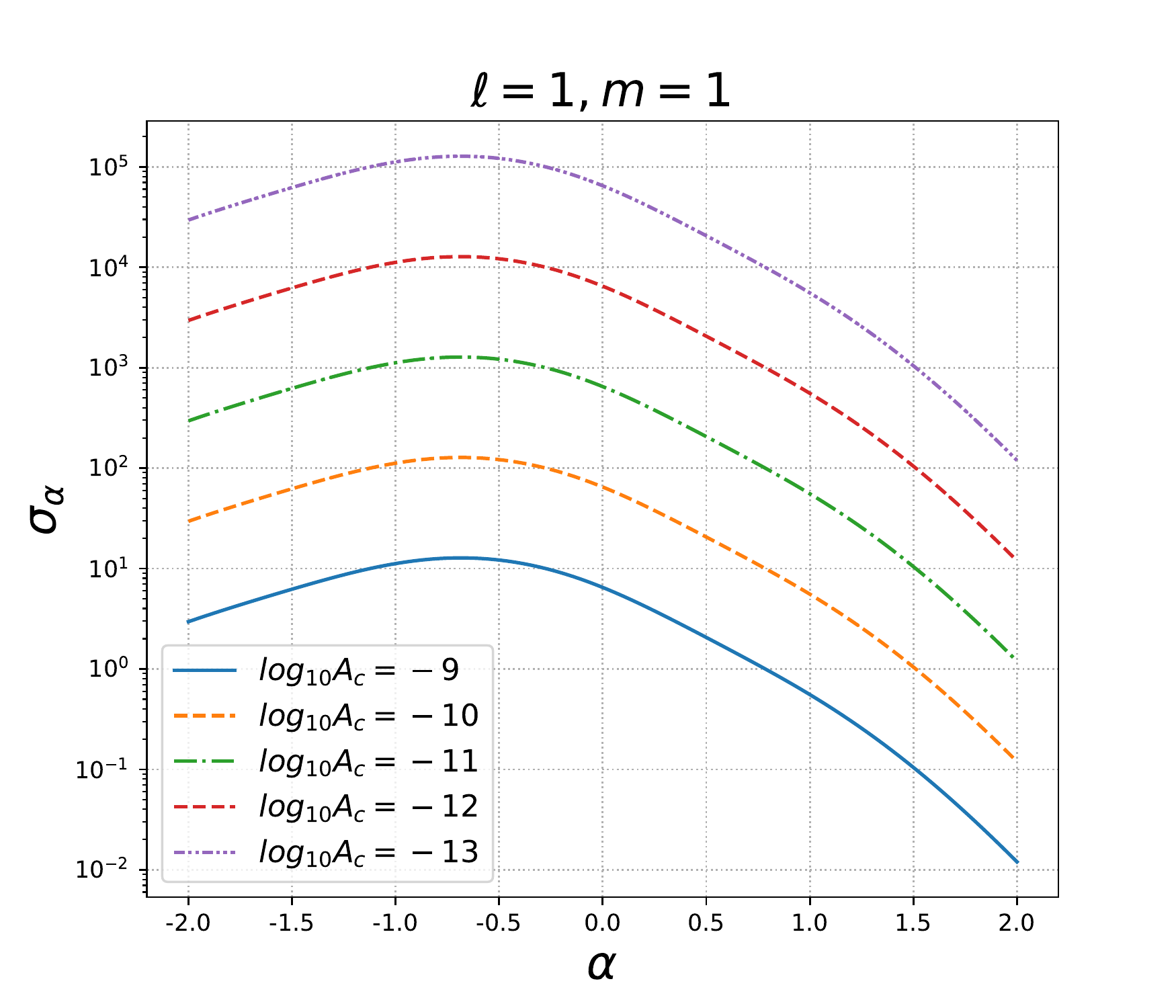}}}
    
    {\resizebox{220pt}{165pt}{\includegraphics{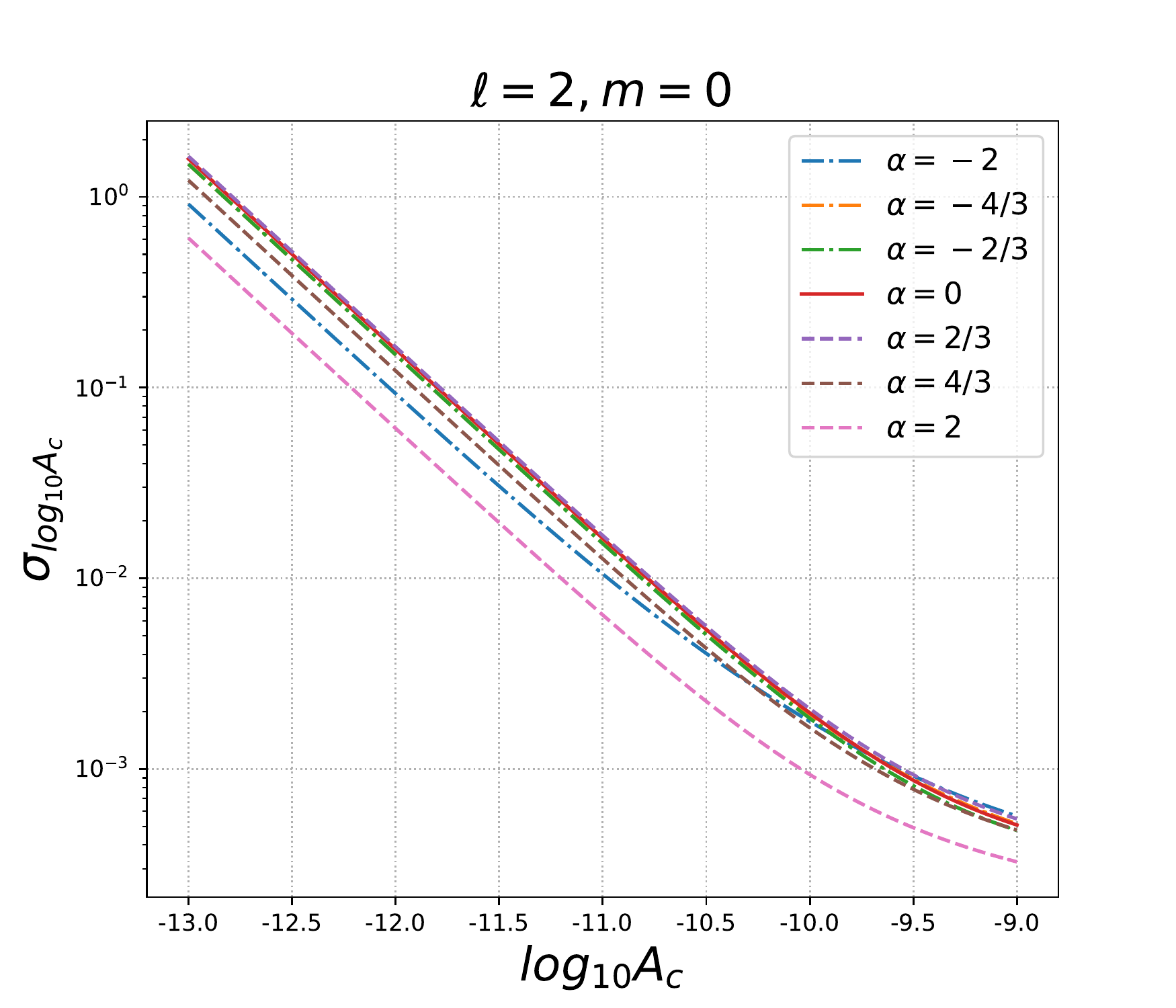}}}{\resizebox{220pt}{165pt}{\includegraphics{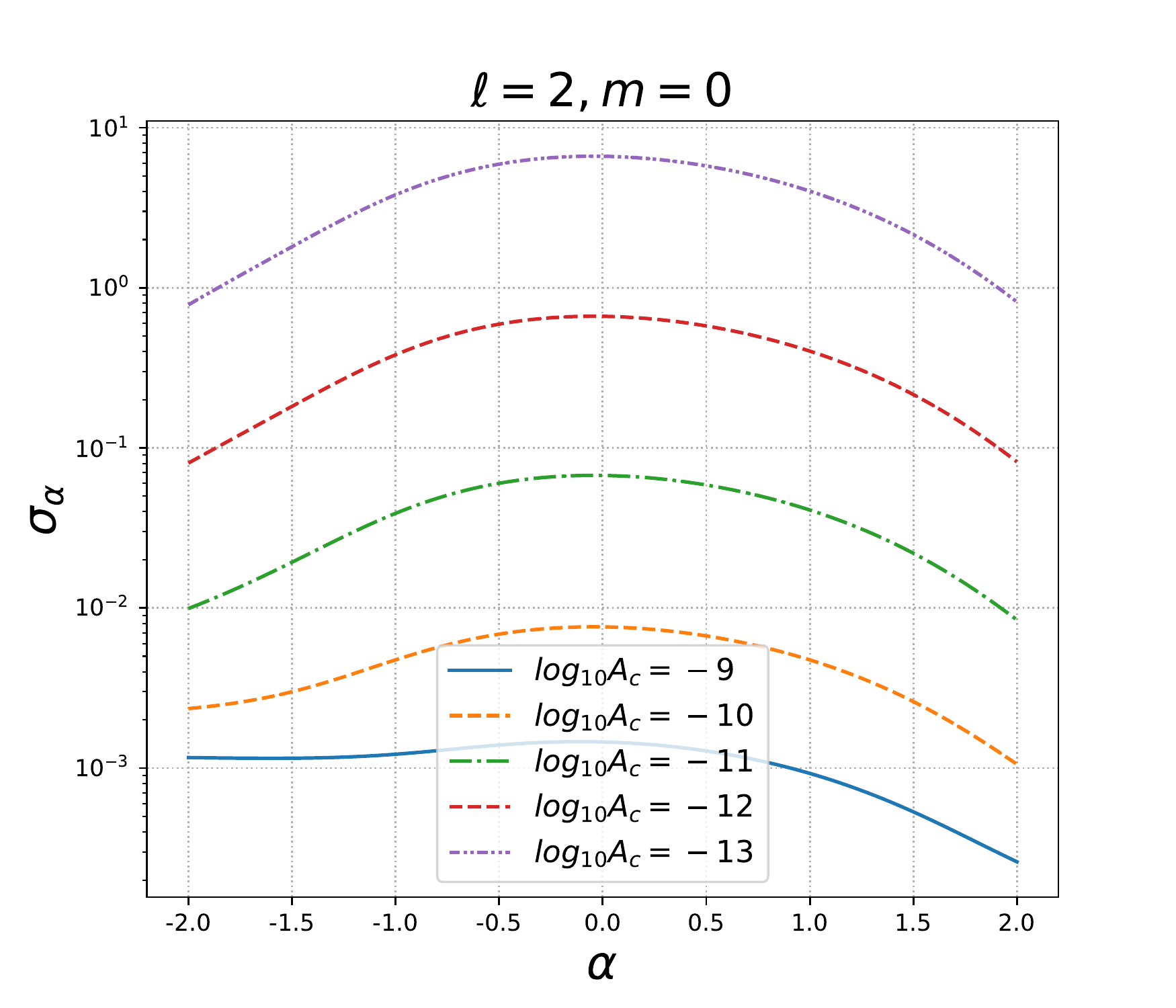}}}
    
    {\resizebox{220pt}{165pt}{\includegraphics{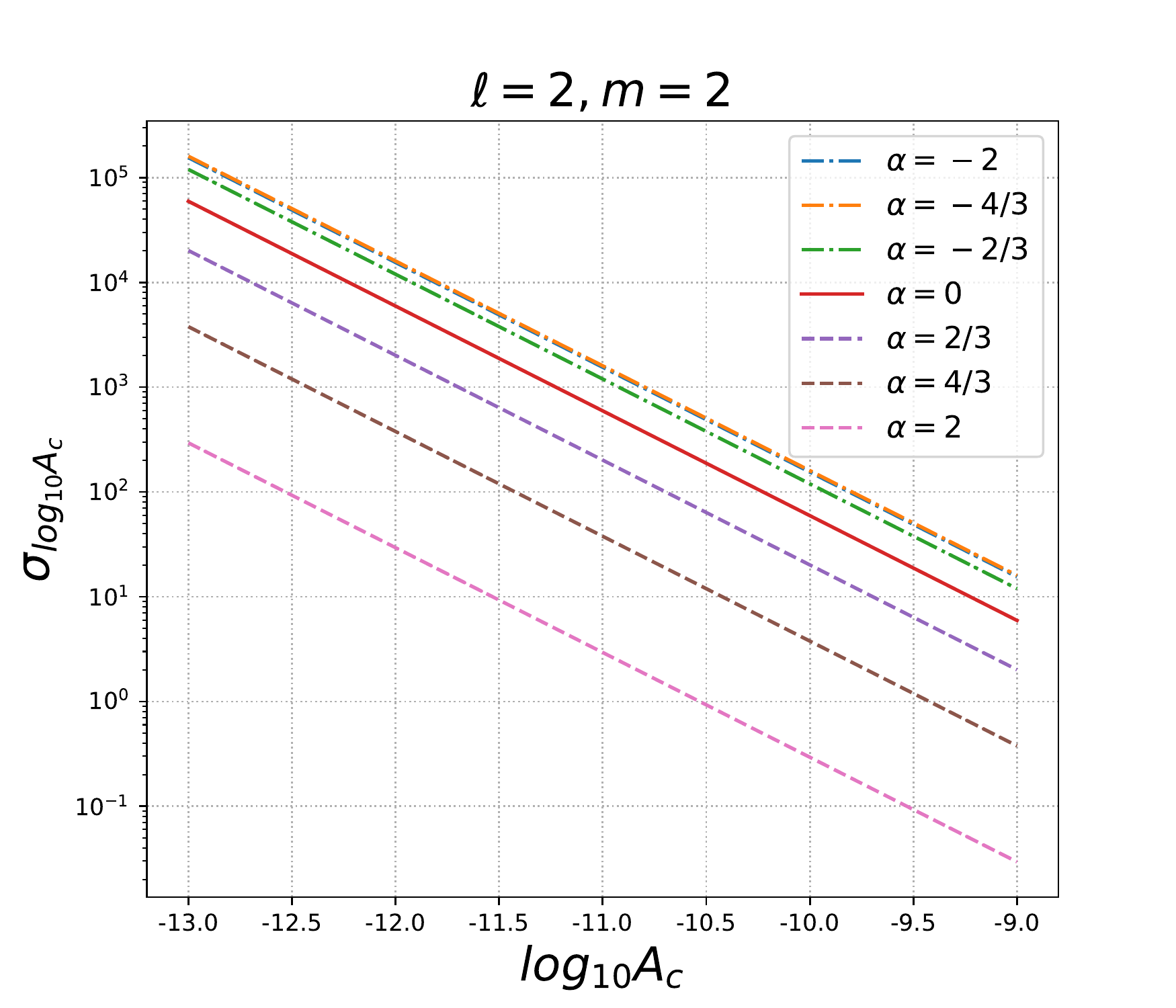}}}{\resizebox{220pt}{165pt}{\includegraphics{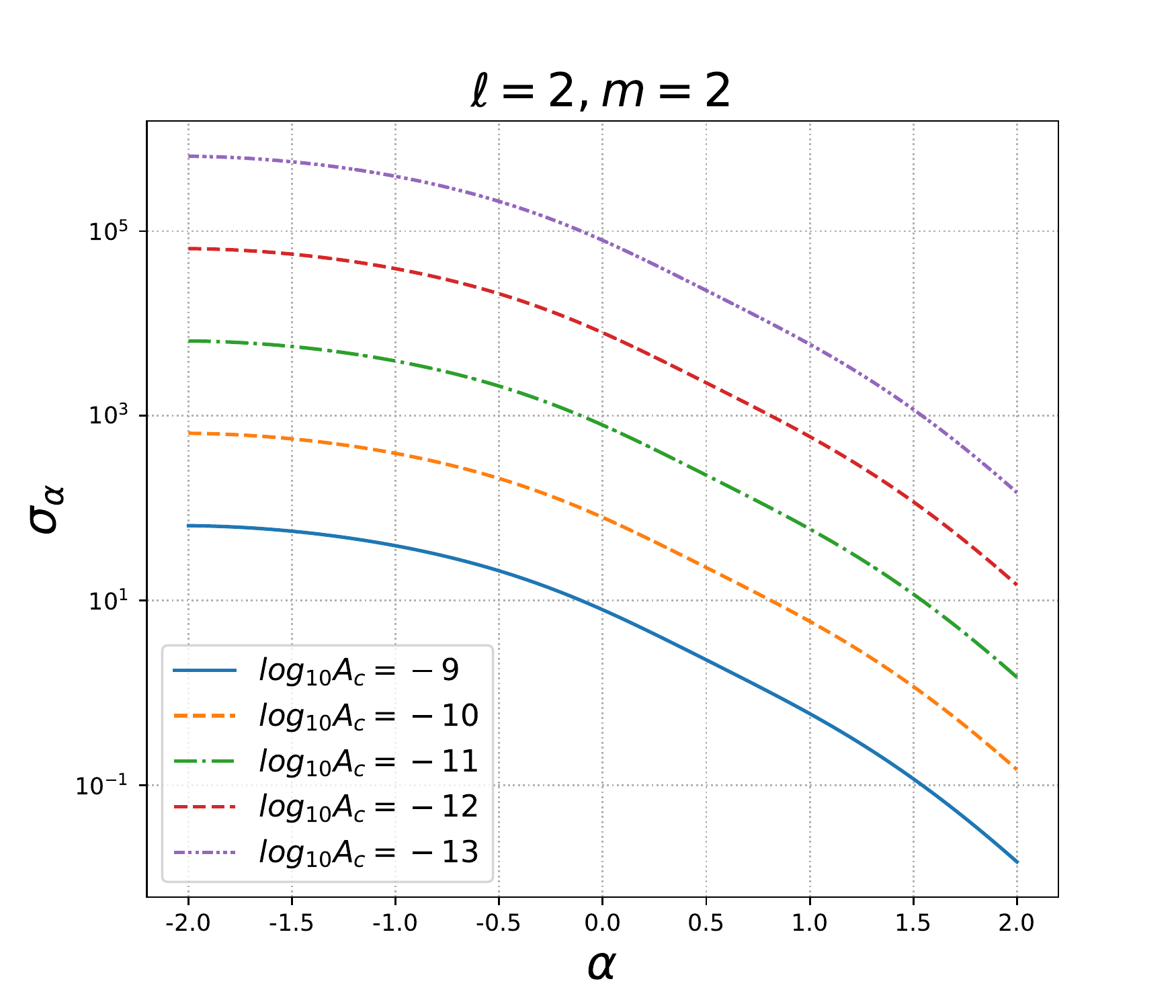}}}
  \end{center}
  \caption{\label{fig:forecasted_sigma_lm}TianQin provides a range of reference values for the SGWB amplitude and spectral index, and the marginalized $1\sigma$ prediction limits for $\log_{10} A_{\mathrm{c}}$ (left) and $\alpha$ (right) are obtained (see equations \eqref{eq:fisher:signal-lm} and \eqref{eq:Cthetaphi-lm}).}
\end{figure*}

Where $\theta$ and $\rho$ are the combined signal model parameters $\log_{10} A_{\mathrm{c}}$ and $\alpha$, the corresponding partial derivatives are given by:
\begin{eqnarray}
  \partial_{\log_{10} A_{\mathrm{c}}} \Omega_{\tmop{GW}}^{\ell m} h^2 &=& \log
  (10) \Omega_{\tmop{GW}}^{\ell} h^2, \nonumber\\
 \partial_{\alpha} \Omega_{\tmop{GW}}^{\ell m} h^2 &=& \log \left( \frac{f}{f_c} \right)
  \Omega_{\tmop{GW}}^{\ell} h^2 . \label{eq:partial-wgwlm}    
\end{eqnarray}
Similar to the previous analysis, for $(\ell, m) = (0, 0), (2, 0)$, as shown in Figure \ref{fig:forecasted_sigma_lm}, the logarithmic amplitude can be determined independently of the sign of the spectral index due to the choice of pivot frequency coinciding with the sensitivity peaks of these two multipoles. On the other hand, for $(\ell, m) = (1, 1), (2, 2)$, a positive spectral index enhances the determination of the amplitude. This is because their corresponding sensitivity peaks occur at slightly higher frequencies relative to $f_c$. Additionally, as shown in Figure \ref{fig:TianQin-sensitivity-lm}, due to the smaller slope at high frequencies for $(\ell, m) = (0, 0), (2, 0)$, $(\ell, m) = (1, 1), (2, 2)$ is closer to the high-frequency noise spectrum compared to the power-law spectrum, especially for lower values of $|\alpha|$.

Similarly to the previous analysis, the higher the logarithmic amplitude, the more effectively the spectral index can be determined for all multipoles. In the optimal case of a signal amplitude of $\Omega_{\mathrm{GW}}(f = f_{\mathrm{c}})h^2 = 10^{-9}$, the spectral index can be determined with uncertainties of approximately $10^{-3}$, $10$, $10^{-3}$, and $10$ for $(\ell, m) = (0, 0)$, $(1, 1)$, $(2, 0)$, and $(2, 2)$, respectively. In the more pessimistic scenario of $\Omega_{\mathrm{GW}}(f = f_{\mathrm{c}})h^2 = 10^{-13}$, for most positive or negative values of the spectral index, the uncertainties in determining the spectral index can exceed $0.1$ or higher for $\ell = 0, 2$. However, for SGWB spectra with spectral indices between $-1$ and $1$, the corresponding uncertainties are on the order of $10$. Therefore, in this low-amplitude scenario, the detector will be more sensitive to models with SGWB spectra that exhibit strong variations. Furthermore, due to the same reasons mentioned earlier, the cases of $(\ell, m) = (1, 1)$ and $(2, 2)$ are more sensitive to positive spectral indices, while the cases of $(\ell, m) = (0, 0)$ and $(2, 0)$ have nearly equal and symmetric precision.

\section{Summary and Discussion}\label{s:Sum_Disc}

We have investigated the unique observational capabilities of the TianQin mission using its $AET$ channel, pointing towards J0806 as the design target. We have calculated the corresponding anisotropic overlap reduction functions (ORFs) and sensitivity curves for different multipole moments, and performed parameter estimation using Fisher analysis for power-law spectra of the multipole moments.

For the TianQin $AET$ channel, considering the anisotropy, the auto-correlation ORFs (such as AA and EE) and cross-correlation ORFs (such as AE and AT) are not necessarily zero. For certain orders, the ORFs are indeed zero, indicating that the detector is insensitive to those specific multipole moments. The ORFs differ for different channel combinations, with the strongest being the auto-correlation of AA and EE, and the weakest being the auto-correlation of TT. TianQin is capable of detecting the multipole moments of the SGWB for different values of $\ell$, with angular sensitivities reaching as low as $10^{-10}$ for quadrupole moments. Due to the orientation of the TianQin mission with the $z$ axis pointing towards J0806, TianQin can distinguish between different $\ell m$ multipole moments, with the best sensitivity achieved for the (2,0) mode reaching $10^{-10}$.

From the perspective of Fisher estimation, considering power-law spectra with $\ell$ multipole moments, in the optimal case with a signal amplitude of $\Omega_{\mathrm{GW}}(f = f_{\mathrm{c}})h^2 = 10^{-9}$, we can estimate the spectral index with uncertainties of $10^{-3}$, $10$, and $10^{-3}$ for $\ell = 0$, $1$, and $2$ multipole moments, respectively. For the $\ell = 1$ case, a positive spectral index enhances the estimation of the logarithmic amplitude. Similarly, considering power-law spectra with $\ell m$ multipole moments, in the optimal case with a signal amplitude of $\Omega_{\mathrm{GW}}(f = f_{\mathrm{c}})h^2 = 10^{-9}$, we can estimate the spectral index with uncertainties of $10^{-3}$, $10$, $10^{-3}$, and $10$ for $(\ell, m) = (0, 0)$, $(1, 1)$, $(2, 0)$, and $(2, 2)$, respectively. For $(\ell, m) = (1, 1)$ and $(2, 2)$, a positive spectral index enhances the estimation of the logarithmic amplitude. Moreover, for all multipole moments, higher logarithmic amplitudes allow for more effective estimation of the spectral index.

In future work, considering that TianQin TDI includes multiple channels besides the $AET$ channel, it would be worthwhile to explore the detection of anisotropy using other TDI channels. Additionally, if we consider the joint detection of multiple detectors, the methodology presented in this paper can be extended to evaluate the detection capabilities of ``TianQin I+II" and ``TianQin+LISA" configurations.
Regarding the type of SGWB, this study focused on a steady-state, Gaussian, non-polarized, and isotropic background. However, by modifying some of these assumptions, such as considering polarized or non-Gaussian backgrounds, it may lead to the discovery of further interesting phenomena and results.

\begin{acknowledgments}

The authors would like to thank Jiandong Zhang and Changfu Shi for useful conversations.

\end{acknowledgments}

\appendix

\section{TianQin's Noise\label{sec:TDI_Noise}}

\subsection{Spectral Density and Power Spectra}\label{sec:power_spectra}

Assuming that all discernible signals have been removed from the data stream $d(t)$, the remaining data stream is purely stochastic, meaning that the residuals are perfect. After this operation, the data stream $d(t)$ will only contain the noise $n(t)$ and the residual random signal $s(t)$. Furthermore, we assume that both the noise and residual signal are stationary, meaning that they have the same statistical properties throughout the observation period. Therefore, we have:
\begin{equation}
d(t) = s(t) + n(t).
\end{equation}
Due to the periodic pointing and intermittent operation of the detector, it is necessary to decompose the data stream into segments of length $T$. For comparison with the research on LISA \cite{Caprini:2019pxz}, a segment length of $T = 11.5$ days can be chosen.

In practical observations, the data stream $d(t) = s(t) + n(t)$ is sampled at a finite rate and can be modeled as a real-valued function over the interval $[-T/2, T/2]$. We assume that the signal and noise are uncorrelated, meaning that
\begin{equation}
  \langle s (t) n (t) \rangle = 0,
\end{equation}
Therefore, we have
\begin{equation}
  \langle d (t) d (t) \rangle = \langle s (t) s (t) \rangle + \langle n (t) n
  (t) \rangle,
\end{equation}
allowing us to treat the signal and noise components separately. We can apply the short-time Fourier transform to the signal part, given by
\begin{equation}
  \tilde{s} (f) = \int_{- T / 2}^{T / 2} \mathrm{d}t \hspace{0.27em} \mathrm{e}^{2 \pi
  ift} s (t) .
\end{equation}
As a result, we have
\begin{equation}
  \hspace{-1cm} \langle \tilde{s}_i (f) \tilde{s}_j^{\ast} (f') \rangle \equiv
  \frac{1}{2} \delta (f - f') S_{ij} (f) \hspace{0.27em},
\end{equation}
where the indices $i, j = {X, Y, Z}$ or ${A, E, T}$, and the power spectral density $S_{ij}(f)$ can be represented by a Hermitian matrix. The matrix elements can be expressed as the product of the response function $\mathcal{R}_{ij}$ and the power spectrum $P_h(f)$, i.e.,
\begin{equation}
  S_{ij} (f) =\mathcal{R}_{ij} (f) P_h (f) \hspace{0.27em} .
  \label{eq:two_signal_correlators}
\end{equation}
Similarly, for the noise component, we also apply the short-time Fourier transform:
\begin{equation}
  \tilde{n} (f) = \int_{- T / 2}^{T / 2} \mathrm{d}t \hspace{0.27em} \mathrm{e}^{2 \pi
  ift} n (t) .
\end{equation}
For stationary and real-valued noise $n(t)$, we have
\begin{equation}
  \hspace{-1cm} \langle \tilde{n}_i (f) \tilde{n}_j^{\ast} (f') \rangle =
  \frac{1}{2} \delta (f - f') N_{ij} (f) \hspace{0.27em},
  \label{eq:singlesidednps}
\end{equation}
where $N_{ij}(f)$ is the one-sided noise power spectral density, also represented by a Hermitian matrix. In the literature, the function $N_{ij}(f)$ is often denoted as $P_n$, and since $\tilde{n}(f)$ and $\delta$ have units of Hz$^{-1}$, $N_{ij}(f)$ has units of Hz$^{-1}$.

\subsection{Noise Model of TianQin}\label{sec:noise_and_signal_model}

Up to this point, the discussion on the power spectral density (PSD) of signals and noise is generally applicable to any gravitational wave interferometer. Now, we specifically focus on TianQin and start with the noise model.
The Time Delay Interferometry (TDI) technique in TianQin aims to mitigate the dominant noise sources caused by the laser frequency fluctuations at the central frequency and the displacements of the optical platforms. In this simplified model, the residual noise components entering each TDI channel can be divided into two effective quantities, referred to as the ``Interferometric Measurement System" (IMS) noise. For example, it includes granular noise and ``acceleration" noise correlated with random displacements of the test masses, such as those caused by local environmental disturbances.

For TianQin, the power spectra of IMS noise and acceleration noise can be obtained from the literature {\cite{TianQin:2015yph}}. The IMS noise power spectrum is given by
\begin{equation}
  P^{\tmop{TQ}}_{\mathrm{I}}  (f, P_{\tmop{TQ}}) = P_{\tmop{TQ}}^2 
  \frac{\mathrm{pm}^2}{\mathrm{Hz}} \left( \frac{2 \pi f}{c} \right)^2
  \hspace{0.27em}, \label{eq:IMS_int_noise-tq}
\end{equation}
and the acceleration noise power spectrum is given by
\begin{eqnarray}
  P^{\tmop{TQ}}_{\mathrm{ac}}  (f, A_{\tmop{TQ}}) &=& A_{\tmop{TQ}}^2 
  \frac{\mathrm{fm}^2}{\mathrm{s}^4 \mathrm{Hz}} \nonumber\\
  & & \left[ 1 + \frac{0.1
  \mathrm{mHz}}{f} \right] \left( \frac{1}{2 \pi f} \right)^4 \left( \frac{2
  \pi f}{c} \right)^2 \hspace{0.27em} .\nonumber\\
  &  & \label{eq:acc_int_noise-tq}    
\end{eqnarray}
where $P_{\text{TQ}} = A_{\text{TQ}} = 1$.

\begin{figure*}
  \begin{center}
    {\resizebox{220pt}{165pt}{\includegraphics{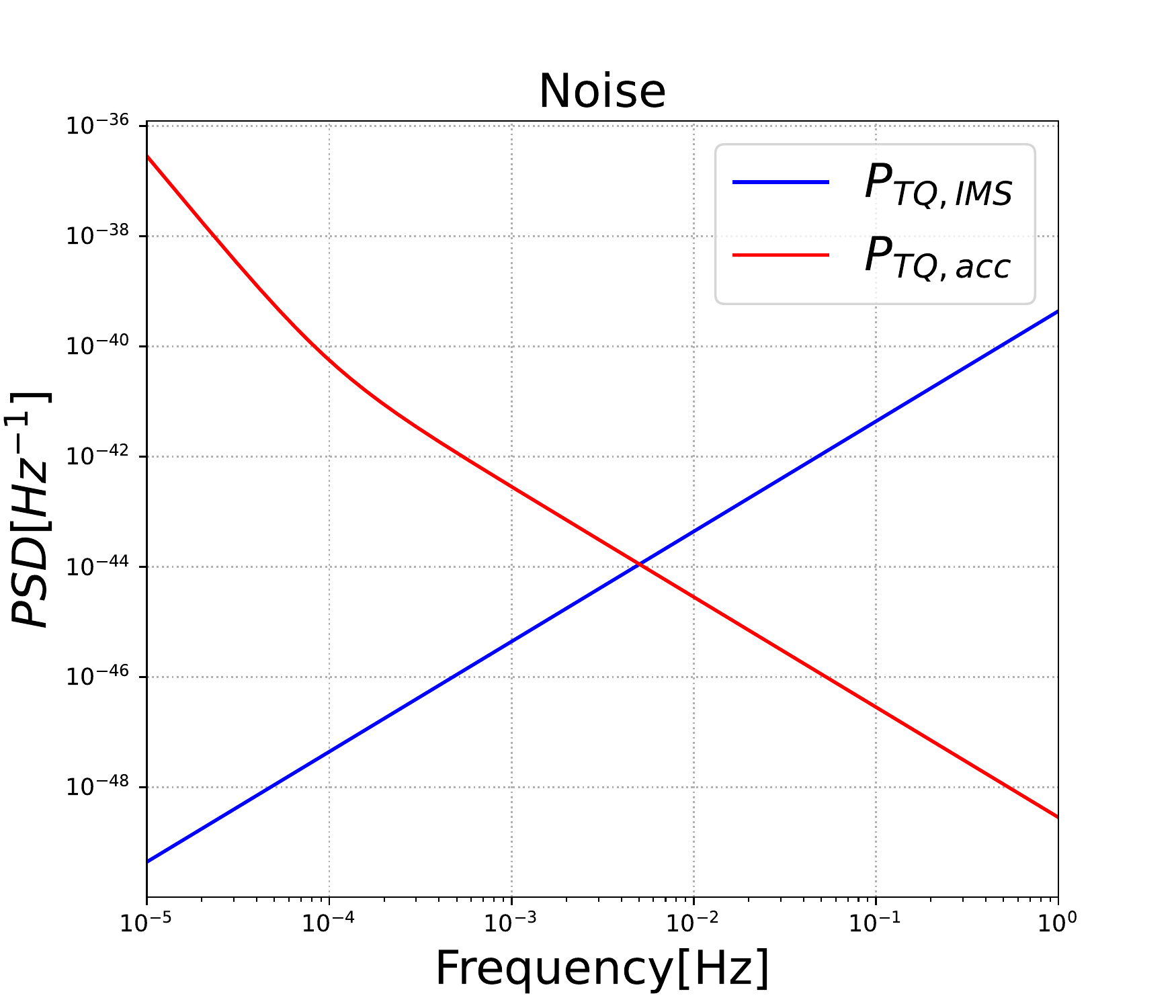}}}{\resizebox{220pt}{165pt}{\includegraphics{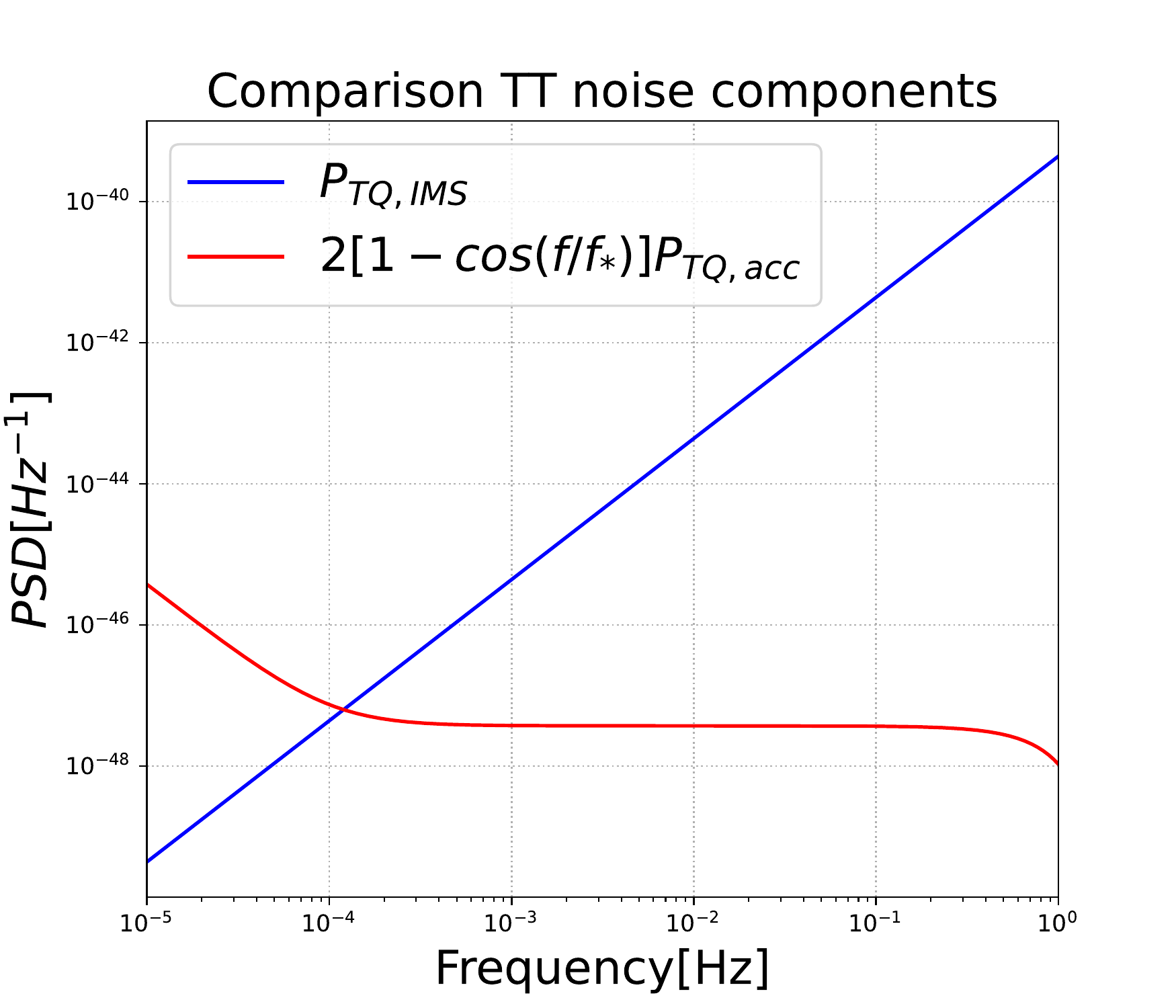}}}
    
  \end{center}
  \caption{\label{fig:PSD_comparisons}On the left-hand side, the power spectra of IMS noise and acceleration noise are given by equations \eqref{eq:IMS_int_noise-tq} and \eqref{eq:acc_int_noise-tq}, respectively. On the right-hand side, the contributions of IMS noise and acceleration noise to $N_{\text{TT}}$ are calculated by weighting them with the factor from equation \eqref{eq:psdTT}, where $P_{\text{TQ}} = A_{\text{TQ}} = 1$.}
\end{figure*}

The figure below (Figure \ref{fig:PSD_comparisons}) shows the $P_{\mathrm{I}}$ and $P_{\mathrm{ac}}$ curves for TianQin.

The precise determination of noise properties remains a major technical challenge for detectors like TianQin, and achieving a thorough understanding of the noise is still an ambitious goal given the current state of research. To simplify the analysis of IMS noise contributions, it is often assumed that the noise spectra of all links are identical, stationary, and uncorrelated. Similarly, for acceleration noise, it is assumed that the fluctuations in detector masses are isotropic, stationary, and uncorrelated, and that the power spectra of different detector masses are equal. Additionally, it is assumed that the three satellites form an equilateral triangle, i.e., $L_1 = L_2 = L_3 = L$ and $L_{\mathrm{TQ}} = \sqrt{3} \times 10^5$ m. Under these assumptions, according to the literature \cite{Flauger:2020qyi}, the total power spectral density of auto-correlated noise for the $XYZ$ channels can be written as:
\begin{widetext}
\begin{equation}
  N_{aa}  (f, A, P) = 16 \sin^2 \left( \frac{f}{f_{\ast}} \right)  \left\{
  \left[ 3 + \cos \left( 2 \frac{f}{f_{\ast}} \right) \right] P_{\mathrm{ac}}
  (f, A) + P_{\mathrm{I}} (f, P) \right\} \hspace{0.27em}, \label{eq:N_XX}
\end{equation}
\end{widetext}
where $a, b \in {\text{X, Y, Z}}$ and $a \neq b$, and the cross-correlation spectra of the noise are given by:
\begin{widetext}
\begin{equation}
\begin{array}{l}
N_{ab}  (f, A, P) = - 8 \sin^2 \left( \frac{f}{f_{\ast}} \right) \cos \left(
\frac{f}{f_{\ast}} \right) [4 P_{\mathrm{ac}} (f, A) + P_{\mathrm{I}} (f,
P)] \hspace{0.27em}, \label{eq:N_XY}
\end{array}
\end{equation}
\end{widetext}
where $a, b \in {\text{X, Y, Z}}$ and $a \neq b$.

\begin{remark}
Here, it is assumed that the covariance matrix elements of the noise are real functions.
\end{remark}

As will be elaborated below, the amplitudes of the IMS noise and acceleration noise power spectra are neglected in the analysis, and the same functional form as the noise model used to generate simulated data is employed. Therefore, any deviation between the instrument noise and the functional form of the noise model used for fitting the data needs to be closely monitored, as it may introduce biases.

\subsection{Noise in the $AET$ Channel of TianQin}\label{sec:three_channels}

\begin{figure*}
  \begin{center}
     {\resizebox{220pt}{165pt}{\includegraphics{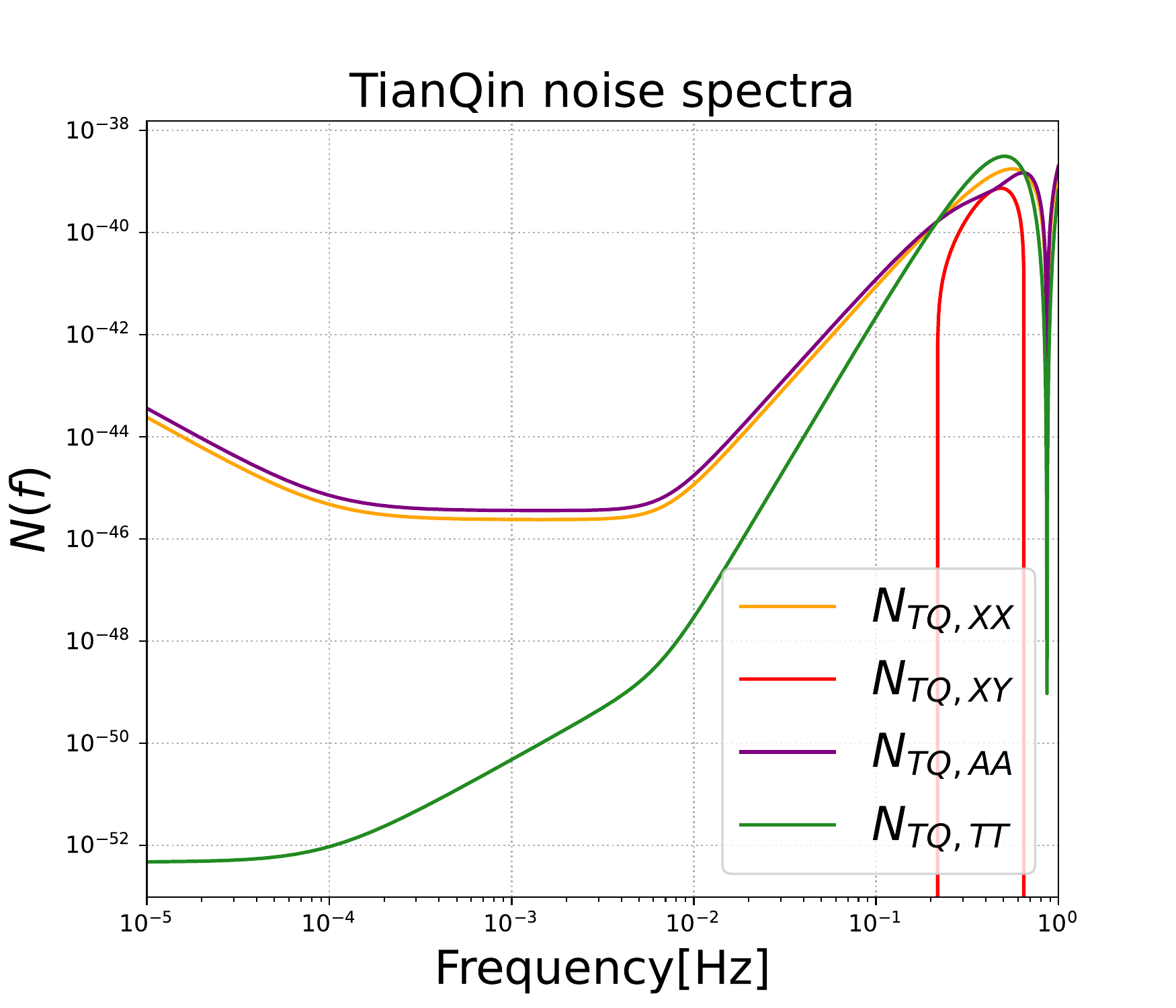}}}{\resizebox{220pt}{165pt}{\includegraphics{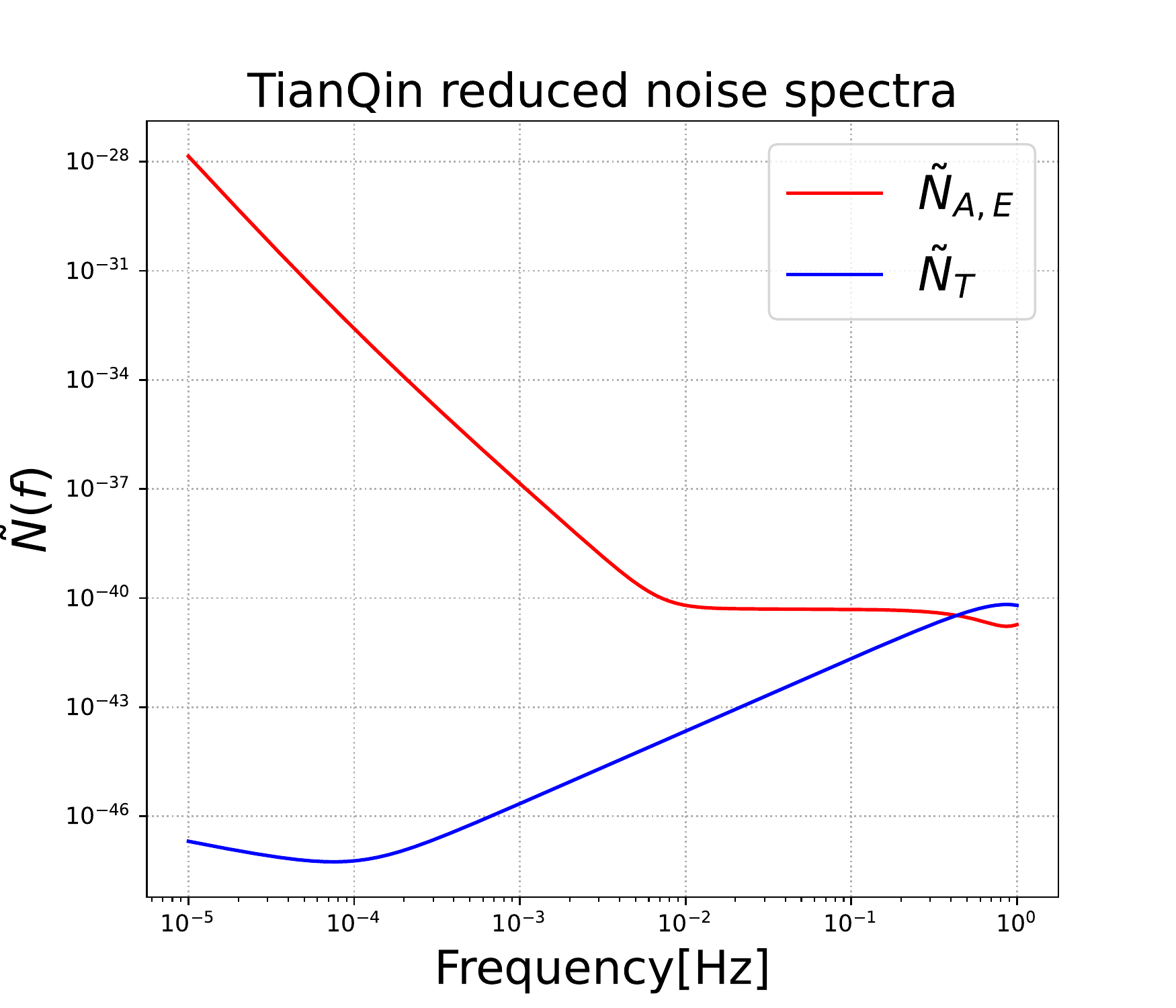}}} 
  \end{center}
  \caption{\label{fig:PSD_strain}The noise spectrum under $XYZ$ and AET channels of TianQin and LISA.}
\end{figure*}

In Section \ref{sec:power_spectra}, we defined the noise power spectra for arbitrary channels, and in Section \ref{sec:noise_and_signal_model}, we introduced the noise power spectra for the $XYZ$ channels of TianQin. Now we will discuss the noise in another commonly used channel, the $AET$ channel, which is related to the $XYZ$ channels through the transformation given by Equation \eqref{DFO}.
Following the analysis in Section \ref{sec:power_spectra}, for a stationary and isotropic SGWB signal and stationary noise (uncorrelated with the signal), we can read the autocorrelation and cross-correlation spectra for different channels as:
\begin{equation}
  \langle \tilde{d}_i  \tilde{d}^{\ast}_j \rangle' =\mathcal{R}_{ij} P_h (f) +
  N_{ij} (f) \hspace{0.27em}, \label{eq:data_correlation_app_XYZ}
\end{equation}
Here, for simplicity, we have removed the frequency delta function and the factor of $1/2$, and we have used Equation \eqref{eq:two_signal_correlators} and Equation \eqref{eq:singlesidednps}. According to the reference {\cite{Flauger:2020qyi}}, we have the following relations:
\begin{equation}
  N_{A A} = N_{E E} = N_{X X} - N_{X Y}
  \hspace{0.27em}, \qquad N_{T T} = N_{X X} + 2 N_{X Y}
  \hspace{0.27em}, \label{eq:responses_PSD}
\end{equation}
\begin{equation}
  \mathcal{R}_{A A} =\mathcal{R}_{E E} =\mathcal{R}_{X X}
  -\mathcal{R}_{X Y} \hspace{0.27em}, \qquad \mathcal{R}_{T T}
  =\mathcal{R}_{X X} + 2\mathcal{R}_{X Y} \hspace{0.27em} .
\end{equation}
By substituting Equations \eqref{eq:N_XX} and \eqref{eq:N_XY} into Equation \eqref{eq:responses_PSD}, we can obtain the autocorrelation and cross-correlation spectra for the $AET$ channel:
\begin{widetext}
\begin{eqnarray}
  N_{A A}  (f, A, P) & = & N_{E E}  (f, A, P) \nonumber\\
  & = & 8 \sin^2 \left( \frac{f}{f_{\ast}} \right)  \left\{ 4 \left[ 1 + \cos
  \left( \frac{f}{f_{\ast}} \right) + \cos^2 \left( \frac{f}{f_{\ast}} \right)
  \right] P_{\mathrm{ac}} (f, A) + [ 2 + \cos \left( \frac{f}{f_{\ast}} \right)
  \right] P_{\mathrm{I}} (f, P) ,  \label{eq:psdAA}
\end{eqnarray}
and
\begin{equation}
  N_{T T}  (f, A, P) = 16 \sin^2 \left( \frac{f}{f_{\ast}} \right) 
  \left\{ 2 \left[ 1 - \cos \left( \frac{f}{f_{\ast}} \right) \right]^2
  P_{\mathrm{ac}} (f, A) \hspace{0.27em} + \left[ 1 - \cos \left(
  \frac{f}{f_{\ast}} \right) \right] P_{\mathrm{I}} (f, P) \right\}
  \label{eq:psdTT},
\end{equation}
\end{widetext}
where $f_{\ast} = \frac{1}{2 \pi
L}$ is the characteristic frequency of the detector. These power spectral densities are shown in Figure \ref{fig:PSD_strain}.

By substituting Equations \eqref{eq:IMS_int_noise-tq} and \eqref{eq:acc_int_noise-tq} into Equations \eqref{eq:psdAA} and \eqref{eq:psdTT}, we can obtain the noise functions for TianQin and LISA in the A and E channels.
For the $A$ and $E$ channels:
\begin{widetext}
\begin{eqnarray}
  \tilde{N}_{A, E}^{\tmop{TQ}} & \equiv & \frac{N^{\tmop{TQ}}_{A, E}}{4 \left(
  \frac{f}{f_{\ast}} \right)^2 |U (f) |^2} \nonumber\\
  & = & \frac{1}{2}  [2 + \cos (\frac{f}{f_{\ast}})] 
  \frac{P_{\tmop{TQ}}^2}{L_{\tmop{TQ}}^2}  \frac{pm^2}{\mathrm{Hz}}
  + 2 [1 + \cos (\frac{f}{f_{\ast}}) + \cos^2 (\frac{f}{f_{\ast}})] 
  \frac{A_{\tmop{TQ}}^2}{L_{\tmop{TQ}}^2}  \frac{fm^2}{\mathrm{s}^4
  \mathrm{Hz}}  \left[ 1 + \left( \frac{0.1 \mathrm{mHz}}{f} \right)^2 \right]
  \left( \frac{1}{2 \pi f} \right)^4 \hspace{0.27em}, \nonumber\\
  &  &  \label{NA,ETQ}
\end{eqnarray}
For the $T$ channel:
\begin{eqnarray}
  \tilde{N}_T^{\tmop{TQ}} & \equiv & \frac{N^{\tmop{TQ}}_{T T}}{4 \left(
  \frac{f}{f_{\ast}} \right)^2 |U (f) |^2} \nonumber\\
  & = & [1 - \cos (\frac{f}{f_{\ast}})] 
  \frac{P_{\tmop{TQ}}^2}{L_{\tmop{TQ}}^2}  \frac{\tmop{pm}^2}{\mathrm{Hz}}  + 2 [1 - \cos (\frac{f}{f_{\ast}})]^2 
  \frac{A_{\tmop{TQ}}^2}{L_{\tmop{TQ}}^2}  \frac{\tmop{fm}^2}{\mathrm{s}^4
  \mathrm{Hz}}  \left[ 1 + \left( \frac{0.1 \mathrm{mHz}}{f} \right)^2 \right]
  \left( \frac{1}{2 \pi f} \right)^4 \hspace{0.27em}, \nonumber\\
  &  &  \label{NTTQ}
\end{eqnarray}
\end{widetext}
Please refer to the reference {\cite{Flauger:2020qyi}} for detailed discussions on these quantities.

\
\nocite{*}
\bibliography{REF_DASGWB_TQ}
\end{document}